%% file: main.tex
\documentclass[a4paper,oneside,11pt]{article}
\usepackage{cprform}
\usepackage{amsmath,amstext,amsfonts,amsbsy,amssymb,amscd,bbm,epsfig,lscape,slashed,color}

\usepackage{booktabs}
\usepackage{multirow}
\usepackage{bbold}
\usepackage{epsfig,cite}
\usepackage{amssymb}
\usepackage{subfigure}
\usepackage{hyperref}
\usepackage{cancel}
\usepackage[font={small}]{caption}
\usepackage[margin=1.1in]{geometry}

\usepackage{tikz}
\usetikzlibrary{arrows,shapes}
\usetikzlibrary{trees}
\usetikzlibrary{matrix,arrows}                          
\usetikzlibrary{positioning}                            
\usetikzlibrary{calc,through}                           
\usetikzlibrary{decorations.pathreplacing}  
\usetikzlibrary{decorations.pathmorphing}       
\usetikzlibrary{decorations.markings}
\tikzset{
    vector/.style={decorate, decoration={snake}, draw},
        provector/.style={decorate, decoration={snake,amplitude=2.5pt}, draw},
        antivector/.style={decorate, decoration={snake,amplitude=-2.5pt}, draw},
    fermion/.style={draw=black, postaction={decorate},
        decoration={markings,mark=at position .55 with {\arrow[draw=black]{>}}}},
    fermionbar/.style={draw=black, postaction={decorate},
        decoration={markings,mark=at position .55 with {\arrow[draw=black]{<}}}},
    fermionnoarrow/.style={draw=black},
    gluon/.style={decorate, draw=black,
        decoration={coil,amplitude=4pt, segment length=5pt}},                           
    scalar/.style={dashed,draw=black, postaction={decorate},
        decoration={markings,mark=at position .55 with {\arrow[draw=black]{>}}}},
    scalarbar/.style={dashed,draw=black, postaction={decorate},
        decoration={markings,mark=at position .55 with {\arrow[draw=black]{<}}}},
    scalarnoarrow/.style={dashed,draw=black},
    electron/.style={draw=black, postaction={decorate},
        decoration={markings,mark=at position .55 with {\arrow[draw=black]{>}}}},
        bigvector/.style={decorate, decoration={snake,amplitude=4pt}, draw},
}

\newcommand{\texp}{{\rm T}\kern-2pt\exp}

\input{macros.tex}

\begin{document}

\input title.tex

\input intro.tex

\input renorm.tex
\input sf.tex
\input latt.tex

\input concl.tex

\newpage
\begin{appendix}
\input app_check.tex

\input app_cutoff.tex

\end{appendix}
\bibliographystyle{JHEPjus}
\bibliography{biblio}
\end{document}

%% file: macros.tex
\newcommand{\ba}{\begin{array}}
\newcommand{\ea}{\end{array}}

\newcommand{\req}[1]{Eq.~(\ref{#1})}

\newcommand{\refig}[1]{Fig.~\ref{#1}}

\newcommand{\dif}{{\rm d}}

\newcommand{\Dslash}{\relax{\kern+.25em / \kern-.70em D}}

\newcommand{\Real}{\relax{\mathsf{\Gamma\kern-.35em R}}}
\newcommand{\Int}{\relax{\mathsf{Z\kern-.40em Z}}}

\newcommand{\half}{{\scriptstyle{{1\over 2}}}}

\newcommand{\ihalf}{{\scriptstyle{{i\over 2}}}}

\newcommand{\NF}{N_\mathrm{\scriptstyle f}}

\newcommand{\SF}{{\rm SF}}

\newcommand{\gbar}{\kern1pt\overline{\kern-1pt g\kern-0pt}\kern1pt}
\newcommand{\mbar}{\kern2pt\overline{\kern-1pt m\kern-1pt}\kern1pt}
\newcommand{\obar}[1]{\kern3pt\overline{\kern-2pt #1\kern-0pt}\kern1pt}
\newcommand{\gren}{g_{\rm R}}
\newcommand{\mren}[1]{m_{{\rm R} #1}}

\newcommand{\Oa}{\mbox{O}(a)}

\newcommand{\icsw}{c_{\rm sw}}
\newcommand{\ict}{c_{\rm t}}

\newcommand{\abar}{\kern1pt\overline{\kern-1pt a\kern-0.5pt}\kern1pt}

\newcommand{\bJ}{{\mathbf{J}}}

\newcommand{\bM}{{\mathbf{M}}}

\newcommand{\bU}{{\mathbf{U}}}

\newcommand{\bW}{{\mathbf{W}}}

\newcommand{\bZ}{{\mathbf{Z}}}

\newcommand{\br}{{\mathbf{r}}}

\newcommand{\cA}{{\cal A}}

\newcommand{\cO}{{\cal O}}

\newcommand{\cQ}{{\cal Q}}

\newcommand{\cS}{{\cal S}}

\newcommand{\cW}{{\cal W}}

\newcommand{\cZ}{{\cal Z}}

\DeclareMathAlphabet\mathbfcal{OMS}{cmsy}{b}{n}

\newcommand{\bcZ}{{\mathbfcal{Z}}}

\newcommand{\bgamma}{{\boldsymbol \gamma}}
\newcommand{\bdelta}{{\boldsymbol \delta}}
\newcommand{\brho}{{\boldsymbol \rho}}
\newcommand{\bsigma}{{\boldsymbol \sigma}}

\newcommand{\bSigma}{{\boldsymbol \Sigma}}

\newcommand{\vx}{\mathbf{x}}
\newcommand{\vy}{\mathbf{y}}
\newcommand{\vz}{\mathbf{z}}

%% file: title.tex

\begin{titlepage}


\vspace*{-30truemm}
\begin{flushright}
IFT-UAM/CSIC-18-005\\
FTUAM-18-2\\[10pt]
{\large January 2018}
\end{flushright}
\vspace{15truemm}


\centerline{\Large Non-Perturbative Renormalisation and Running}
\vskip 3 true pt
\centerline{\Large of BSM Four-Quark Operators in $\NF=2$ QCD}
\vskip 3 true pt
\vskip 7 true mm
\begin{center}
\includegraphics[width=25mm]{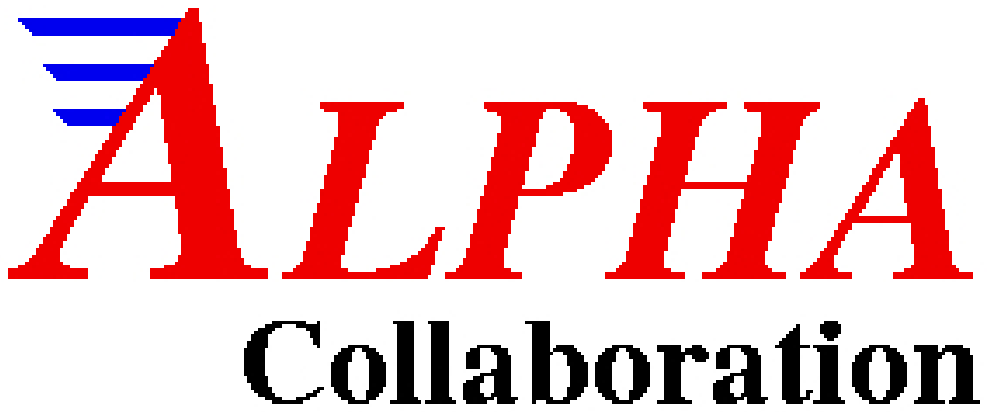}
\end{center}
\centerline{\bigrm  P.~Dimopoulos$^{a,b}$, G.~Herdo\'iza$^{c,d}$, M.~Papinutto$^e$,  C.~Pena$^{c,d}$, D.~Preti$^{c,f}$, and A.~Vladikas$^g$}
\vskip 4 true mm
\centerline{\it $^a$ Centro Fermi - Museo Storico della Fisica e Centro Studi e Ricerche ``Enrico Fermi",}
\centerline{\it Compendio del Viminale, Piazza del Viminale 1, 00184 Rome, Italy}
\vskip 3 true mm
\centerline{\it $^b$ Dipartimento di Fisica, Universit\`a di Roma Tor Vergata,}
\centerline{\it Via della Ricerca Scientifica 1, I-00133 Rome, Italy}
\vskip 3 true mm
\centerline{\it $^c$ Instituto de F\'{\i}sica Te\'orica UAM/CSIC,}
\centerline{\it c/Nicol\'as Cabrera 13-15, Universidad Aut\'onoma de Madrid,}
\centerline{\it Cantoblanco E-28049 Madrid, Spain}
\vskip 3 true mm
\centerline{\it $^d$ Departamento de F\'{\i}sica Te\'orica, Universidad Aut\'onoma de Madrid,}
\centerline{\it Cantoblanco E-28049 Madrid, Spain}
\vskip 3 true mm
\centerline{\it $^e$ Dipartimento di Fisica, ``Sapienza'' Universit\`a di Roma, and INFN, Sezione di Roma}
\centerline{\it Piazzale A. Moro 2, I-00185 Roma, Italy}
\vskip 3 true mm
\centerline{\it $^f$ INFN Sezione di Torino,}
\centerline{\it Via Pietro Giuria 1, I-10125 Turin, Italy}
\vskip 3 true mm
\centerline{\it $^g$ INFN-Sezione di Tor Vergata, c/o Dipartimento di Fisica,}
\centerline{\it Universit\`a di Roma Tor Vergata, Via della Ricerca Scientifica 1,}
\centerline{\it I-00133 Rome, Italy}
\vskip 15 true mm


\noindent{\tenbf Abstract:}
{\tenrm
We perform a non-perturbative study of the scale-dependent
renormalisation factors of a complete set of dimension-six
four-fermion operators. The renormalisation-group (RG) running is
determined in the continuum limit for a specific Schr\"odinger Functional (SF) renormalisation scheme
in the framework of lattice QCD with two dynamical flavours ($\NF = 2$). The theory is regularised on a lattice with
a plaquette Wilson action and $\Oa$-improved Wilson fermions.
For one of these operators, the computation had been performed in ref.~\cite{Dimopoulos:2007ht}; the present work completes the study for the rest of the
operator basis, on the same simulations (configuration ensembles).
The related weak matrix elements arise in several operator product expansions; in $\Delta F = 2$ transitions they contain the QCD
long-distance effects, including contributions from beyond-Standard Model (BSM) processes. Some of these operators mix under renormalisation
and their RG-running is governed by anomalous dimension matrices. In ref.~\cite{Papinutto:2016xpq} the RG formalism 
for the operator basis has been worked out in full generality and the anomalous dimension matrix has been calculated in NLO perturbation theory.
Here the discussion is extended to the matrix step-scaling functions (matrix-SSFs), which are used in finite-size recursive techniques. We rely on these matrix-SSFs to obtain
non-perturbative estimates of the operator anomalous dimensions for scales ranging from $\cO(\Lambda_{\rm QCD})$ to $\cO(M_{\rm W})$.
}
\vspace{10truemm}

\eject
\end{titlepage}

%% file: intro.tex
\section{Introduction}
\label{sec:intro}

In lattice QCD, the renormalisation of composite operators
is an important step towards obtaining estimates of hadronic
low-energy quantities in the continuum limit. Quark masses,
decay constants, form factors, etc. are extracted from matrix elements
of such operators; see ref.~\cite{Aoki:2016frl} for a recent review
of lattice flavour phenomenology. Of interest to the present work is
the class of dimension-six, four-fermion composite fields,
arising in operator product expansions (OPE),
in which the heavier quark degrees of freedom are integrated out.
For $\Delta F = 2$ and many  $\Delta F =1$ transitions
($F$ stands for flavour here), the resulting weak matrix elements
of these operators govern long-distance QCD effects.
They can be reliably evaluated by applying an intrinsically
non-perturbative approach. Lattice QCD is our regularisation of choice
which, by combining theoretical
and computational methods, allows for an evaluation of these
quantities with errors that can be reliably estimated and systematically
improved.

Here we address the problem of calculating the renormalisation parameters
and their renormalisation group (RG)-running for the operators 
defined in \req{eq:rel_ops} below. We opt for the lattice regularisation consisting in
the Wilson plaquette gauge action and the $\Oa$-improved Wilson quark action. We renormalise the bare operators in the
Schr\"odinger Functional (SF) renormalisation scheme.
This problem has first been studied with Wilson fermions for the relatively simple 
case of the multiplicatively renormalisable operators $\cQ_1^\pm$
of  \req{eq:rel_ops}, both perturbatively~\cite{Palombi:2005zd} and
non-perturbatively in the quenched approximation~\cite{Guagnelli:2005zc}.
Subsequently results for $\cQ_1^\pm$ have also been obtained 
with $\NF=2$ dynamical sea quarks~\cite{Dimopoulos:2007ht}.
(An analogous study with quenched Neuberger fermions
may be found in ref.~\cite{Dimopoulos:2006ma}.)
Recently the perturbative calculations have been extended in ref.~\cite{Papinutto:2016xpq}
for the rest of the operator basis $\cQ_k^\pm (k=2, \ldots, 5)$ of \req{eq:rel_ops}.
The present is a companion paper of this work,
complementing it by providing non-perturbative results for the operators
$\cQ_k^\pm (k=2, \ldots, 5)$, computed in $\NF=2$ lattice QCD.

As stressed in refs.~\cite{Donini:1999sf,Papinutto:2016xpq}, 
these operators, treated here in full generality, become relevant for a number of
interesting processes, once specific physical flavours are assigned to their fermion fields.
For example, with $\psi_1 = \psi_3 = s$ and $\psi_2 = \psi_4 = d$ (cf. \req{eq:gen_4f}), 
the weak matrix element $\langle \bar K^0 \vert Q_1^+ \vert K^0 \rangle$ comprises 
leading long-distance contributions in the effective Hamiltonian formalism for neutral
$K$-meson oscillations in the Standard Model (SM).
Allowing for beyond-Standard-Model (BSM) interactions introduces similar matrix elements
of the remaining operators $Q_2^+, \ldots, Q_5^+$. In some lattice regularisations,
the corresponding bare matrix elements are expressed in terms of the operators $\cQ_1^+, \ldots, \cQ_5^+$,
with some important simplifications in their renormalisation properties~\cite{Donini:1999sf,
Pena:2004gb,Frezzotti:2004wz}. In refs.~\cite{Papinutto:2016xpq} other flavour
assignments are listed, leading to four-fermion operators related to
the low-energy effects of $\Delta B = 2$ transitions ($B^0$--$\bar
B^0$ and $B_s^0$--$\bar B_s^0$ mixing) and to the
$\Delta S=1$ effective weak Hamiltonian with an active charm quark.
In the present work we will concentrate on the renormalisation
and RG-runing of $\cQ_2^\pm, \ldots, \cQ_5^\pm$.

It is important to keep in mind that the sets $\{ Q_2^\pm, \ldots, Q_5^\pm \}$ and
$\{ \cQ_2^\pm, \ldots, \cQ_5^\pm \}$ are parity-even and parity-odd components
of operators with chiral structures (such as ``left-left'' or ``left-right'') which ensure their transformation
under specific irreducible chiral representations. Chiral symmetry may be broken by the regularisation
(e.g. lattice Wilson fermions) but it is recovered by the continuum theory. An important consequence is
that our results, obtained for the continuum RG-evolution of the parity-odd bases $\{ \cQ_2^\pm, \ldots, \cQ_5^\pm \}$,
are also valid for the parity-even ones $\{ Q_2^\pm, \ldots, Q_5^\pm \}$.

The paper is organised as follows:
In section~\ref{sec:renorm} we list the operators we
are studying and their basic renormalisation pattern. We also
derive their RG-equations, and define the evolution matrices and the renormalisation-group
invariant operators, which are scale- and scheme-independent quantities.
This is an abbreviated version of section~2 of ref.~\cite{Papinutto:2016xpq}. The interesting
feature of the renormalisation pattern of operators $\cQ_k^\pm (k=2,
\ldots, 5)$ is that
they mix in pairs\footnote{This is not to be confused with the operator mixing of $Q_1^\pm$
(the operator arising in $\Delta F = 2$ transitions in the Standard Model), which
mixes with $Q_k^\pm (k=2, \ldots, 5)$ when Wilson lattice fermions are used, and chiral
symmetry is broken by the regularisation.}. In the case of $\cQ_k^\pm
(k=2, \ldots, 5)$, mixing is not an artefact of the lattice
regularisation, as it also happens in schemes where all symmetries of the continuum target
theory (QCD) are preserved; cf. ref.~\cite{Donini:1999sf}. An important consequence of this property is that
the RG-running of these operators is governed by anomalous-dimension and RG-evolution  
{\it matrices}, rather than scalar functions. The RG-evolution matrices are well known in NLO perturbation
theory; cf. refs.~\cite{Ciuchini:1997bw,Buras:2000if}. Here, following ref.~\cite{Papinutto:2016xpq},
we use them in closed form, suitable for non-perturbative evaluations.\\
In section~\ref{sec:sf} we outline our strategy. First we
define the SF renormalisation conditions for the operators $\cQ_k^\pm
(k=1, \ldots, 5)$; again this is
an abridged version of subsection~3.3 of ref.~\cite{Papinutto:2016xpq}. Next, we define 
in the SF scheme
the {\it matrix} step-scaling functions (SSFs) as the RG-evolution matrices for a change of renormalisation scale by a fixed arbitrary factor; this factor is 2 in the present work. These are our basic lattice quantities,
computed for a sequence of lattice spacings, at fixed renormalised gauge coupling. They have a well defined continuum limit, which is obtained by extrapolation, as explained in the same section. Repeating the calculation for a range of renormalisation scales (i.e. a range of renormalised couplings) and interpolating our data points,
we finally have the SSFs as continuous polynomials of the gauge coupling, from which we obtain the anomalous dimension matrices, with NLO perturbation theory taking over only at $\cO(M_{\rm W})$ scales. \\
In section~\ref{sec:latt} we present our results. Besides the aforementioned SSFs, RG-evolution matrices and anomalous-dimension matrices, we also compute the renormalisation matrices for
values of the gauge coupling corresponding to low-energy scales. These renormalisation factors
are needed, in order to renormalise the corresponding bare matrix elements at these hadronic scales.
The computation of the latter requires independent simulations on large physical lattices (of about
3-5 fm), which is beyond the scope of this work.\\
Appendix~\ref{sec:checks} collects additional tests of the comparision
between perturbative and non-perturbative RG evolution, including the
specific renormalisation scale range, [2\,GeV,~3\,GeV], considered
in~ref.~\cite{Boyle:2017skn}. Further details about one-loop cutoff
effects in the SSFs are presented in Appendix~\ref{sec:cutoff}.

%% file: renorm.tex
\section{Renormalisation of four-quark operators}
\label{sec:renorm}

This section is an abridged version of sect.~2 of ref.~\cite{Papinutto:2016xpq},
which we repeat here for completeness. 

\subsection{Renormalisation and mixing of four-quark operators}
\label{subsec:op-bases}

We recapitulate the main renormalisation properties of the four-fermion
operators under study. These results have been obtained
in full generality in ref.~\cite{Donini:1999sf}.
The absence of subtractions
is elegantly implemented by using a formalism in which the operators
consist of quark fields with four distinct flavours. A complete set of Lorentz-invariant
operators is
\begin{gather}
\label{eq:rel_ops}
\ba{l@{}l@{}l@{\hspace{20mm}}l@{}l@{}l}
Q_1^\pm &\,\,=\,\,& \cO^\pm_{\rm VV+AA}\,, \quad &\cQ_1^\pm &\,\,=\,\,& \cO^\pm_{\rm VA+AV}\,,\\[1.0ex]
Q_2^\pm &\,\,=\,\,& \cO^\pm_{\rm VV-AA}\,, \quad &\cQ_2^\pm &\,\,=\,\,& \cO^\pm_{\rm VA-AV}\,,\\[1.0ex]
Q_3^\pm &\,\,=\,\,& \cO^\pm_{\rm SS-PP}\,, \quad &\cQ_3^\pm &\,\,=\,\,& \cO^\pm_{\rm PS-SP}\,,\\[1.0ex]  
Q_4^\pm &\,\,=\,\,& \cO^\pm_{\rm SS+PP}\,, \quad &\cQ_4^\pm &\,\,=\,\,& \cO^\pm_{\rm PS+SP}\,,\\[1.0ex] 
Q_5^\pm &\,\,=\,\,& -2 \,\, \cO^\pm_{\rm TT}\,,    \quad &\cQ_5^\pm &\,\,=\,\,& -2 \,\, \cO^\pm_{\rm T\tilde{T}}\,,
\ea
\end{gather}
where
\begin{gather}
\label{eq:gen_4f}
\cO^\pm_{\Gamma_1\Gamma_2} = \frac{1}{2}\left[
(\bar\psi_1\Gamma_1\psi_2)(\bar\psi_3\Gamma_2\psi_4)\,\pm\,
(\bar\psi_1\Gamma_1\psi_4)(\bar\psi_3\Gamma_2\psi_2)
\right]\,,
\end{gather}
$\cO^\pm_{\Gamma_1\Gamma_2\pm\Gamma_2\Gamma_1}\equiv\cO^\pm_{\Gamma_1\Gamma_2}\pm\cO^\pm_{\Gamma_2\Gamma_1}$.
The operator subscripts obviously correspond to the labelling
${\rm V}\to\gamma_\mu$, ${\rm A}\to\gamma_\mu\gamma_5$, ${\rm S}\to\mathbf{1}$,
${\rm P}\to\gamma_5$, ${\rm T}\to\sigma_{\mu\nu}$, ${\rm \tilde{T}}\to\half\varepsilon_{\mu\nu\rho\tau}\sigma_{\rho\tau}$,
with $\sigma_{\mu\nu}\equiv\ihalf [\gamma_\mu,\gamma_\nu]$. Repeated Lorentz indices, such as $\gamma_\mu \gamma_\mu$ and
$\sigma_{\mu\nu} \sigma_{\mu\nu}$ are summed over.
In the above expression round parentheses indicate spin and colour traces and the subscripts $1, \ldots, 4$ of the fermion fields
are flavour labels.
Note that operators $Q_k^\pm$ are parity-even, and $\cQ_k^\pm$ are parity-odd.

In the following we will assume a mass-independent renormalisation scheme.
Renormalised operators can be written as
\begin{gather}
\label{eq:ren_relativistic}
\begin{split}
\bar Q_k^\pm &= Z_{kl}^\pm(\delta_{lm}+\Delta_{lm}^\pm) Q_m^\pm\,,\\
\bar \cQ_k^\pm &= \cZ_{kl}^\pm(\delta_{lm}+\mbox{\textcyr{D}}_{lm}^\pm)\cQ_m^\pm
\end{split}
\end{gather}
(summations over $l,m$ are implied), where the renormalisation matrices $\bZ^\pm, {\boldsymbol \cZ}^\pm$ are scale-dependent and reabsorb
logarithmic divergences, while ${\boldsymbol \Delta}^\pm,\boldsymbol{\textcyr{\bf D}}^\pm$ are (possible) matrices of finite subtraction coefficients
that only depend on the bare coupling. 
Throughout this work  we use boldface symbols for the column vectors of four-fermion operator and the matrices which act on these vectors
(e.g. ${\boldsymbol Q}, {\boldsymbol \cQ}, {\bZ}, {\boldsymbol
  \Delta}, {\boldsymbol \cZ}, \boldsymbol{\textcyr{\bf D}}$, etc.) while their elements are indicated with explicit indices (e.g. $Q_k, \cQ_k, Z_{kl}, \Delta_{kl}, \cZ_{kl}, \mbox{\textcyr{D}}_{kl}$, etc.).
We also introduce a simplification in our notation, by dropping the $\pm$ superscripts , wherever no
ambiguity arises. This should not be a problem as the symmetric operator bases $\{ Q_k^+ \}$ and$ \{ \cQ_k^+ \}$
(symmetric under flavour exchange $2 \leftrightarrow 4$) never mix with the antisymmetric ones
$\{ Q_k^-\}$ and $\{ \cQ_k^- \}$, and thus equations are valid separately for each basis.

The renormalisation matrices  have the generic structure
\begin{gather}
\label{eq:ren_pattern}
{\bf Z}=\left(\ba{ccccc}
Z_{11} & 0 & 0 & 0 & 0 \\
0 & Z_{22} & Z_{23} & 0 & 0 \\
0 & Z_{32} & Z_{33} & 0 & 0 \\
0 & 0 & 0 & Z_{44} & Z_{45} \\
0 & 0 & 0 & Z_{54} & Z_{55}
\ea\right)\,,\qquad
{\boldsymbol \Delta}=\left(\ba{ccccc}
0 & \Delta_{12} & \Delta_{13} & \Delta_{14} & \Delta_{15} \\
\Delta_{21} & 0 & 0 & \Delta_{24} & \Delta_{25} \\
\Delta_{31} & 0 & 0 & \Delta_{34} & \Delta_{35} \\
\Delta_{41} & \Delta_{42} & \Delta_{43} & 0 & 0 \\
\Delta_{51} & \Delta_{52} & \Delta_{53} & 0 & 0
\ea\right)\,.
\end{gather}
Analogous expressions hold for $\bcZ$ and $\boldsymbol{\textcyr{\bf D}}$.
If chiral symmetry is preserved by the regularisation, both ${\boldsymbol \Delta}$ 
and $\boldsymbol{\textcyr{\bf D}}$ vanish. In the case of Wilson fermions, 
with chiral symmetry explicitly broken, 
we have ${\boldsymbol \Delta} \neq 0$, whereas due to residual discrete
flavour symmetries $\boldsymbol{\textcyr{\bf D}}=0$; this is the 
main result of ref.~\cite{Donini:1999sf}. Therefore the left-left operators
$\cQ_1 = \cO_{\rm VA+AV}$, which mediate Standard Model-allowed 
transitions, renormalise multiplicatively,
while operators $\cQ_2, \ldots , \cQ_5$, which appear as effective interactions
in extensions of the Standard Model, mix in pairs: $\{ \cQ_2, \cQ_3 \}$ and 
$\{ \cQ_4, \cQ_5 \}$.

In conclusion, with Wilson fermions the parity-odd basis $\{ \cQ_k \}$ renormalises in a pattern
analogous to that of a chirally symmetric regularisation, while the parity-even one $\{ Q_k \}$
has a more complicated renormalisation pattern due to the non-vanishing of ${\boldsymbol \Delta}$.
We will henceforth concentrate on the non-perturbative
renormalisation of the parity-odd basis $\{ \cQ_k \}$ with Wilson fermions.

\subsection{Renormalisation group equations}
\label{subsec:op-RG}

The scale dependence of renormalised quantities is governed by renormalisation
group evolution. Denoting as $\mu$ the running momentum scale and $\mu$ the renormalisation
scale where mass-independent renormalisation conditions are imposed, we have the
following Callan-Symanzik equations for the gauge coupling and quark masses respectively:
\begin{align}
\label{eq:betadef}
\mu \frac{\dif}{\dif \mu}\gbar(\mu) &= \beta(\gbar(\mu)) \,,\\
\mu\frac{\dif}{\dif \mu}\mbar_{\rm f}(\mu) &= \tau(\gbar(\mu))\mbar_{\rm f}(\mu) \,,
\end{align}
where ${\rm f}$ is a flavour label. The scheme mass-independence implies that
the Callan-Symanzik function $\beta$ and the mass anomalous dimension $\tau$ depend
only on the coupling. Asymptotic perturbative expansions read
\begin{align}
\label{eq:beta-PT}
\beta(g) &\underset{g \sim 0}{\approx} -g^3(b_0+b_1g^2+\ldots)\,,\\
\tau(g)  &\underset{g \sim 0}{\approx} -g^2(d_0+d_1g^2+\ldots)\,.
\end{align}

Let us now turn to Euclidean correlation functions of gauge-invariant composite operators,
of the form\footnote{To simplify the notation,
we have omitted the dependence of $G_k$ on the coupling, the masses and the UV cutoff (e.g. the lattice spacing).}
\begin{gather}
G_k(x;y_1,\ldots,y_n) = \langle \cQ_k(x) \cO_1(y_1)\cdots\cO_n(y_n))\rangle\,,
\end{gather}
with $x \neq y_j~\forall j,~y_j \neq y_k~\forall j \neq k$. For concreteness we have
opted for correlation functions of the parity-odd operators $\cQ_k$, which are the subject of the present
work. Nevertheless, the results of this section apply to any set of operators that mix under renormalisation.
The operators $\cO_l (l=1, \cdots ,n)$ may be any convenient,  multiplicatively renormalisable source field. For example
they could be currents or densities (e.g. $V_\mu(y), A_\mu(y), S(y)$ and/or $P(y)$), or Schr\"odinger functional
sources at the time-boundaries. The latter will be explicitly discussed in Sect.~\ref{sec:sf}.
Renormalised correlation functions satisfy the system of Callan-Symanzik 
equations
\begin{gather}
\label{eq:rge_compact}
\mu\frac{\dif}{\dif\mu} \bar G_j =
\sum_k\left[\gamma_{jk}(\gren) + \Big ( \sum_{l=1}^n\tilde\gamma_l(\gren) \Big ) \delta_{jk} \right]\bar G_k
\end{gather}
or, expanding the total derivative,
\begin{gather}
\label{eq:rge_detailed}
\left\{
\mu\frac{\partial}{\partial \mu} +
\beta(\gren)\frac{\partial}{\partial\gren} +
\sum_{{\rm f}=1}^{\NF}\tau(\gren)\mren{,\rm f}\frac{\partial}{\partial\mren{,\rm f}} -
\sum_{l=1}^n\tilde\gamma_l(\gren)
\right\}\bar G_j =
\sum_k\gamma_{jk}(\gren)\,\bar G_k \,,
\end{gather}
where $\gamma$ is a matrix of anomalous dimensions describing the mixing of $\{\cQ_k \}$
(cf. \req{eq:rg_op} below),
and $\tilde\gamma_l$ is the anomalous dimension of $\cO_l$ (defined in a way analogous to \req{eq:rg_op}).
A possible term arising from the running of the gauge parameter $\lambda$ of the action
is omitted here, for reasons explained in ref.~\cite{Papinutto:2016xpq}.
A convenient shorthand notation for the anomalous dimension matrix
of the operators $\bar \cQ_k$ is thus
\begin{gather}
\label{eq:rg_op}
\mu\frac{\dif}{\dif \mu}\bar\cQ_j(\mu) = \sum_{k=1}^5\gamma_{jk}(\gbar(\mu)) \bar \cQ_k(\mu) \,.
\end{gather}
The operator anomalous dimensions admit perturbative expansions of the form
\begin{gather}
\label{eq:gamma-PT}
\gamma_{jk}(g) \underset{g \sim 0}{\approx} -g^2(\gamma_{jk}^{(0)}+\gamma_{jk}^{(1)}g^2+\ldots)\,.
\end{gather}
In standard fashion we can then derive
\begin{gather}
\label{eq:rg_Z}
\mu \sum_{l=1}^5 \frac{\dif}{\dif \mu} \cZ_{jl}  (\cZ^{-1})_{lk} = \gamma_{jk} \,.
\end{gather}
This result implies that the block-diagonal form of the renormalisation matrices ${\boldsymbol \cZ}$ (and $\bZ$) of \req{eq:ren_pattern} 
induces the same block-diagonal structure for the anomalous dimension matrix~${\boldsymbol \gamma}$. Thus the sums  in Eqs.~(\ref{eq:rg_op})
and (\ref{eq:rg_Z}) simplify: for operator $\bar \cQ_1$ and its anomalous dimension $\gamma_{11}$ there is no summation;
for operators $\left  \{ \bar \cQ_2, \bar \cQ_3 \right \}$ summations run over indices 2 and 3 only, and similarly for the operator
sub-basis $\left  \{ \bar \cQ_4, \bar \cQ_5 \right \}$.

\subsection{Evolution matrices and renormalisation group invariants}
\label{sec:rgi}

In order to obtain a solution of~\req{eq:rg_op} in closed form, it is convenient to
introduce the renormalisation group evolution matrix $\bU(\mu_2,\mu_1)$
that evolves renormalised operators between scales\footnote{Restricting the evolution operator
to run towards the IR avoids unessential algebraic technicalities below. The running towards
the UV can be trivially obtained by taking $\left[U(\mu_2,\mu_1)\right]^{-1}$.}
$\mu_1$ and $\mu_2<\mu_1$:
\begin{gather}
\label{eq:evol}
\obar{\cQ}_i(\mu_2) = U_{ij}(\mu_2,\mu_1) \obar{\cQ}_j(\mu_1)\,.
\end{gather}
Substituting the above into~\req{eq:rg_op} we obtain for the running of $\bU(\mu_2,\mu_1)$
\begin{gather}
\label{eq:rg_evol}
\mu_2 \, \frac{\dif}{\dif \mu_2} \,\bU(\mu_2,\mu_1) = {\boldsymbol \gamma}[\gbar(\mu_2)]\bU(\mu_2,\mu_1)\,,
\end{gather}
with initial condition $\bU(\mu_1,\mu_1)=\mathbf{1}$. Note that the r.h.s. is a matrix product.
Following a standard procedure, the above expression can be converted into a Volterra-type integral
equation and solved iteratively, viz.
\begin{gather}
\bU(\mu_2,\mu_1) = \texp\left\{
\int_{\gbar(\mu_1)}^{\gbar(\mu_2)}\kern-8pt\dif g\,\frac{1}{\beta(g)}\,{\boldsymbol \gamma}(g)
\right\}\,,
\label{eq:text-closed}
\end{gather}
where as usual the notation $\texp$ denotes a Taylor expansion of the exponent, in which
each term is an ordered (here $g$-ordered) product. Explicitly, for a generic matrix function $\bM(x)$, we have
\begin{gather}
\label{eq:texp}
\begin{split}
\texp\left\{\int_{x_-}^{x_+}\kern-8pt\dif x\,\bM(x)\right\} \equiv \mathbf{1}
&+ \int_{x_-}^{x_+}\kern-8pt\dif x\,\bM(x) \\
&+ \int_{x_-}^{x_+}\kern-8pt\dif x_1\,\bM(x_1)\int_{x_-}^{x_1}\kern-8pt\dif x_2\,\bM(x_2)\\
&+ \int_{x_-}^{x_+}\kern-8pt\dif x_1\,\bM(x_1)\int_{x_-}^{x_1}\kern-8pt\dif x_2\,\bM(x_2)\int_{x_-}^{x_2}\kern-8pt\dif x_3\,\bM(x_3)\\
&+ \ldots\\
= \mathbf{1}
&+ \int_{x_-}^{x_+}\kern-8pt\dif x\,\bM(x) \\
&+ \frac{1}{2!}\int_{x_-}^{x_+}\kern-8pt\dif x_1\int_{x_-}^{x_+}\kern-8pt\dif x_2\,\Big\{\theta(x_1-x_2)\bM(x_1)\bM(x_2)+\\
&~~~~~~~~~~~~~~~~~~~~~~~~~~~~~\theta(x_2-x_1)\bM(x_2)\bM(x_1)\Big\}\\
&+ \ldots
\end{split}
\end{gather}
In the specific case of interest, $\bM(g) = {\boldsymbol \gamma}(g)/\beta(g)$, with ${\boldsymbol \gamma}(g)$ a matrix function and $\beta (g)$ a real function.
To leading order (LO) we have that $\bM(g) = {\boldsymbol \gamma}^{(0)}/(b_0 g)$ and the independence of the matrix ${\boldsymbol \gamma}^{(0)}$ from the coupling $g$
simplifies eq.~(\ref{eq:texp}), so that the $\texp$ becomes a standard exponential.
One can then easily integrate the exponent in eq.~(\ref{eq:text-closed}) and obtain the LO approximation of the evolution matrix:
\begin{gather}
\bU(\mu_2,\mu_1) \underset{\rm LO}{=} \left[\frac{\gbar^2(\mu_2)}{\gbar^2(\mu_1)}\right]^{\frac{{\boldsymbol \gamma}^{(0)}}{2b_0}}
\equiv \bU_{\rm LO}(\mu_2,\mu_1)\,.
\end{gather}
When next-to-leading order corrections are included, the T-exponential becomes non-trivial.
Further insight is gained upon realising that the associativity property of the evolution matrix
$\bU(\mu_3,\mu_1) = \bU(\mu_3,\mu_2) \bU(\mu_2,\mu_1)$ implies that it can actually be factorised
in full generality as
\begin{gather}
\label{eq:Urun}
\bU(\mu_2,\mu_1) = \left[\tilde \bU(\mu_2)\right]^{-1} \tilde \bU(\mu_1)\,,
\end{gather}
and the matrix  $\tilde \bU(\mu)$ can be expressed in terms of a matrix $\bW(\mu)$,
defined through
\begin{gather}
\label{eq:Utilde}
\tilde \bU(\mu) \equiv \left[\frac{\gbar^2(\mu)}{4\pi}\right]^{-\frac{{\boldsymbol \gamma}^{(0)}}{2b_0}}\bW(\mu) \,\, .
\end{gather}
The matrix $\bW$ can be interpreted as the piece of the evolution operator
containing contributions beyond the leading perturbative order.
Putting everything together, we see that
\begin{gather}
\label{eq:def_W}
\bU(\mu_2,\mu_1) \equiv \left[\bW(\mu_2)\right]^{-1}\, \bU_{\rm LO}(\mu_2,\mu_1) \bW(\mu_1)\,,
\end{gather}
and thus we make contact with the literature (see e.g.~\cite{Buras:2000if,Ciuchini:1997bw}).

Upon inserting~\req{eq:def_W} in~\req{eq:rg_evol} we obtain for $\bW$ the RG equation
\begin{gather}
\label{eq:rg_W}
\begin{split}
\mu\frac{d}{d \mu}\bW(\mu) &= -\bW(\mu){\boldsymbol \gamma}(\gbar(\mu))+\beta(\gbar(\mu))\frac{{\boldsymbol \gamma}^{(0)}}{b_0\gbar(\mu)}\bW(\mu) \\
&= [{\boldsymbol \gamma}(\gbar(\mu)),\bW(\mu)] - \beta(\gbar(\mu))\left(
\frac{{\boldsymbol \gamma}(\gbar(\mu))}{\beta(\gbar(\mu))}-\frac{{\boldsymbol \gamma}^{(0)}}{b_0\gbar(\mu)}
\right)\bW(\mu)\,.
\end{split}
\end{gather}
Expanding perturbatively  we can check~\cite{Papinutto:2016xpq}
that $\bW$ is regular in the UV, and all the logarithmic divergences
in the evolution operator are contained in $\bU_{\rm LO}$; in particular,
\begin{gather}
\label{eq:Winfty}
\bW(\mu)\underset{\mu\to \infty}{=}\mathbf{1}\,.
\end{gather}
Rewriting~\req{eq:evol} as
\begin{gather}
\left[\frac{\gbar^2(\mu_2)}{4\pi}\right]^{-\frac{{\boldsymbol \gamma}^{(0)}}{2b_0}}\bW(\mu_2)\overline{{\boldsymbol \cQ}}(\mu_2)
= \left[\frac{\gbar^2(\mu_1)}{4\pi}\right]^{-\frac{{\boldsymbol \gamma}^{(0)}}{2b_0}}\bW(\mu_1)\overline{{\boldsymbol \cQ}}(\mu_1)\,,
\end{gather}
and observing that the l.h.s. (respectively r.h.s.) is obviously independent of $\mu_1$ (respectively $\mu_2$),
we conclude that these are scale-independent expressions. Thus we can define the vector of RGI operators as
\begin{gather}
\label{eq:rgi_mix}
\hat{\boldsymbol \cQ} \equiv \left[\frac{\gbar^2(\mu)}{4\pi}\right]^{-\frac{{\boldsymbol {\boldsymbol \gamma}}^{(0)}}{2b_0}}\bW(\mu)\overline{\boldsymbol \cQ}(\mu)  \,\, = \,\, \lim_{\mu \to \infty}  \left[\frac{\gbar^2(\mu)}{4\pi}\right]^{-\frac{{\boldsymbol \gamma}^{(0)}}{2b_0}}\overline{\boldsymbol \cQ}(\mu) \,, 
\end{gather}
where in the last step we use~\req{eq:Winfty}.

%% file: sf.tex
\section{Schr\"odinger Functional renormalisation setup}
\label{sec:sf}

In this section we introduce the finite volume Schr\"odinger Functional (SF)
renormalisation schemes and the RG evolution
matrix between scales separated by a fixed factor (i.e. the matrix-step-scaling function).

\subsection{Renormalisation conditions}
\label{subsec:sf}

We first define Schr\"odinger Functional renormalisation
schemes for the operator basis of \req{eq:rel_ops}.
This section is an abridged version of sec.~3.3 of ref.~\cite{Papinutto:2016xpq}.
We use the standard SF setup as described in~\cite{Luscher:1996sc}, where the reader
is referred for full details including unexplained notation.

We work with lattices of spatial extent $L$ and time extent $T$; here we opt for $T=L$.
Source fields are made up of boundary quarks and antiquarks,
\begin{align}
\cO_{\alpha\beta}[\Gamma]  &\equiv a^6\sum_{\vy,\vz}\bar\zeta_\alpha(\vy)\Gamma\zeta_\beta(\vz)\,,\\
\cO'_{\alpha\beta}[\Gamma] &\equiv a^6\sum_{\vy,\vz}\bar\zeta'_\alpha(\vy)\Gamma\zeta'_\beta(\vz)\,,
\end{align}
where $\alpha,\beta$ are flavour indices, unprimed (primed) fields live at the
$x_0=0$ ($x_0=T$) boundary, and $\Gamma$ is a Dirac matrix. The boundary
fields $\zeta, \bar \zeta$ are constrained to satisfy the conditions
\begin{gather}
\label{eq:boundary-zeta}
\zeta(\vx)=\half(\mathbf{1}-\gamma_0)\zeta(\vx)\,,~~~~~~~~~~
\bar\zeta(\vx)=\bar\zeta(\vx)\half(\mathbf{1}+\gamma_0)\,,
\end{gather}
and similarly for primed fields. This implies that the Dirac matrices $\Gamma$ must
anticommute with $\gamma_0$, otherwise the boundary operators $\cO_{\alpha\beta}[\Gamma]$
and $\cO'_{\alpha\beta}[\Gamma]$ vanish; thus $\Gamma$ may be either
$\gamma_5$ or $\gamma_k$ ($k = 1,2,3$).

\begin{figure}[t!]
\begin{center}
\includegraphics[width=100mm]{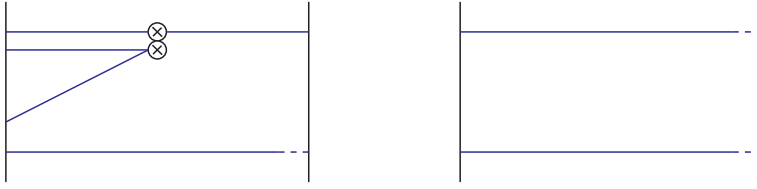}
\end{center}
\vspace{-5mm}
\caption{Four-quark correlation functions $F_{k;s}$ (left)
and the boundary-to-boundary correlators $f_1,k_1$ (right), depicted in terms of quark propagators.
Euclidean time goes from left to right.
The double blob indicates the four-quark operator insertion, and dashed lines indicate
the explicit time-like link variable involved in boundary-to-boundary quark propagators.}
\label{fig:diagtl}
\end{figure}

Renormalisation conditions are imposed in the massless theory, in order
to obtain a mass-independent scheme by construction. They are furthermore
imposed on the parity-odd four-quark operators $\{ \cQ_k^\pm \}$ 
of ~\req{eq:rel_ops}, since working in the parity-even $\{ Q_k^\pm \}$
sector would entail dealing with the extra mixing due to explicit chiral
symmetry breaking with Wilson fermions, cf.~\req{eq:ren_pattern}.
In order to obtain non-vanishing SF correlation functions, we then
need a product of source operators with overall negative parity;
taking into account the above observation about boundary fields, and
the need to saturate flavour indices, the minimal structure involves
three boundary bilinear operators and the introduction of an extra,
``spectator'' flavour (labeled as number 5, keeping with the notation in~\req{eq:gen_4f}).
We thus end up with correlation functions of the generic form
\begin{align}
\label{eq:F}
F_{k;s}(x_0) &\equiv \langle \cQ_k(x) \cS_s \rangle\,,\\
G_{k;s}(T-x_0) &\equiv \eta_k\langle \cQ_k(x) \cS'_s \rangle\,,
\end{align}
where $\cS_s$ is one of the five source operators
\begin{align}
\cS_1 &\equiv \cW[\gamma_5,\gamma_5,\gamma_5]\,,\\
\cS_2 &\equiv \frac{1}{6}\sum_{k,l,m=1}^3\epsilon_{klm}\cW[\gamma_k,\gamma_l,\gamma_m]\,,\\
\cS_3 &\equiv \frac{1}{3}\sum_{k=1}^3\cW[\gamma_5,\gamma_k,\gamma_k]\,,\\
\cS_4 &\equiv \frac{1}{3}\sum_{k=1}^3\cW[\gamma_k,\gamma_5,\gamma_k]\,,\\
\cS_5 &\equiv \frac{1}{3}\sum_{k=1}^3\cW[\gamma_k,\gamma_k,\gamma_5]
\end{align}
with
\begin{gather}
\cW[\Gamma_1,\Gamma_2,\Gamma_3] \equiv L^{-3}\cO'_{21}[\Gamma_1]\cO'_{45}[\Gamma_2]\cO_{53}[\Gamma_3]\,,
\end{gather}
and similarly for $\cS'_s$, which is defined with the boundary fields exchanged between time boundaries; e.g
$\cO_{53} \leftrightarrow \cO_{53}^\prime$ etc.
The constant $\eta_k$ is a sign that ensures
$F_{k;s}(x_0)=G_{k;s}(x_0)$ for all possible indices\footnote{This time reversal property,
besides being a useful numerical cross check of our codes,
allows taking the average of $F_{k;s}(x_0)$ and $G_{k;s}(x_0)$ so as to reduce statistical fluctuations.
From now on $F_{k;s}(x_0)$ denotes this average.}; it is easy
to check that $\eta_2=-1,~\eta_{s \neq 2}=+1$.We also use the two-point functions of boundary sources
\begin{align}
f_1 &\equiv -\,\frac{1}{2L^6}\langle\cO'_{21}[\gamma_5]\cO_{12}[\gamma_5]\rangle\,,\\
k_1 &\equiv -\,\frac{1}{6L^6}\sum_{k=1}^3\langle\cO'_{21}[\gamma_k]\cO_{12}[\gamma_k]\rangle\,.
\end{align}
Finally, we define the ratios
\begin{gather}
\label{eq:corr_ratio}
\cA_{k;s,\alpha} \equiv \frac{F_{k;s}(T/2)}{f_1^{\scriptscriptstyle{\frac{3}{2}}-\alpha}k_1^\alpha}\,,
\end{gather}
where $\alpha$ is an arbitrary real parameter.
The geometry of $F_{k;s}, f_1$, and $k_1$ is illustrated in~\refig{fig:diagtl}.

We can now impose Schr\"odinger functional renormalisation conditions on the ratio of correlation
functions defined in \req{eq:corr_ratio},
at fixed bare coupling $g_0$, vanishing quark mass, and scale $\mu=1/L$.
For the renormalisable multiplicative operators $\cQ_1$ we set
\begin{gather}
\label{eq:mult_ren}
\cZ_{11;s,\alpha}\,\cA_{1;s,\alpha} =
\left.\cA_{1;s,\alpha}\right|_{g_0^2=0}\,.
\end{gather}
For operators that mix in doublets, we impose\footnote{S.~Sint, private communication.}
\begin{gather}
\label{eq:mat_ren}
\left(\ba{cc}
\cZ_{22;s_1,s_2,\alpha} & \cZ_{23;s_1,s_2,\alpha} \\
\cZ_{32;s_1,s_2,\alpha} & \cZ_{33;s_1,s_2,\alpha} \\
\ea\right)
\left(\ba{cc}
\cA_{2;s_1,\alpha} & \cA_{2;s_2,\alpha} \\
\cA_{3;s_1,\alpha} & \cA_{3;s_2,\alpha} \\
\ea\right)
=
\left(\ba{cc}
\cA_{2;s_1,\alpha} & \cA_{2;s_2,\alpha} \\
\cA_{3;s_1,\alpha} & \cA_{3;s_2,\alpha} \\
\ea\right)_{g_0^2=0}
\,,
\end{gather}
and similarly for $\cQ_{4,5}$.
The product of boundary-to-boundary correlators in the denominator of~\req{eq:corr_ratio}
cancels the renormalisation of the boundary operators in $F_{k;s}$, and therefore
$\cZ_{jk;s_1,s_2,\alpha}$ only contains anomalous dimensions of four-fermion operators.
Following~\cite{Capitani:1998mq,Guagnelli:2005zc,Dimopoulos:2007ht}, conditions are imposed on renormalisation functions
evaluated at $x_0=T/2$, and the phase that parameterises spatial boundary
conditions on fermion fields is fixed to $\theta=0.5$.
Together with the $L=T$ geometry of our finite box, this fixes the
renormalisation scheme completely, up to the choice of boundary source,
indicated by the index $s$, and the parameter $\alpha$. The latter can
in principle take any value, but we restrict our choice to $\alpha=0,1,3/2$.

One still has to check that the above renormalisation conditions are well-defined
at tree-level. This is straightforward for~\req{eq:mult_ren}, but not
for~\req{eq:mat_ren}: it is still possible that
the matrix of ratios $\cA$ has zero determinant at tree-level, rendering
the system of equations for the renormalisation matrix
ill-conditioned. This is indeed obviously the case for $s_1 = s_2$,
but the determinant vanishes also for other non-trivial
choices of $s_1 \neq s_2$. In practice, out of the ten possible schemes
one is only left with six, viz.\footnote{Note that schemes obtained by
exchanging $s_1 \leftrightarrow s_2$ are trivially related to each other.}
\begin{gather}
(s_1,s_2) \in \{(1,2),(1,4),(1,5),(2,3),(3,4),(3,5)\}\,.
\end{gather}
This property is independent of the choice
of $\theta$ and $\alpha$. Thus, we are left with a total of 15 schemes for $\cQ_1$, and 18 for each of the pairs $(\cQ_2,\cQ_3)$ and $(\cQ_4,\cQ_5)$.

Given the strong scheme dependence of the matrices ${\boldsymbol \gamma}^{(1); \rm SF}$ (cf.~\req{eq:gamma-PT}), a criterion has been devised in
ref.~\cite{Papinutto:2016xpq} in order to single out the scheme with the smallest NLO anomalous dimension.
This consists in choosing the scheme with the smallest determinant and trace of the matrix
$16 \pi^2 {\boldsymbol \gamma}^{(1); \rm SF} [ {\boldsymbol \gamma}^{(0)} ]^{-1}$ for each non-trivial $2 \times 2$ anomalous dimension
matrix. It turns out that the scheme defined by $\alpha = 3/2$ and $(s_1,s_2) = (3,5)$ satisfies
these requirements in all cases (i.e. for the matrices related to $(\cQ_2, \cQ_3)$ and $(\cQ_4, \cQ_5)$).
In the following we will present non-perturbative results for this scheme only\footnote{Although we have completed our analyses
in all schemes discussed here, for reasons of economy of presentation we will not show these results. In any case, 
the $\alpha = 3/2$ and $(s_1,s_2) = (3,5)$ scheme displays the most reliable matching to perturbative RG-running at the electorweak scale.}.

\subsection{Matrix-step-scaling functions and non-perturbative computation of RGI operators}

In order to trace the RG evolution non-perturbatively, we introduce matrix-step-scaling functions (matrix-SSFs), 
defined as\footnote{The relative factor between
the scales is arbitrary; one could introduce a $\bsigma(s,u)$ that evolves
from scale $\mu$ to scale $\mu/s$. In this notation, our choice corresponds to $s=2$.}
\begin{gather}
\label{eq:ssf}
\bsigma(u) \equiv \left.\bU(\mu/2,\mu)\right|_{\gbar^2(\mu)=u}=\left[\bW(\mu/2)\right]^{-1}\bU_{\rm LO}(\mu/2,\mu)\bW(\mu)\,.
\end{gather}
The above definition generalises the step-scaling functions (SSFs) defined for quark masses~\cite{Capitani:1998mq}
and multiplicatively renormalisable four-fermion operators~\cite{Guagnelli:2005zc} such as $\cQ_1^\pm$.
Just like the anomalous dimension matrix $\bgamma$, the matrix-SSF $\bsigma$ has a block-diagonal structure. So the above
definition either refers to one of the two multiplicative operators $\cQ_1^\pm$,
or to one of the four pairs of operators that mix under renormalisation; i.e.
$(\cQ_2^\pm,\cQ_3^\pm)$ or $(\cQ_4^\pm,\cQ_5^\pm)$. In the former cases
$\bsigma$ is a real function, whereas in the latter cases it is a $2 \times 2$
matrix of real functions. Again in what follows the $\pm$ superscripts will be suppressed.

The advantage of working with step-scaling functions is that they can be computed
on the lattice with all systematic uncertainties under control. More concretely, we
define the lattice matrix-SSF $\bSigma$ in a finite $(L/a)^3 \times (T/a)$ lattice;
as repeatedly stated previously, in this work we set $L=T$. Working in the chiral limit, 
at a given bare coupling $g_0$ (i.e. at a given finite UV cutoff $a^{-1}$) , $\bSigma$ 
is defined as the following
``ratio" of renormalisation matrices at two renormalisation scales
$\mu = 1/L$ and $\mu/2 = 1/(2L)$: 
\begin{gather}
\bSigma(g_0^2,a/L) \equiv {\boldsymbol \cZ}\left(g_0^2,\frac{a}{2L}\right)\left[{\boldsymbol \cZ}\left(g_0^2,\frac{a}{L}\right)\right]^{-1}\,.
\label{SIGMA}
\end{gather}
This quantity has a well defined continuum limit. For a sequence of lattice sizes $L/a$, we tune the bare coupling $g_0(a)$ (and thus the
corresponding lattice spacing $a$) to a sequence of values which correspond to a constant renormalised squared
coupling $\bar g^2(1/L) = u$. Keeping $u$ fixed implies that the renormalisation scale $\mu = 1/L$ is also held fixed.
It is then straightforward to check that $\bSigma$ satisfies
\begin{gather}
\label{eq:ssf_cl}
\bsigma(u) = \lim_{a \to 0} \bSigma(g_0^2,a/L) \Big \vert_{\bar g^2(1/L) = u} \,\, .
\end{gather}
Thus, the computation of the renormalisation matrices ${\boldsymbol \cZ}$ at a fixed value of the renormalised
squared coupling $u$ and various values of the lattice sizes $L/a$ and $2L/a$, allows for a controlled extrapolation of
the matrix-SSFs to the continuum limit.

The strategy for obtaining non-perturbative estimates of RGI operators proceeds in standard fashion:
We start from a low-energy scale $\mu_{\rm had}=1/L_{\rm max}$, implicitly defined by $\bar g^2(1/L_{\rm max}) = u_0$.
The SSF $\sigma(u)$ for the coupling, defined as $\sigma(\bar g^2(1/L)) = g^2(1/(2L))$, is known for $\NF =2$ from ref.~\cite{DellaMorte:2004bc}. Thus we generate a sequence of squared couplings $(u_1,\ldots,u_N)$ through the recursion $\sigma^{-1}(u_{n-1}) = u_n$,
and compute recursively the matrix-SSFs $(\bsigma(u_1),\ldots,\bsigma(u_N))$
which correspond to a sequence of physical lattice lengths (inverse renormalisation scales)
$(L_{\rm max}/2, \ldots, L_{\rm max}/2^N)$. 
This is followed by the computation of
\begin{gather}
\bU(\mu_{\rm had},\mu_{\rm pt}) = \bsigma(u_1)\cdots\bsigma(u_N)\,,
\label{eq:ssf_prod}
\end{gather}
with $\mu_{\rm had}=2^{-N}\mu_{\rm pt}=L_{\rm max}^{-1}$. Here $\mu_{\rm pt} \sim \cO(M_{\rm W})$ is thought of
as a high-energy scale, safely into the perturbative regime, and $\mu_{\rm had} \sim \cO(\Lambda_{\rm QCD})$
as a low-energy scale, characteristic of hadronic physics.
The RGI operators of~\req{eq:rgi_mix} can finally be constructed as follows:
\begin{gather}
\hat {\boldsymbol \cQ} = \left[\frac{\gbar^2(\mu_{\rm pt})}{4\pi}\right]^{-\frac{\bgamma^{(0)}}{2b_0}}\bW(\mu_{\rm pt})
\big [ \bU(\mu_{\rm had},\mu_{\rm pt}) \big ]^{-1}
\overline{\boldsymbol \cQ}(\mu_{\rm had})\,.
\label{eq:rgi_prod}
\end{gather}
In other words, once we know the column of renormalised operators $\overline{\boldsymbol \cQ}(\mu_{\rm had})$ at a hadronic scale from a standard computation on a lattice of ``infinite" physical volume (which is beyond the scope of the present paper), we can combine it with the non-perturbative evolution matrix $[ U(\mu_{\rm had},\mu_{\rm pt}) \big ]$ (which is the result of this work) and the remaining $\mu_{\rm pt}$-dependent factors at scale $\mu_{\rm pt}$ (known
in NLO perturbation theory from ref.~\cite{Papinutto:2016xpq}), to obtain the RGI operators\footnote{The computation of operators $\overline{\boldsymbol \cQ}(\mu_{\rm had})$ (i.e. their physical matrix elements) must be known with a precision similar to that of the evolution matrix.}. 
All factors on the r.h.s. must be known in the same renormalisation scheme, which here is the SF. 
The scheme dependence should cancel in the product of the r.h.s., since $\hat {\boldsymbol \cQ}$ is scheme-independent.
In practice a residual dependence remains due to the fact that $\bW(\mu_{\rm pt})$ is only known in perturbation theory (typically to NLO). Finally we stress that $\hat {\boldsymbol \cQ}$
depends, through the operators $\overline{\boldsymbol \cQ}(\mu)$, on the values of the quark masses; of course the
result also depends on the  flavour content of the QCD model under
scrutiny (i.e. $\NF$).

We mentioned above that the matrix $\bW(\mu_{\rm pt})$ is known in NLO perturbation theory from ref.~\cite{Papinutto:2016xpq}. This statement requires a brief elucidation: $\bW(\mu_{\rm pt})$ is obtained by numerically integrating \req{eq:rg_W}, using
the NLO (2-loop) perturbative result for $\bgamma$ and the NNLO (3-loop)  perturbative result for $\beta$. In what follows this will be
abbreviated as NLO-2/3PT. In line with ref.~\cite{Papinutto:2016xpq}, also
the present work devotes considerable effort to the investigation of the reliability of NLO-2/3PT at the scale $\mu_{\rm pt}$.

\subsection{Matrix-step-scaling functions and continuum extrapolations}

We now turn to some practical considerations concerning the extrapolation of $\bSigma(u,a/L)$ to the continuum
limit $a/L \to 0$, from which we obtain $\bsigma(u)$; cf. \req{eq:ssf_cl}. We stress that although fermionic and gauge actions are
Symanzik-improved by the presence of bulk and boundary counter-terms, correlation functions with dimension-six operators in 
the bulk of the lattice, such as those defined in Eqs.~(\ref{eq:F}) and~(\ref{eq:corr_ratio}), are subject to linear discretisation errors.
Their removal could be achieved in principle by the subtraction
of dimension-7 counter-terms, but their coefficients are not easy to determine in practice.
We therefore expect linear cutoff effects and consequently fit with the Ansatz
\begin{gather}
\label{eq:Sigma-CL}
\bSigma(u,a/L)=\bsigma(u)+\brho(u)(a/L)\, .
\end{gather}

In analogy to ref.~ \cite{Sint:1998iq}, we explore the reliability of the above extrapolations with the help of
the lowest-order perturbative expression for $\Sigma_{ij}$, which includes $\cO(ag_0^2)$ terms. In general the perturbative series for the operator renormalisation matrices has the form~\cite{Sint:1998iq}
\begin{gather}
{\boldsymbol \cZ} (g_0,L/a) = {\bf 1} + \sum_{l=1}^\infty {\boldsymbol \cZ}^{(l)}(L/a) g_0^{2l} \,\, ,
\end{gather}
where in the limit $a/L  \to 0$ the coefficients ${\boldsymbol \cZ}^{(l)}$ are $l$-degree polynomials in $\ln (L/a)$ up to corrections of $\cO(a/L)$.
In particular the coefficient of the logarithmic divergence in ${\boldsymbol \cZ}^{(1)}$ is given by the one-loop anomalous dimension
$\bgamma^{(0)}$, and thus we parametrise ${\boldsymbol \cZ}^{(1)}$ as
\begin{gather}
{\boldsymbol \cZ}^{(1)}  = C_F \,\, {\bf z}(\theta,T/L) - \bgamma^{(0)}  \ln(L/a) + \cO(a/L) \,\, ,
\end{gather}
with $\theta = 0.5$ and $T/L = 1$. It is now easy to see that the
one-loop perturbative expression for the matrix-SSF is given by
\begin{gather}
\bSigma(g^2_{\rm R},a/L)=\mathbf{1}+ {\bf k}(L/a) g_{\rm R}^2 + \mathcal{O}(g_{\rm R}^2) \,\, ,
\end{gather}
with
\begin{gather}
{\bf k}(L/a) = {\boldsymbol \cZ}^{(1)}(2L/a) - {\boldsymbol \cZ}^{(1)}(L/a) \,\, .
\end{gather}
In the continuum limit ($a/L \to 0$ with $\bar g^2 = u$ fixed) we have
\begin{gather}
{\bf k}(\infty) = \bgamma^{(0)} \log(2). 
\end{gather}
The quantity
\begin{gather}
\label{eq:def-delta}
\bdelta_k(L/a) \equiv {\bf k}(L/a) [{\bf k}(\infty)]^{-1} - \mathbf{1}. 
\end{gather}
contains all lattice artefacts at $\cO(g_0^2)$. Results for $\bdelta_k(L/a)$ are reported in Appendix~\ref{sec:cutoff}.

The ``subtracted'' matrix-SSF, defined as
\begin{gather}
\label{eq:Sigma-tilde}
\tilde{\bSigma}(u,a/L) \equiv \left . \bSigma(u,a/L) \,\,  \Big [ \mathbf{1} + u \log(2) \bdelta_k(a/L) \bgamma^{(0)} \Big ]^{-1} \right | _{u=\bar g^2(L)}
\end{gather}
also tends to $\bsigma$ in the continuum limit, but has the $\mathcal{O}(a\bar g^2)$ effects removed.
We will also use this quantity when studying the reliability of the linear continuum extrapolations below.

\subsection{Perturbative expansion of matrix-step-scaling functions}
\label{subsect:pert_ssf}

Once the continuum matrix-SSF $\bsigma(u)$ has been computed for $N$
discrete values of the renormalised coupling $\bar g^2(1/L) = u$, it is
useful to interpolate the data so as to obtain $\bsigma(u)$ as a continuous function. 
This is done by fitting the $N$ points by a suitably truncated
polynomial
\begin{gather}
\label{eq:sigmaQ-PT}
\bsigma(u) = \mathbf{1} + \br_1 u + \br_2 u^2 + \br_3 u^3 + \cdots \,.
\end{gather}
With only a few ($N$) points at our disposal, the fit stability is greatly facilitated
by fixing the first two coefficients (matrices) $\br_1$ and $\br_2$ respectively to
their LO and NLO perturbative values, leaving $\br_3$ as the only free fit parameter.
We will now derive the perturbative coefficients $\br_1$ and $\br_2$.

Since the operator RG-running is coupled to that of the strong coupling,
we also need the LO and NLO coefficients of its step-scaling function (SSF); i.e.
\begin{gather}
\label{eq:sigma-PT}
\sigma (u) = u [ 1 + s_1 u + s_2 u^2 + \cdots ] \,.
\end{gather}
Given the strong coupling value $\bar g^2(1/L) = u$ at a renormalisation scale
$\mu = 1/L$, its SSF is defined as $\sigma(u) = \bar g^2(1/2L)$; cf.~ref.\cite{Luscher:1992zx}.
Combining this definition with that of the Callan-Symanzik $\beta$-function
of \req{eq:betadef}, we find that 
\begin{gather}
-\ln 2 = \int_{\sqrt{u}}^{\sqrt{\sigma(u)}}\frac{\dif g}{\beta(g)} \, .
\end{gather}
Plugging the NLO expansion of~\req{eq:beta-PT} in the above and taking~\req{eq:sigma-PT} into
account, we obtain the coefficients of the coupling SSF 
\begin{align}
s_1 &= 2b_0\ln 2\,,\\
s_2 &= 2b_1\ln 2 + 4b_0^2\ln^2 2\,.
\end{align}

Matrix-SSFs for four-quark operators have been introduced in~\req{eq:ssf}. In order to calculate
the coefficients $\br_1$ and $\br_2$ of its perturbative expansion~\req{eq:sigmaQ-PT}, 
we first write down the LO evolution matrix as
\begin{gather}
\begin{split}
\label{eq:U_LO}
& \left.\bU_{\rm LO}(\mu/2,\mu)\right \vert_{\gbar^2(\mu)=u}
= \left[\frac{\sigma(u)}{u}\right]^{\frac{\bgamma^{(0)}}{2b_0}}
=  \exp\left\{\frac{\bgamma^{(0)}}{2b_0}\ln\left[\frac{\sigma(u)}{u}\right]\right\} \\
&= \mathbf{1} + u\bgamma^{(0)}\ln 2 + u^2\left[\left( b_0\ln 2+\frac{b_1}{b_0}\right)\bgamma^{(0)}\ln 2 + \frac{\ln^2 2}{2}\left(\bgamma^{(0)}\right)^2\right] + \ldots \,\,\, \,.
\end{split}
\end{gather}
Furthermore, the matrix $\bW(\mu)$ of~\req{eq:ssf} has the NLO perturbative expansion (cf. ref.~\cite{Papinutto:2016xpq} and references therein)
\begin{gather}
\begin{split}
\label{eq:W_NLO}
\bW(\mu) &= \mathbf{1} + u\bJ_1 + u^2 \bJ_2 + \ldots \,, \\
\end{split}
\end{gather}
from which the inverse matrix is readily obtained:
\begin{gather}
\begin{split}
\label{eq:Winv_NLO}
\left[\bW(\mu/2)\right]^{-1} &= \mathbf{1}-\sigma(u)\bJ_1 + (\bJ_1^2 - \bJ_2) \sigma(u)^2 + \ldots = \mathbf{1} - u\bJ_1+ u^2 (\bJ_1^2 - s_1 \bJ_1  - \bJ_2) + \ldots \,\, .
\end{split}
\end{gather}
We arrive at the last expression on the rhs by inserting the power-series expansion of $\sigma(u)$ form~\req{eq:sigma-PT}.
Substituting the various terms in~\req{eq:ssf} by the perturbative series~(\ref{eq:U_LO}), (\ref{eq:W_NLO}) and (\ref{eq:Winv_NLO}), we find
\begin{align}
\br_1 &= \bgamma^{(0)}\ln 2 \,,\\
\br_2 &= [\bgamma^{(0)}, \bJ_1] \ln 2 -  2 b_0 \bJ_1 \ln 2 +(b_0 \ln 2 + \dfrac{b_1}{b_0}) \bgamma^{(0)} \ln 2 + \frac{1}{2}\,\left(\bgamma^{(0)}\right)^2\ln^2 2 
\nonumber \\
& = \bgamma^{(1)}\ln 2 + b_0\bgamma^{(0)}\ln^2 2 + \frac{1}{2}\,\left(\bgamma^{(0)}\right)^2\ln^2 2 \,.
\label{eq:r2_pt}
\end{align}
From the first expression obtained for $\br_2$ we see explicitly that $\cO(u^2)$ corrections
to $\bW$ do not contribute  (i.e. terms with $\bJ_2$ are absent), in accordance
with the fact that the $\cO(u)$ term of $\bW$ already contains all NLO contributions.
The second expression for $\br_2$ is obtained by using the property (cf. ref.~\cite{Papinutto:2016xpq} and references therein)
\begin{align}
2 b_0 \bJ_1 - [\bgamma_0,\bJ_1] = \dfrac{b_1}{b_0} \bgamma^{(0)} - \bgamma^{(1)} \,\, .
\end{align}
Remarkably, the final result for $\br_2$ is the exact analogue of the  one found for operators that
renormalise multiplicatively, cf. e.g.~Eq.~(6.6) in~\cite{Palombi:2005zd}.

%% file: latt.tex
\section{Non-perturbative computations}
\label{sec:latt}
Our simulations are performed using the lattice regularisation of QCD consisting of
the standard plaquette Wilson action for the gauge fields and the
non-perturbatively $\Oa$ improved Wilson action for $\NF=2$ dynamical fermions.
The fermion action is Clover-improved
with the Sheikoleslami-Wohlert (SW) coefficient $\icsw$ determined in~\cite{Jansen:1998mx}.
The matrix-SSFs are computed at six different values of the~\SF~renormalised coupling, corresponding to six physical lattice extensions $L$ (i.e six values of the renormalisation scale $\mu$).
For each physical volume three different values of the lattice spacing $a$ are simulated, corresponding to lattices with $L/a=6,8,12$; this is achieved by tuning the bare coupling $g_0(a)$ so that the renormalised coupling (and thus $L$) is approximately fixed. At the same
$g_0(a)$ we also generate configuration ensembles at twice the lattice volume; i.e.
$2L/a=12,16,24$ respectively. We compute ${\boldsymbol \cZ}(g_0,a/L)$ and ${\boldsymbol \cZ}(g_0,a/(2L))$ and thus $\bSigma(g_0^2,a/L)$; cf. \req{SIGMA}.
The gauge configuration ensembles used in the present work and the tuning of the lattice parameters $(\beta,\kappa)$ are taken over from ref.~\cite{DellaMorte:2005kg} where all technical details concerning these dynamical fermion simulations are discussed.
As pointed out in~\cite{DellaMorte:2005kg}, the gauge configurations at the three weakest couplings have been produced using the one-loop perturbative estimate of~$\ict$~\cite{Luscher:1992an}, except for $(L/a = 6, \beta = 7.5420)$ and $(L/a = 8, \beta = 7.7206)$. For these two cases and for the three strongest couplings the two-loop value of~$\ict$~\cite{Bode:1999sm} has been used. 

Statistical errors are computed by blocking (binning) the measurements of each renormalisation
parameter and calculating the bootstrap error on the binned averages. In order to take their autocorrelation length into account,
we determine the block-size for which the bootstrap error of a given renormalisation
parameter reaches a plateau. This varies for each of the four matrix elements of a given $2 \times 2$ renormalisation matrix.
We conservatively fix our preferred block-size to the maximum of {\it all} four cases, and estimate our statistical error accordingly.
We crosscheck our results by also applying the Gamma method error analysis of ref.~\cite{Wolff:2003sm}, and by varying the 
summation-window size. The results from the two methods agree within the (relevant) uncertainties.

Numerical results for $[{\boldsymbol \cZ}(g_0,a/L)]^{-1}$ and ${\boldsymbol \cZ}(g_0,a/(2L) )$, computed from~\req{eq:mat_ren}, are collected in Tabs.~\ref{tab:Z23} and \ref{tab:Z45}. The reason we prefer quoting the inverse of ${\boldsymbol \cZ}(g_0,a/L)$ is that it is this quantity which is required
for the computation of the matrix-SSFs; cf.~\req{SIGMA}.

\subsection{Lattice computation of matrix-functions}

We perform linear extrapolations in $a/L$ of both $\bSigma$ and $\tilde{\bSigma}$ (cf. Eqs.~(\ref{eq:Sigma-CL}) and~(\ref{eq:Sigma-tilde})), so as
to crosscheck the reliability of the continuum value $\bsigma(u)$. The extrapolation results can be found in Tabs.~\ref{tab:sigma23} and~\ref{tab:sigma45}, as well as in Figs.~\ref{fig:cont23+},\ref{fig:cont45+},\ref{fig:cont23-}, and \ref{fig:cont45-}.
In most cases both extrapolations agree; at worst the agreement is within two standard deviations (e.g. in Fig.~\ref{fig:cont23+} 
the difference between off-diagonal elements of the matrices $\bSigma$ and $\tilde{\bSigma}$ is sizeable).
We quote, as our best results, those obtained from linear extrapolations in $a/L$, involving all three data-points of the ``subtracted'' matrix-SSFs. We estimate the systematic error as the difference between the value of $\bsigma$ obtained by extrapolating $\bSigma$ and $\tilde{\bSigma}$. This error is added in quadrature to the one from the fit.

Similar checks with another two definitions of ``subtracted'' matrix-SSFs, namely:
\begin{align}
\label{eq:Sigma-tilde2}
\bSigma^{\prime}(u,a/L) & \equiv \Big [ \mathbf{1} + u \log(2) \bdelta_k(a/L) \bgamma^{(0)} \Big ]^{-1}  \bSigma(u,a/L) \, \, ,  \\
\bSigma^{\prime \prime}(u,a/L) & \equiv \bSigma(u,a/L) - u \log(2) \bdelta_k(a/L) \bgamma^{(0)} \,\, , 
\end{align}
which differ at $\mathcal{O}(u^2)$ have not revealed any substantial differences in the results.

\begin{table}[t!]
\begin{scriptsize}
\begin{center}
\begin{tabular}{ccc}
\toprule
$u$ & $\bsigma_{(2,3)}^+(u)$ & $\bsigma_{(2,3)}^-(u)$ \\ 
\midrule
0.9793 & $\begin{pmatrix}
 1.0112(71) & 0.067(21) \\
 0.0095(40) & 0.9227(100)\\
 \end{pmatrix}$
 & $\begin{pmatrix}
 1.0003(74) & -0.074(11) \\
 -0.0094(41) & 0.918(11)\\
 \end{pmatrix}$ \\
1.1814 & $\begin{pmatrix}
 1.0167(90) & 0.054(23) \\
 0.0073(44) & 0.919(10)\\
 \end{pmatrix}$
 & $\begin{pmatrix}
 1.0098(83) & -0.059(11) \\
 -0.0055(40) & 0.918(12)\\
 \end{pmatrix}$ \\
1.5078 & $\begin{pmatrix}
 1.016(12) & 0.065(30) \\
 0.0116(57) & 0.882(14)\\
 \end{pmatrix}$
 & $\begin{pmatrix}
 1.007(12) & -0.089(17) \\
 -0.0106(60) & 0.883(18)\\
 \end{pmatrix}$ \\
2.0142 & $\begin{pmatrix}
 1.0061(100) & 0.101(33) \\
 0.0186(55) & 0.829(11)\\
 \end{pmatrix}$
 & $\begin{pmatrix}
 0.9952(85) & -0.117(11) \\
 -0.0213(55) & 0.835(14)\\
 \end{pmatrix}$ \\
2.4792 & $\begin{pmatrix}
 0.988(20) & 0.087(42) \\
 0.0171(76) & 0.794(22)\\
 \end{pmatrix}$
 & $\begin{pmatrix}
 0.986(14) & -0.095(14) \\
 -0.0200(75) & 0.812(21)\\
 \end{pmatrix}$ \\
3.3340 & $\begin{pmatrix}
 0.990(30) & 0.138(55) \\
 0.049(11) & 0.691(20)\\
 \end{pmatrix}$
 & $\begin{pmatrix}
 0.950(19) & -0.141(21) \\
 -0.0500(95) & 0.716(22)\\
 \end{pmatrix}$ \\
\bottomrule\end{tabular}
\caption{Continuum matrix-SSFs for the operator bases $\{ \cQ^\pm_2, \cQ_3^\pm \}$.}
\label{tab:sigma23}
\end{center}
\end{scriptsize}
\end{table}

\begin{table}[t!]
\begin{scriptsize}
\begin{center}
\begin{tabular}{ccc}
\toprule
$u$ & $\bsigma_{(4,5)}^+(u)$ & $\bsigma_{(4,5)}^-(u)$ \\ 
\midrule
0.9793 & $\begin{pmatrix}
 0.9554(90) & -0.00212(78) \\
 -0.256(41) & 1.0479(76)\\
 \end{pmatrix}$
 & $\begin{pmatrix}
 0.8870(94) & -0.00092(79) \\
 0.093(37) & 1.0040(66)\\
 \end{pmatrix}$ \\
1.1814 & $\begin{pmatrix}
 0.957(12) & -0.0005(10) \\
 -0.195(56) & 1.076(11)\\
 \end{pmatrix}$
 & $\begin{pmatrix}
 0.883(11) & -0.0024(10) \\
 0.009(46) & 1.0012(95)\\
 \end{pmatrix}$ \\
1.5078 & $\begin{pmatrix}
 0.930(16) & -0.0016(15) \\
 -0.252(76) & 1.089(16)\\
 \end{pmatrix}$
 & $\begin{pmatrix}
 0.833(18) & -0.0026(13) \\
 0.022(62) & 0.994(10)\\
 \end{pmatrix}$ \\
2.0142 & $\begin{pmatrix}
 0.896(14) & -0.0034(11) \\
 -0.355(67) & 1.105(12)\\
 \end{pmatrix}$
 & $\begin{pmatrix}
 0.763(11) & -0.0021(12) \\
 0.046(55) & 0.988(12)\\
 \end{pmatrix}$ \\
2.4792 & $\begin{pmatrix}
 0.874(18) & -0.0020(14) \\
 -0.288(82) & 1.136(17)\\
 \end{pmatrix}$
 & $\begin{pmatrix}
 0.718(19) & -0.0039(18) \\
 -0.066(67) & 0.959(17)\\
 \end{pmatrix}$ \\
3.3340 & $\begin{pmatrix}
 0.812(25) & -0.0098(32) \\
 -0.52(13) & 1.204(36)\\
 \end{pmatrix}$
 & $\begin{pmatrix}
 0.587(20) & 0.0012(23) \\
 -0.056(92) & 0.948(22)\\
 \end{pmatrix}$ \\
\bottomrule\end{tabular}
\caption{Continuum matrix-SSFs for the operator bases $\{ \cQ^\pm_4, \cQ_5^\pm \}$.}
\label{tab:sigma45}
\end{center}
\end{scriptsize}
\end{table}

\subsection{RG running in the continuum}
\begin{figure}
\begin{center}
\vspace*{-5\baselineskip}
\begin{minipage}[t]{0.4\textwidth}
\includegraphics[width=7cm]{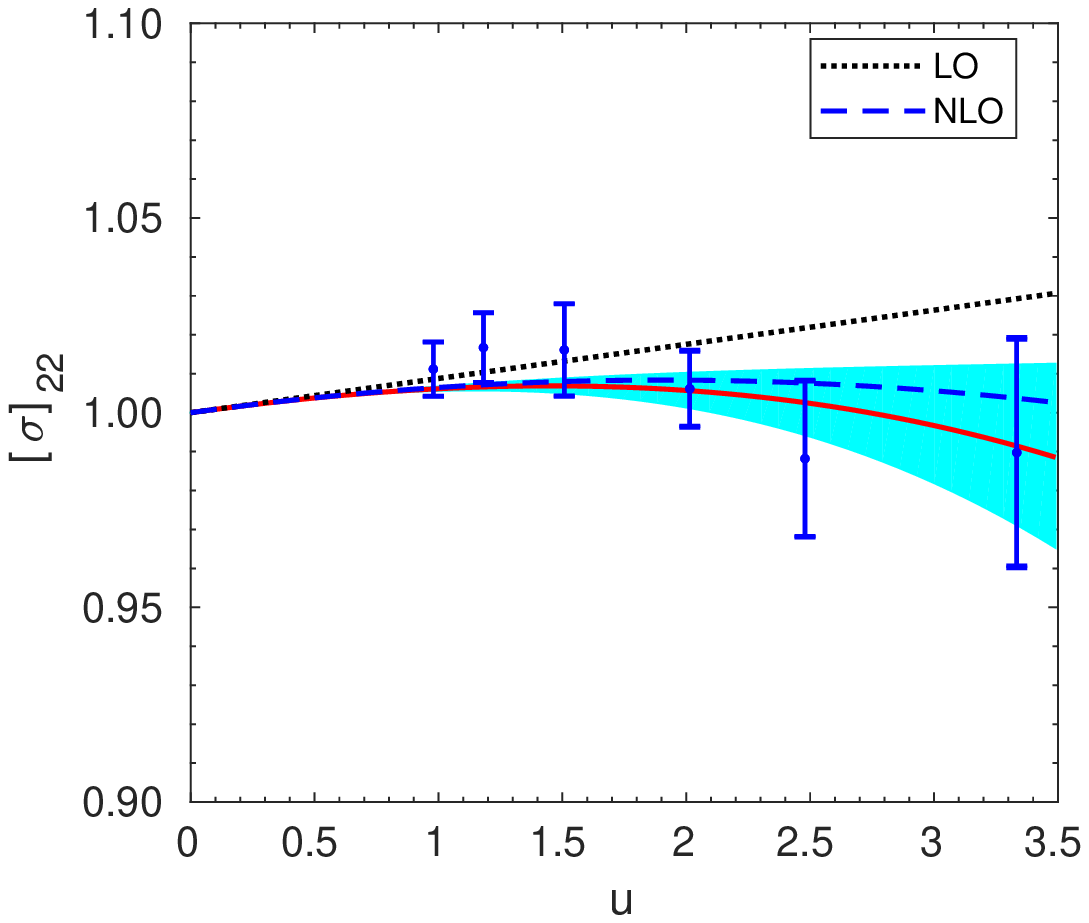}
\end{minipage}\hspace{1cm}
\begin{minipage}[t]{0.4\textwidth}
\includegraphics[width=7cm]{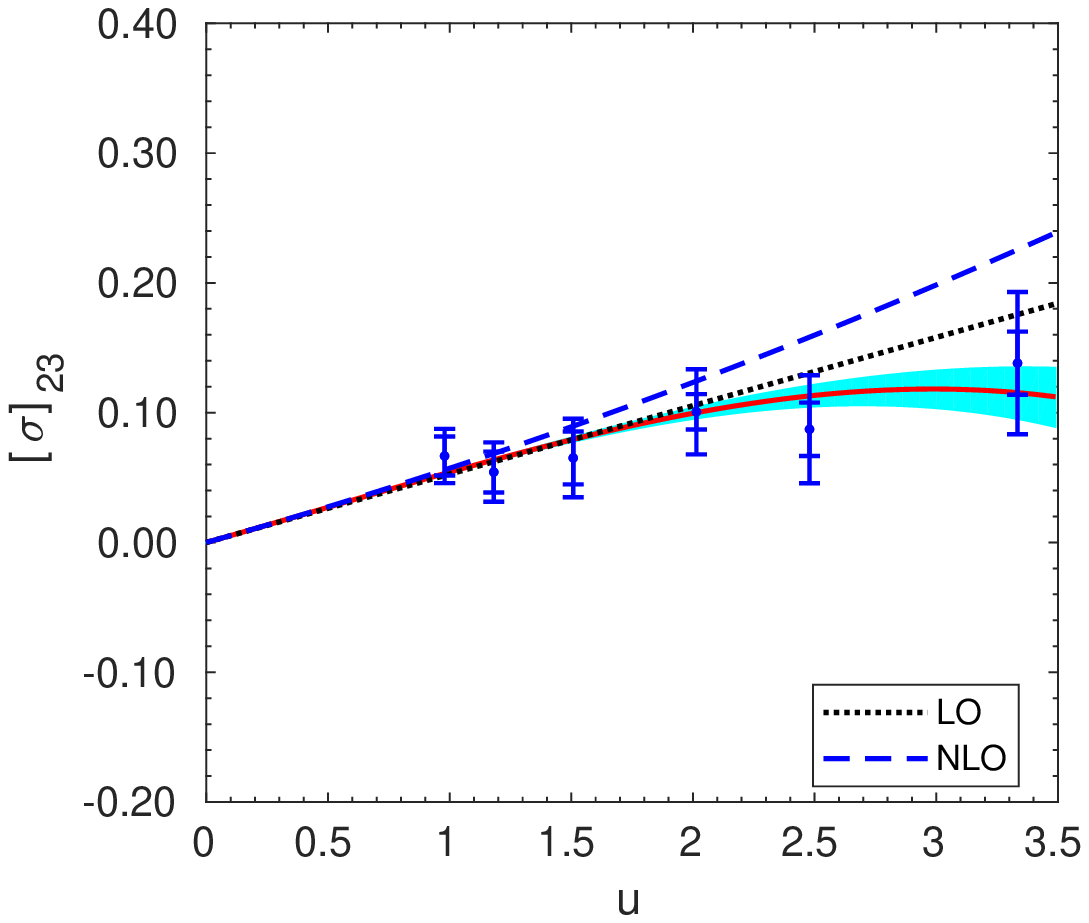}
\end{minipage}

\vspace{-0mm}

\begin{minipage}[t]{0.4\textwidth}
\includegraphics[width=7cm]{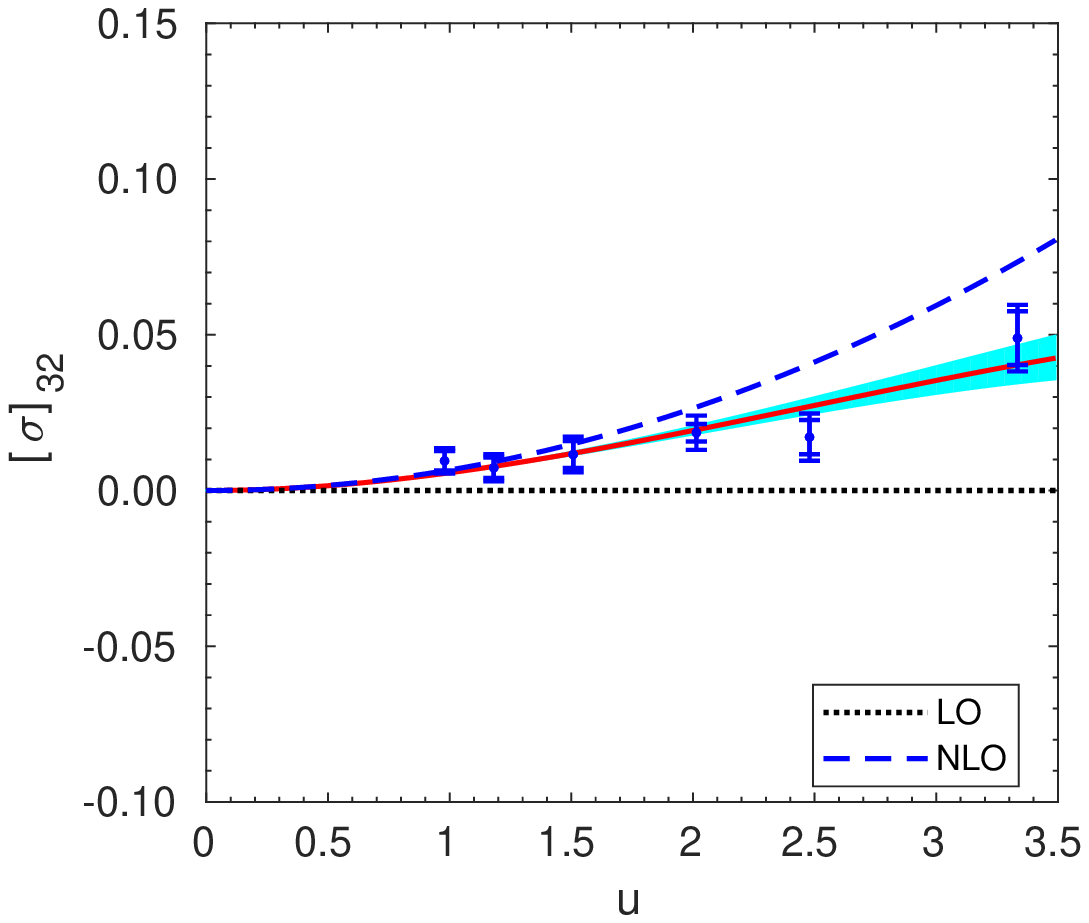}
\end{minipage}\hspace{1cm}
\begin{minipage}[t]{0.4\textwidth}
\includegraphics[width=7cm]{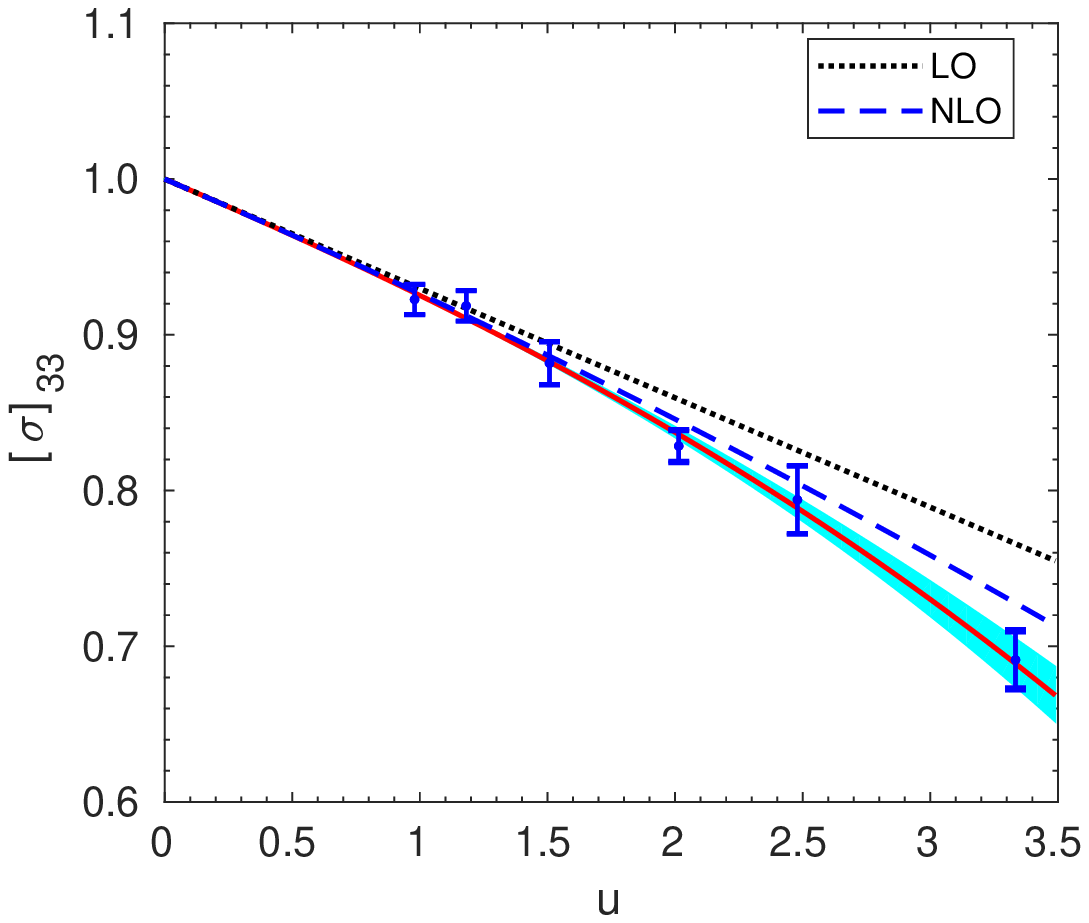}
\end{minipage}

\hspace{2mm}

\begin{minipage}[t]{0.4\textwidth}
\includegraphics[width=7cm]{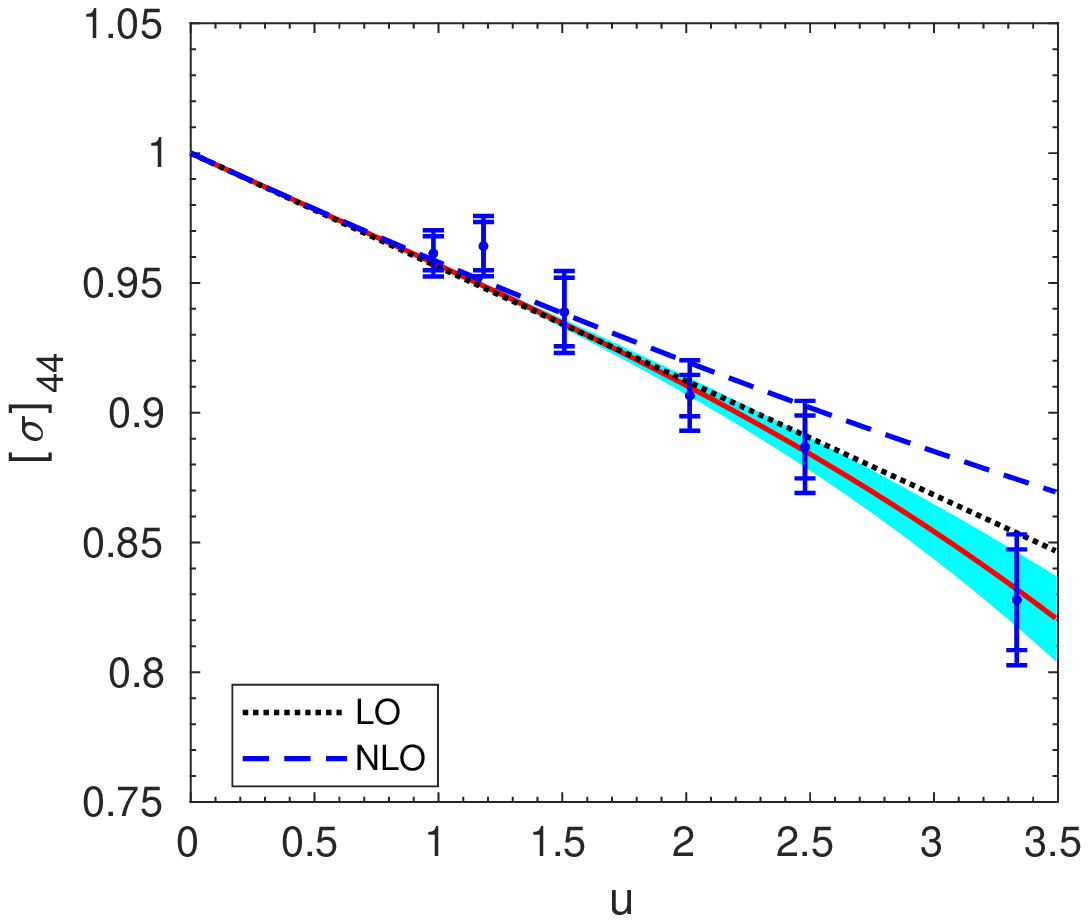}
\end{minipage}\hspace{1cm}
\begin{minipage}[t]{0.4\textwidth}
\includegraphics[width=7cm]{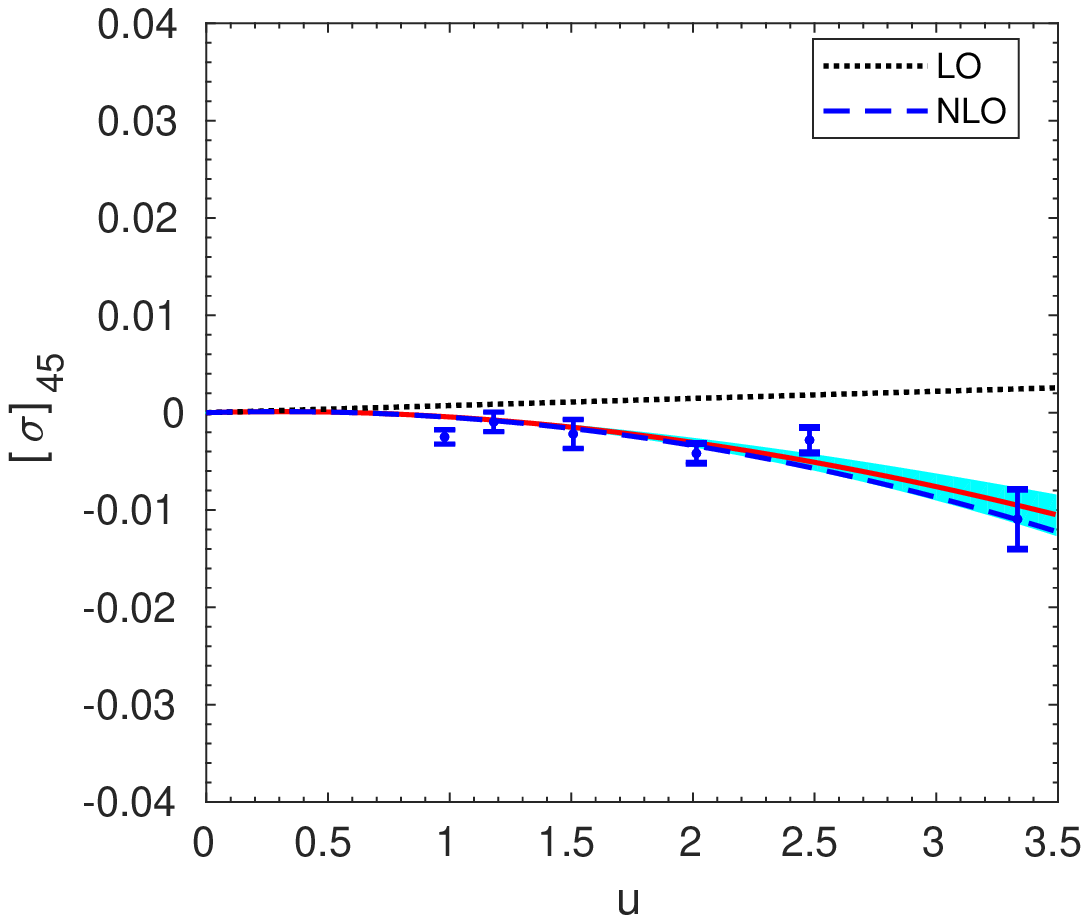}
\end{minipage}

\vspace{-0mm}

\begin{minipage}[t]{0.4\textwidth}
\includegraphics[width=7cm]{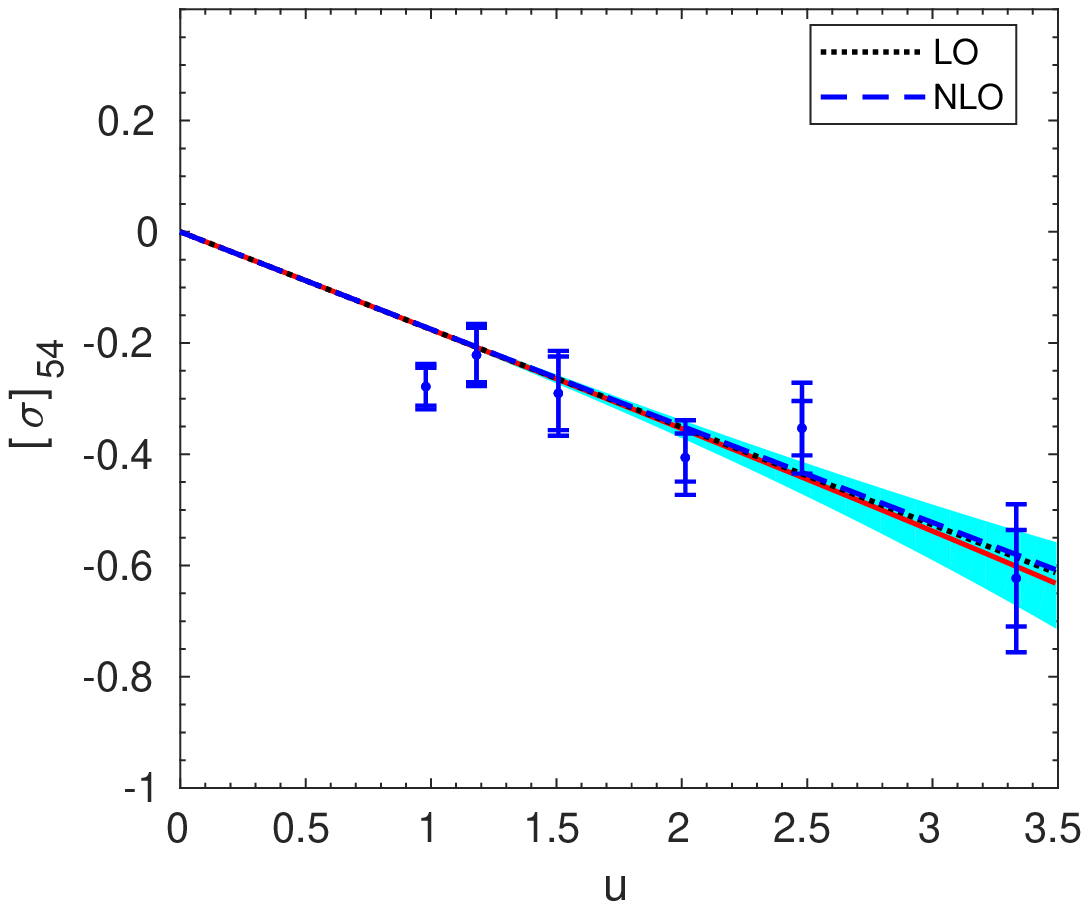}
\end{minipage}\hspace{1cm}
\begin{minipage}[t]{0.4\textwidth}
\includegraphics[width=7cm]{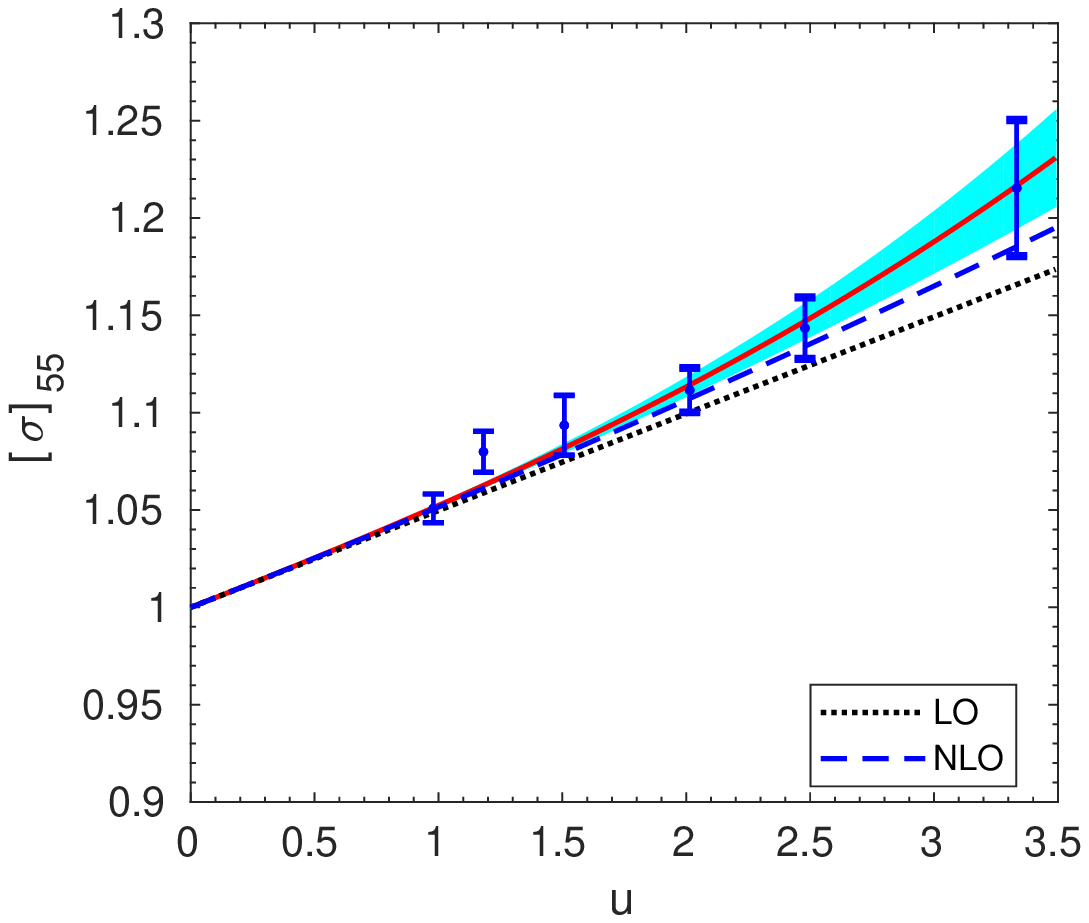}
\end{minipage}
\caption{Continuum matrix-SSFs for operator bases $\{ \cQ_2^+, \cQ_3^+ \}$ (top) and $\{ \cQ_4^+, \cQ_5^+ \}$ (bottom). 
The LO perturbative result is shown by the dotted black line, while the NLO one by the dashed blue line. The red line (with error band) is the non-perturbative result from the $\mathcal{O}(u^3)$ fit as described in the text.
The two error bars on each data point are the statistical and total uncertainties; the systematic error contributing to the latter
has been estimated as explained in the text.}
\label{fig:SSFs+}
\end{center}
\end{figure}

\begin{figure}
\begin{center}
\vspace*{-5\baselineskip}
\begin{minipage}[t]{0.4\textwidth}
\includegraphics[width=7cm]{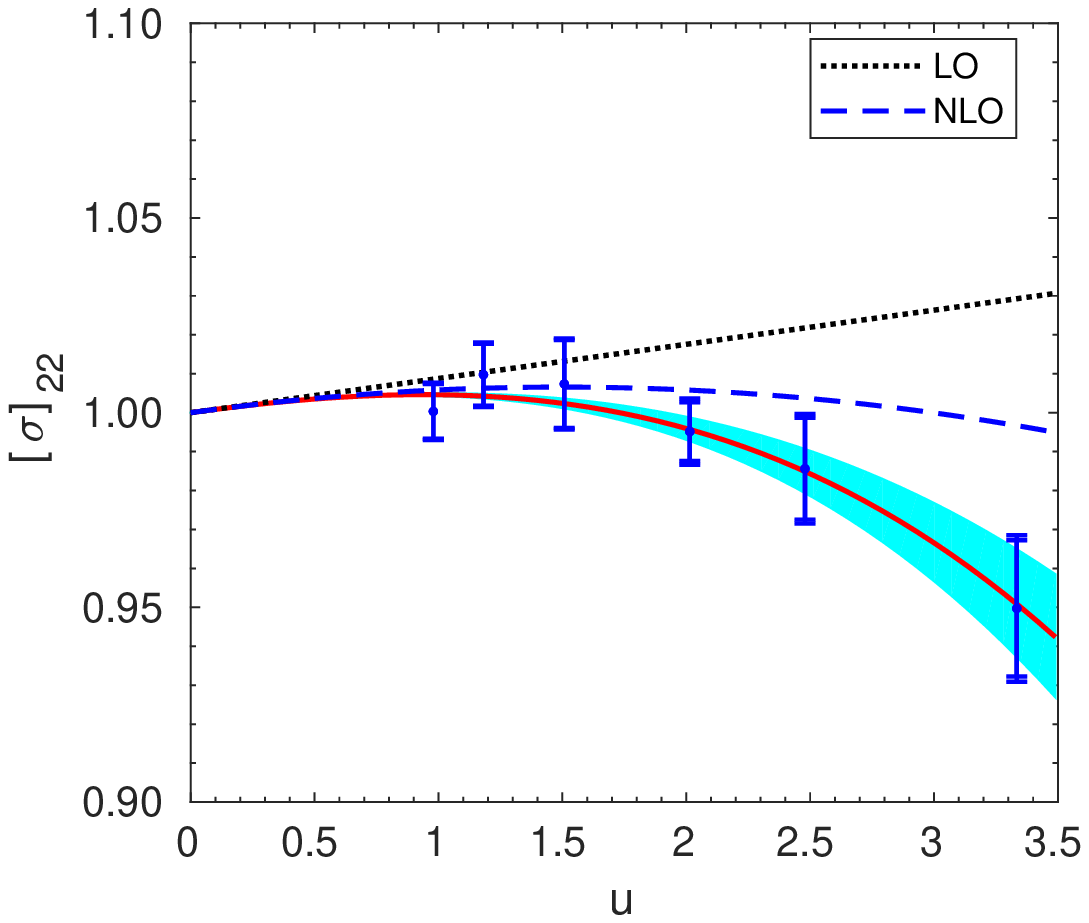}
\end{minipage}\hspace{1cm}
\begin{minipage}[t]{0.4\textwidth}
\includegraphics[width=7cm]{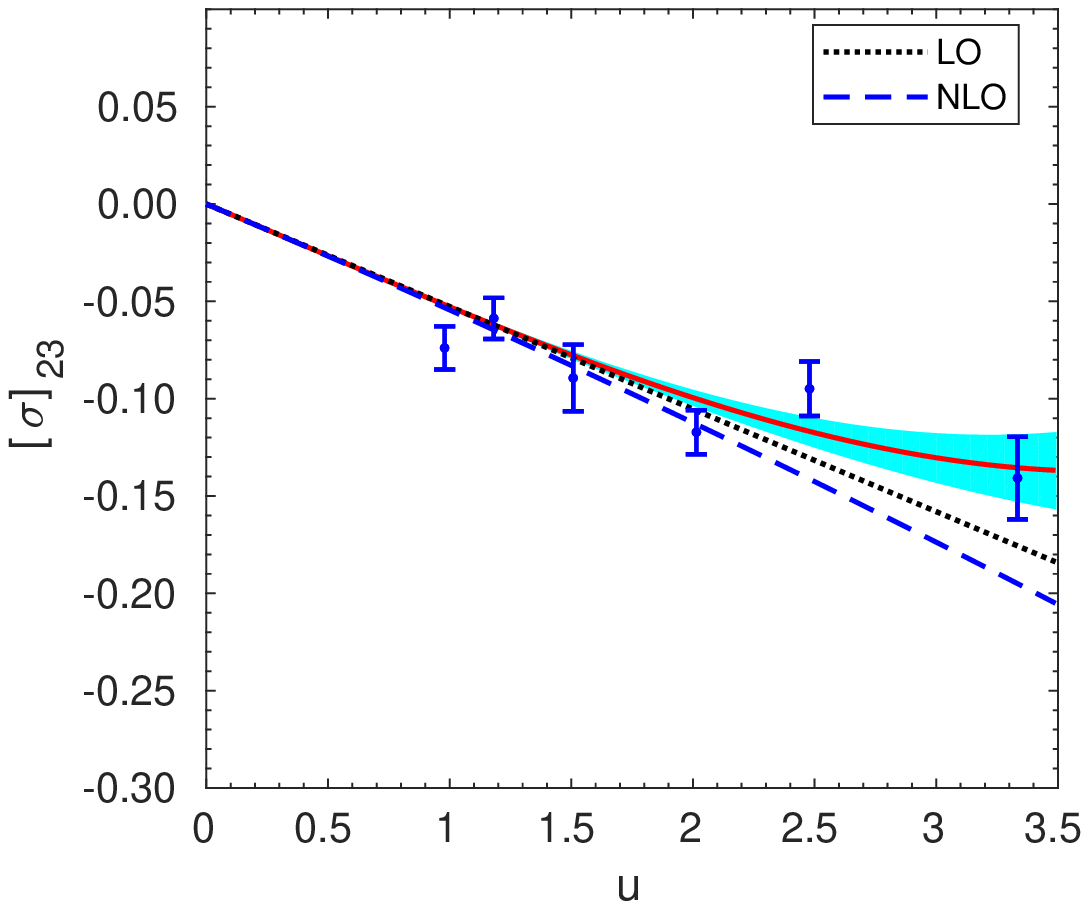}
\end{minipage}

\vspace{-0mm}

\begin{minipage}[t]{0.4\textwidth}
\includegraphics[width=7cm]{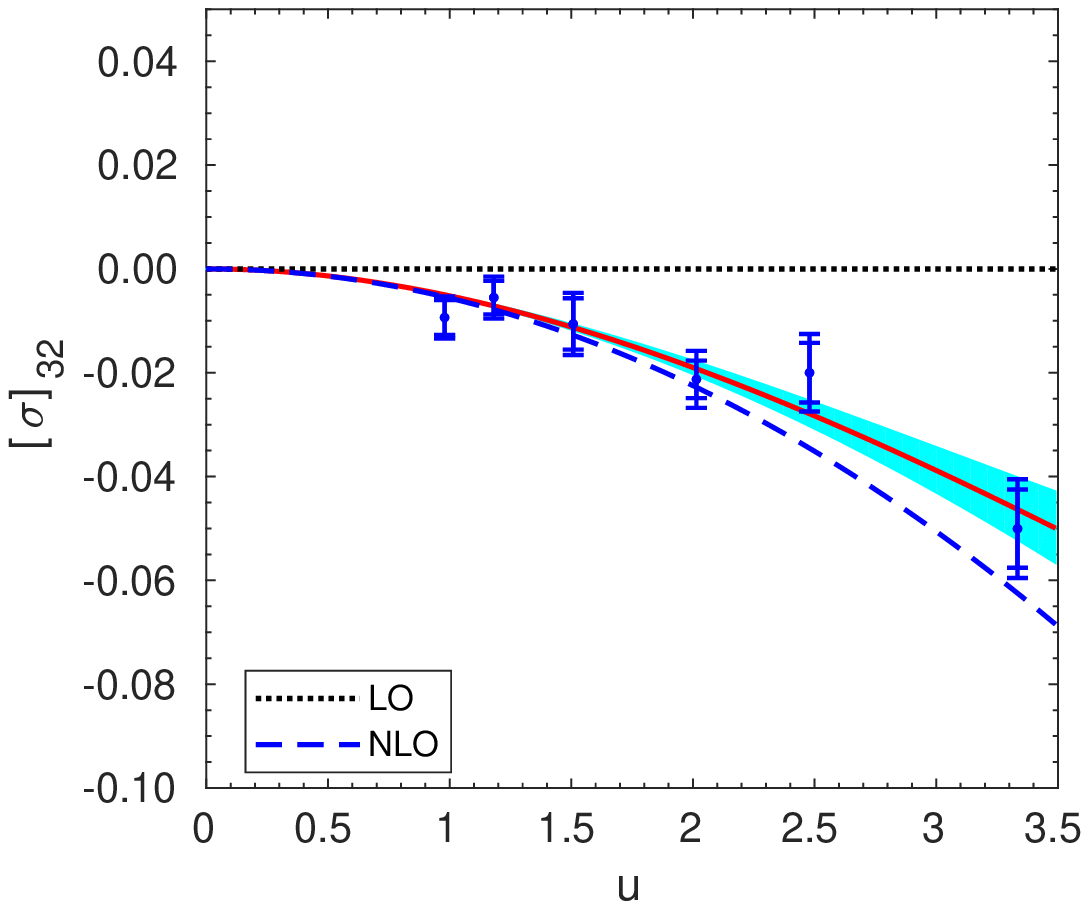}
\end{minipage}\hspace{1cm}
\begin{minipage}[t]{0.4\textwidth}
\includegraphics[width=7cm]{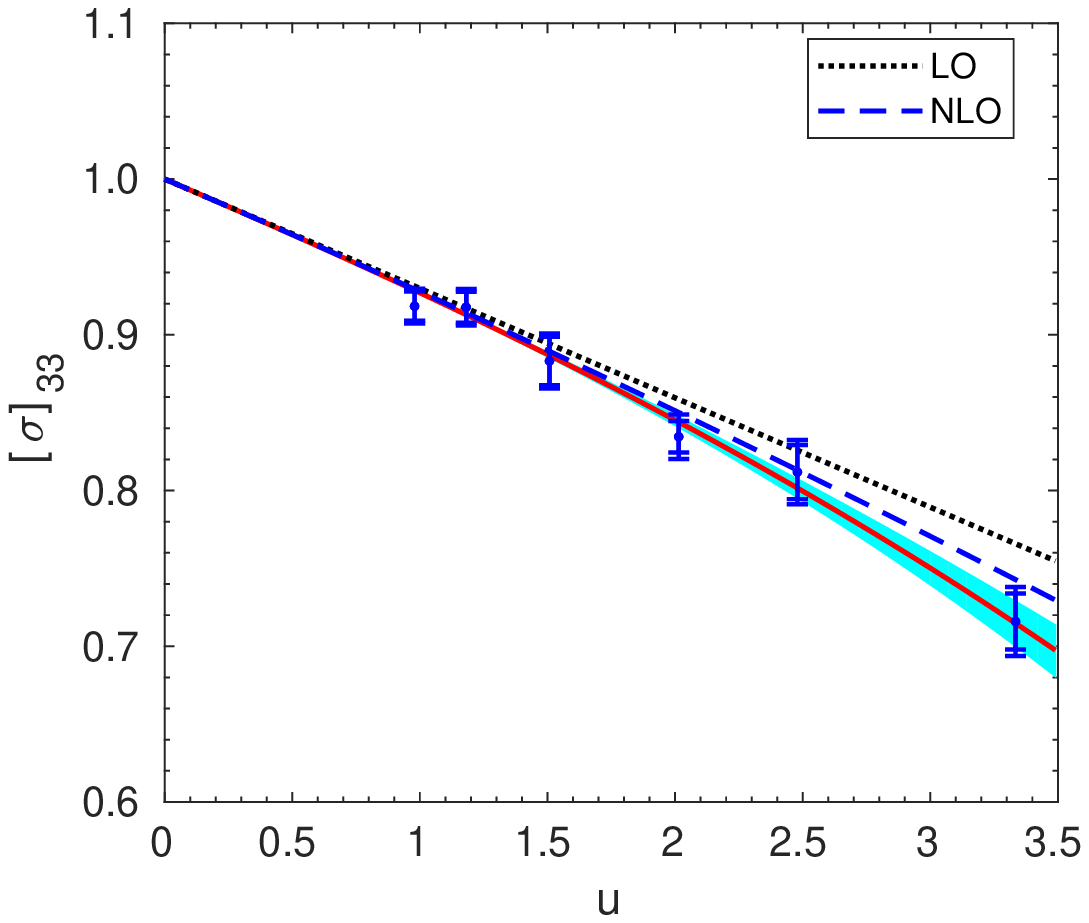}
\end{minipage}

\vspace{2mm}

\begin{minipage}[t]{0.4\textwidth}
\includegraphics[width=7cm]{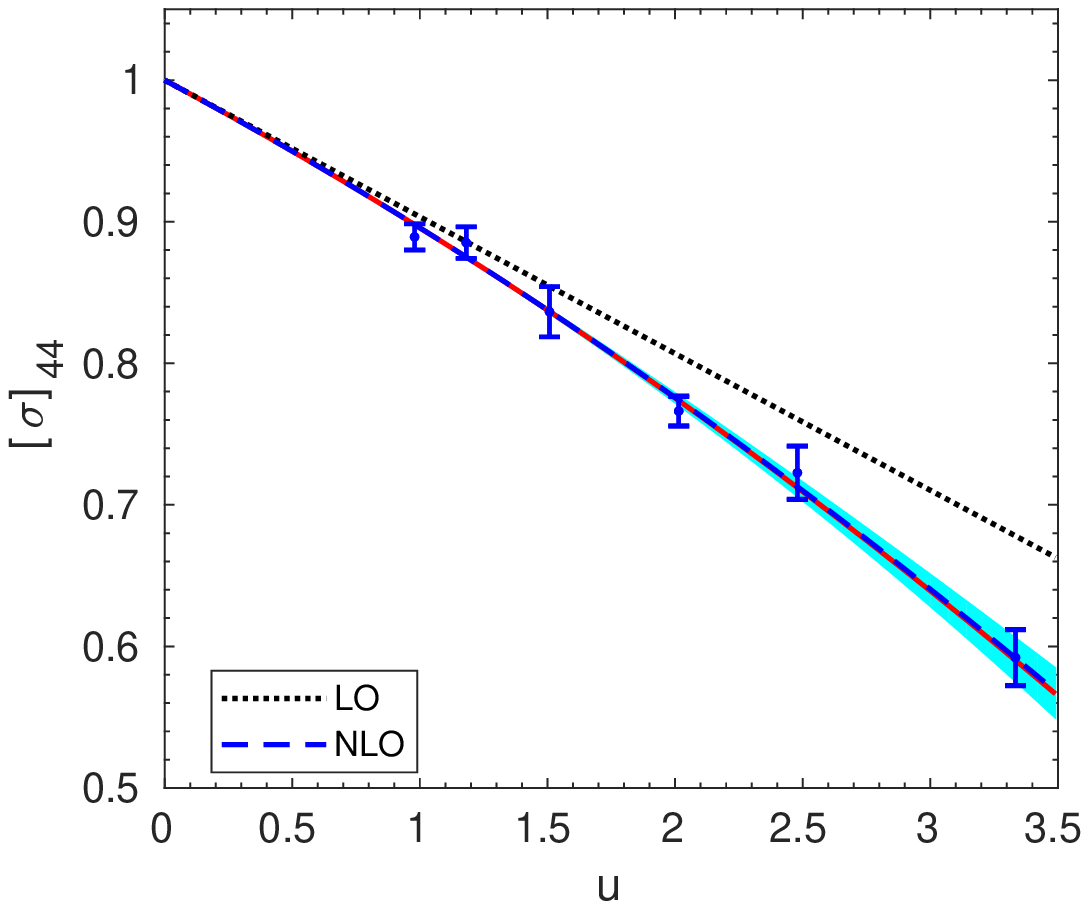}
\end{minipage}\hspace{1cm}
\begin{minipage}[t]{0.4\textwidth}
\includegraphics[width=7cm]{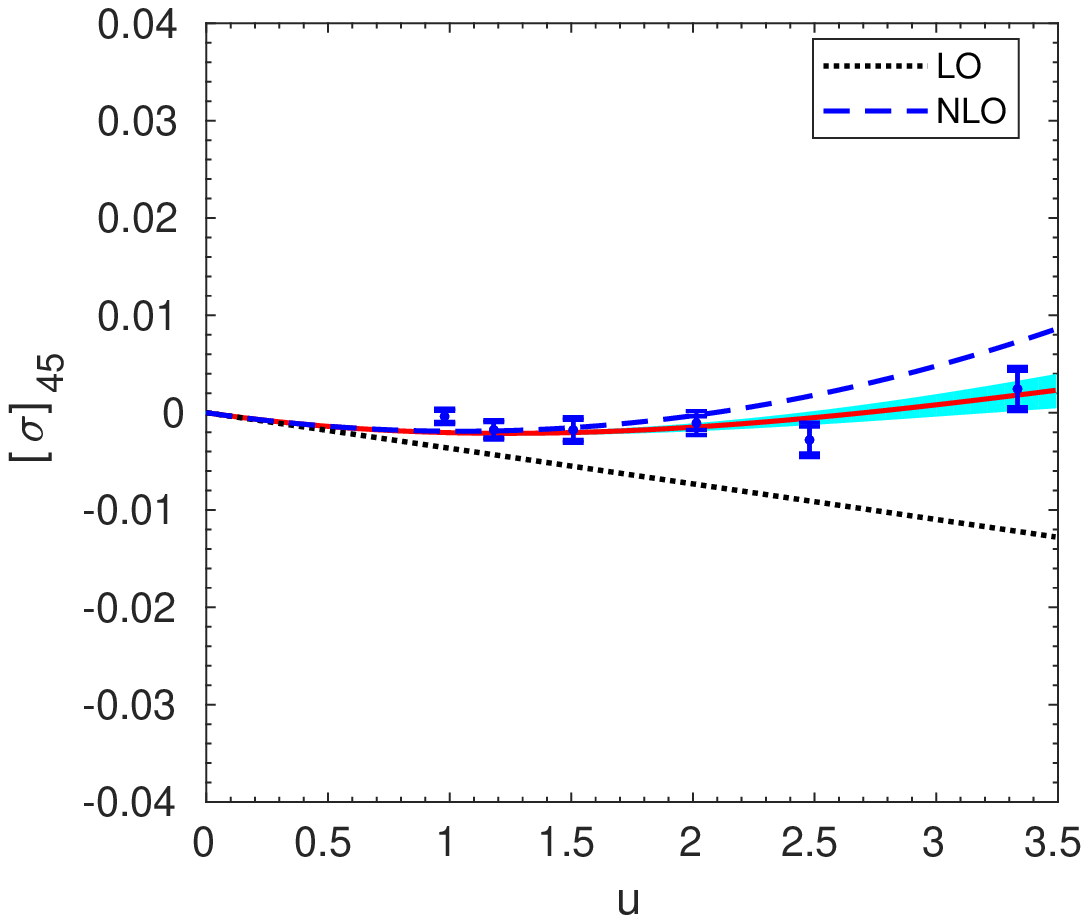}
\end{minipage}

\vspace{-0mm}

\begin{minipage}[t]{0.4\textwidth}
\includegraphics[width=7cm]{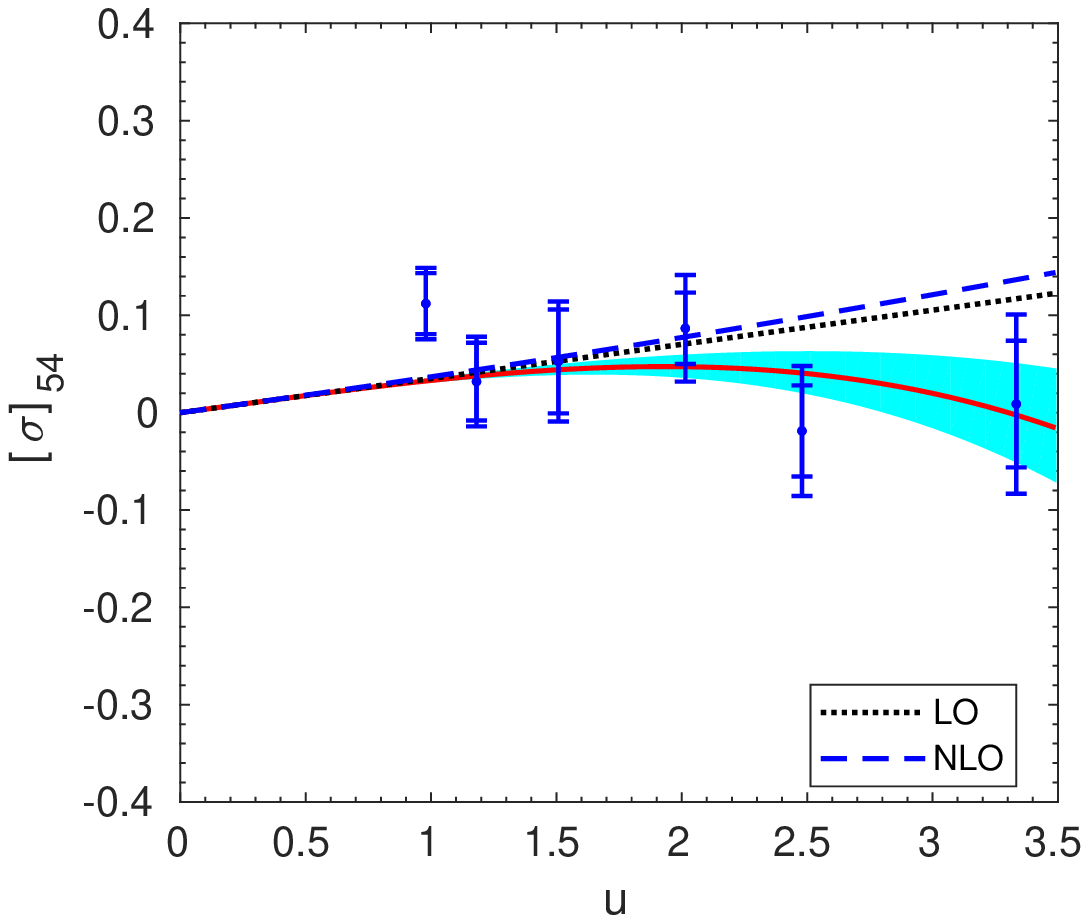}
\end{minipage}\hspace{1cm}
\begin{minipage}[t]{0.4\textwidth}
\includegraphics[width=7cm]{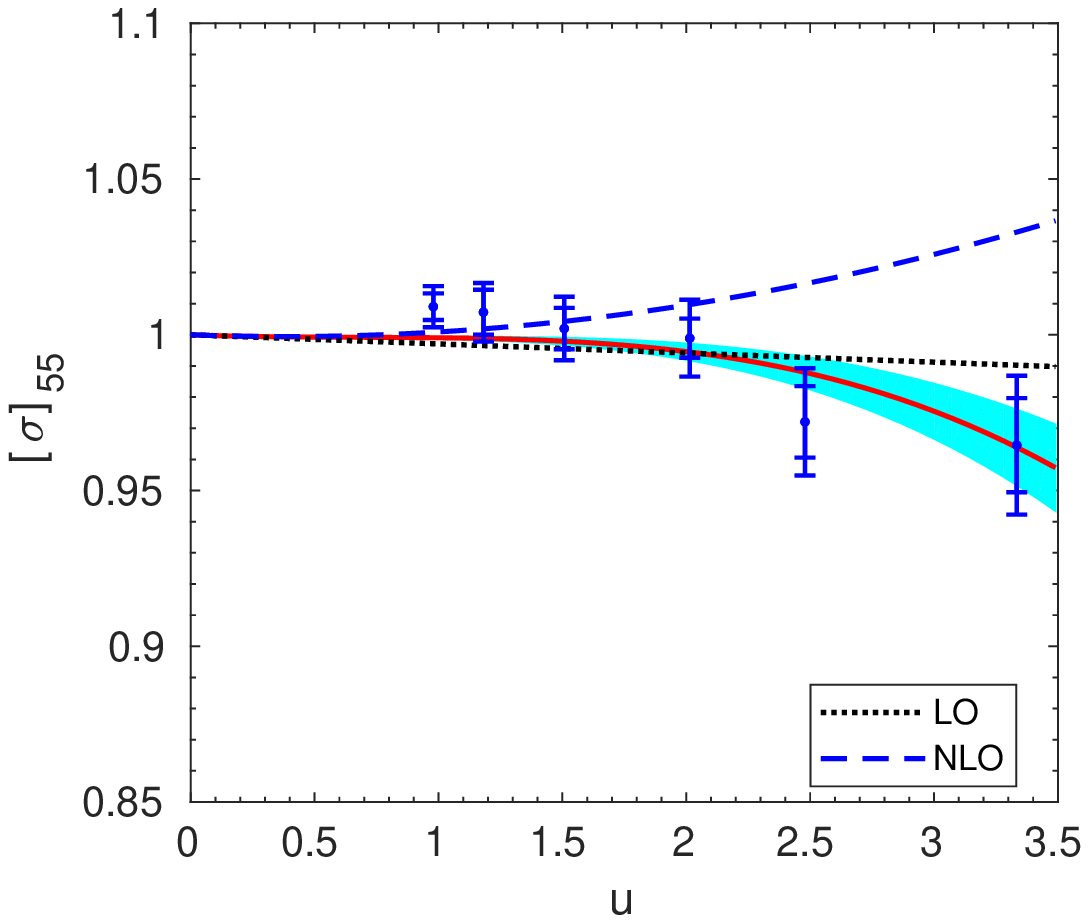}
\end{minipage}
\caption{Continuum matrix-SSFs for operator bases $\{ \cQ_2^-, \cQ_3^- \}$ (top) and $\{ \cQ_4^-, \cQ_5^- \}$ (bottom). 
The LO perturbative result is shown by the dotted black line, while the NLO one by the dashed blue line. The red line (with error band) is the non-perturbative result from the $\mathcal{O}(u^3)$ fit as described in the text.
The two error bars on each data point are the statistical and total uncertainties; the systematic error contributing to the latter
has been estimated as explained in the text.}

\label{fig:SSFs-}
\end{center}
\end{figure}

In order to compute the RG running of the operators in the continuum limit, matrix-SSFs have to be fit to the functional form shown
in~\req{eq:sigmaQ-PT}. Several fits have been tried out, with different orders in the polynomial expansion and  
$\br_2$ either kept fixed to its perturbative value or allowed to be a free fit parameter.
Fits with $\br_1$ fixed by perturbation theory and $\br_2$ the only free fit parameter do not describe the data well.
This is understandable, as deviations from LO are large for some matrix elements (for $\bsigma_{54}^+$ in particular) and knowledge of the NLO anomalous dimension $\bgamma^{(1)}$ (and therefore $\br_2$; cf. \req{eq:r2_pt}) is necessary for 
a well-converging fit. It is however an encouraging crosscheck that the $\br_2$ value returned by the fit is close to the perturbative prediction of Eq.~(\ref{eq:r2_pt}).
If, besides $\br_2$, we also include $\br_3$ as a free fit parameter, the results have large errors. 
The best option turns out to be the one with the polynomial expansion of~\req{eq:sigmaQ-PT} truncated at $\cO(u^4)$, $\br_1$ and $\br_2$
fixed to their perturbative values and $\br_3$ left as free fit parameter. 
The plots of the matrix-SSFs are collected in Figs.~\ref{fig:SSFs+} and \ref{fig:SSFs-}. 

In the same Figures we also show the LO and NLO perturbative results, calculated from \req{eq:sigmaQ-PT},
truncated at $\cO(u)$ and $\cO(u^2)$ respectively.
The comparison between the non-perturbative, the LO, and the NLO results provides
a useful assessment of the reliability of the perturbative series.
There is coincidence of all three curves at very small (perturbative) values of the squared gauge coupling $u$,
but this is obviously guaranteed by the form of our fit function, as described above. At larger $u$-values 
one would ideally hope to see the NLO curves lying closer to the non-perturbative ones, compared to the LO curves.
For $\bsigma^+$ this is mostly the case, as shown in Fig.~\ref{fig:SSFs+}, the only exception being $[\sigma^+]_{23}$
and $[\sigma^+]_{44}$. For the operator basis $\{ \cQ_2^+, \cQ_3^+ \}$, non-perturbative and NLO curves seem in
good agreement for the diagonal elements $[\sigma^+]_{22}$ and $[\sigma^+]_{33}$. This is less so for the
non-diagonal  $[\sigma^+]_{23}$ and $[\sigma^+]_{32}$. For the operator basis $\{ \cQ_4^+, \cQ_5^+ \}$,
non-perturbative and NLO curves mostly agree, with the exception of $[\sigma^+]_{44}$.
We also note that the non-perturbative $[\sigma^+]_{23}$
tends to decrease at large $u$, unlike the monotonically increasing perturbative predictions.
For $\bsigma^-$ the NLO curves lie closer to the non-perturbative results  compared to the LO ones,
in all cases but $[\sigma^-]_{23}$ and $[\sigma^-]_{55}$ (for $[\sigma^-]_{54}$ LO and NLO are very close to each other).
In several cases non-perturbative and NLO curves are in fair, or even excellent, agreement also at large $u$-values
(cf. $[\sigma^-_{32}]$, $[\sigma^-_{33}]$, $[\sigma^-_{44}]$ and $[\sigma^-_{45}]$). In other cases this comparison in less satisfactory.
Note that the NLO $[\sigma^-]_{54}$ and $[\sigma^-]_{55}$ curves
are monotonically increasing, as opposed to the non-perturbative ones.
In conclusion the overall picture in the renormalisation scheme under investigation is in
accordance with our general expectations, although there are signs of slow or bad convergence of the perturbative
results to the non-perturbative ones.

Once the matrix-SSFs are known as continuum functions of the renormalised coupling, we can obtain the RG-running matrix $\bU(\mu_{\rm had},2^n \mu_{\rm had}) = \bsigma(u_1) \ldots \bsigma(u_n)$; cf.~\req{eq:ssf_prod}. We check the reliability of our results by writing~\req{eq:Urun} as
\begin{align}
\tilde \bU(\mu_{\rm had}) & =  \tilde \bU(2^n \mu_{\rm had})  \left [ \bU(\mu_{\rm had},2^n \mu_{\rm had}) \right ]^{-1} \nonumber  \\
& = \left[\frac{\gbar^2(2^n \mu_{\rm had})}{4\pi}\right]^{-\frac{\bgamma^{(0)}}{2b_0}}\,\bW(2^n \mu_{\rm had})
\left[\bU(\mu_{\rm had},2^n \mu_{\rm had})\right]^{-1} \, .
\label{eq:oprun}
\end{align}
The matrix $\tilde \bU(\mu_{\rm had})$ does not depend on the higher-energy scale $2^n\mu_{\rm had}$, so the $n$-dependence on the rhs should in principle cancel out.
We check this by computing the second line for varying $n$, using our non-perturbative result for $\bU(\mu_{\rm had},2^n \mu_{\rm had})$ and the perturbative one for $\tilde \bU(2^n \mu_{\rm had})$. As explained in the comments following \req{eq:rgi_prod}, the latter is obtained as the NLO-2/3PT $\bW(2^n \mu_{\rm had})$, multiplied by $\left[ \gbar^2(2^n \mu_{\rm had})/(4\pi)\right]^{-(\bgamma^{(0)}/2b_0)}$. The scale $\mu_{\rm had}$ is held fixed through $\gbar^2(\mu_{\rm had})=4.61$, which defines $L_{\rm max}$; see the $\NF=2$ running coupling computation of ref.~\cite{DellaMorte:2005kg} for details. The higher-energy scale $2^n \mu_{\rm had}$ is varied over a range of values $n = 0, \ldots , 8$; for each of these
$\tilde \bU(\mu_{\rm had})$ is computed. Our results are shown in 
Tabs.~\ref{tab:mupt23} and \ref{tab:mupt45}. As expected, $2^n \mu_{\rm had}$-independence sets in with increasing $n$. 

More specifically, taking $\log(\Lambda_{\rm SF}/\mu_{\rm had}) = -1.298(58)$ from ref.~\cite{DellaMorte:2005kg} and $r_0 \Lambda_{\rm SF} = 0.30(3)$ from ref.~\cite{DellaMorte:2004bc} with $r_0 = 0.50 \, \rm{fm}$, we obtain the hadronic matching energy scale $\mu_{\rm had} \approx 432(50) \, \rm{MeV}$. Our final results for the non-perturbative running at $\mu_{\rm had}$ are obtained from \req{eq:oprun} and for $n=8$. They are:
\begin{eqnarray}
\label{eq:U+23}
\tilde \bU_{(2,3)}^+(\mu_{\rm had}) &=& \begin{pmatrix} 1.2028(436)(3)  & 0.1202(692)(180)  \\ -0.0423(36)(2) & 0.4572(152)(8) \end{pmatrix}  \, \, , \\  
\label{eq:U+45}
\tilde \bU_{(4,5)}^+(\mu_{\rm had}) &=& \begin{pmatrix} 0.5657(158)(2) & 0.0224(11)(0) \\ 1.7245(4070)(627) & 2.1317(679)(25) \end{pmatrix} \,\, ,
\end{eqnarray}
for the operator bases $\{\cQ^{+}_2,\cQ^{+}_3\}$,$\{\cQ^{+}_4,\cQ^{+}_5\}$ and 
\begin{eqnarray}
\label{eq:U-23}
\tilde \bU_{(2,3)}^-(\mu_{\rm had}) &=& \begin{pmatrix} 1.2377(281)(19) & -0.8289(486)(69)  \\ 0.0420(42)(2) & 0.4192(131)(8) \end{pmatrix} \, \, , \\
\label{eq:U-45}
\tilde \bU_{(4,5)}^-(\mu_{\rm had}) &=& \begin{pmatrix} 0.4297(195)(5) & -0.03145(88)(1) \\ -1.6825(2182)(387) & 0.8976(176)(29) \end{pmatrix} \,\, ,
\end{eqnarray} 
for $\{\cQ^{-}_2,\cQ^{-}_3\}$,$\{\cQ^{-}_4,\cQ^{-}_5\}$. The first error refers to the statistical uncertainty, while the second is the systematic one due to the use of NLO-2/3PT at the higher scale $2^n \mu_{\rm had}$.
We estimate the systematic error as the difference between the final result, obtained with perturbation theory setting in at
scale $2^{8}\mu_{\rm had}$, and the one where perturbation theory sets in at $2^{7}\mu_{\rm had}$ 
 (cf. Tabs.~\ref{tab:mupt23},\ref{tab:mupt45}).

We note that systematic errors are almost negligible compared to statistical ones, the latter being the result of error propagation in the product of matrix-SSFs from $\mu_{\rm had}$ to $2^{8}\mu_{\rm had}$. This however does not tell us much about the accuracy of NLO-2/3PT around the scale $\mu_{\rm pt}=2^n \mu_{\rm had}$. We investigate this issue 
in Appendix~\ref{sec:checks}, where we compare $\bsigma(u_{n})$, calculated in NLO-2/3PT and non-perturbatively.
For several matrix elements of $\bsigma(u_{n})$ we see that NLO-2/3PT is
not precise enough, even at the largest scale we can reach (corresponding to $n=8$).

\begin{figure}
\begin{center}
\vspace*{-3\baselineskip}
\includegraphics[width=0.9\textwidth]{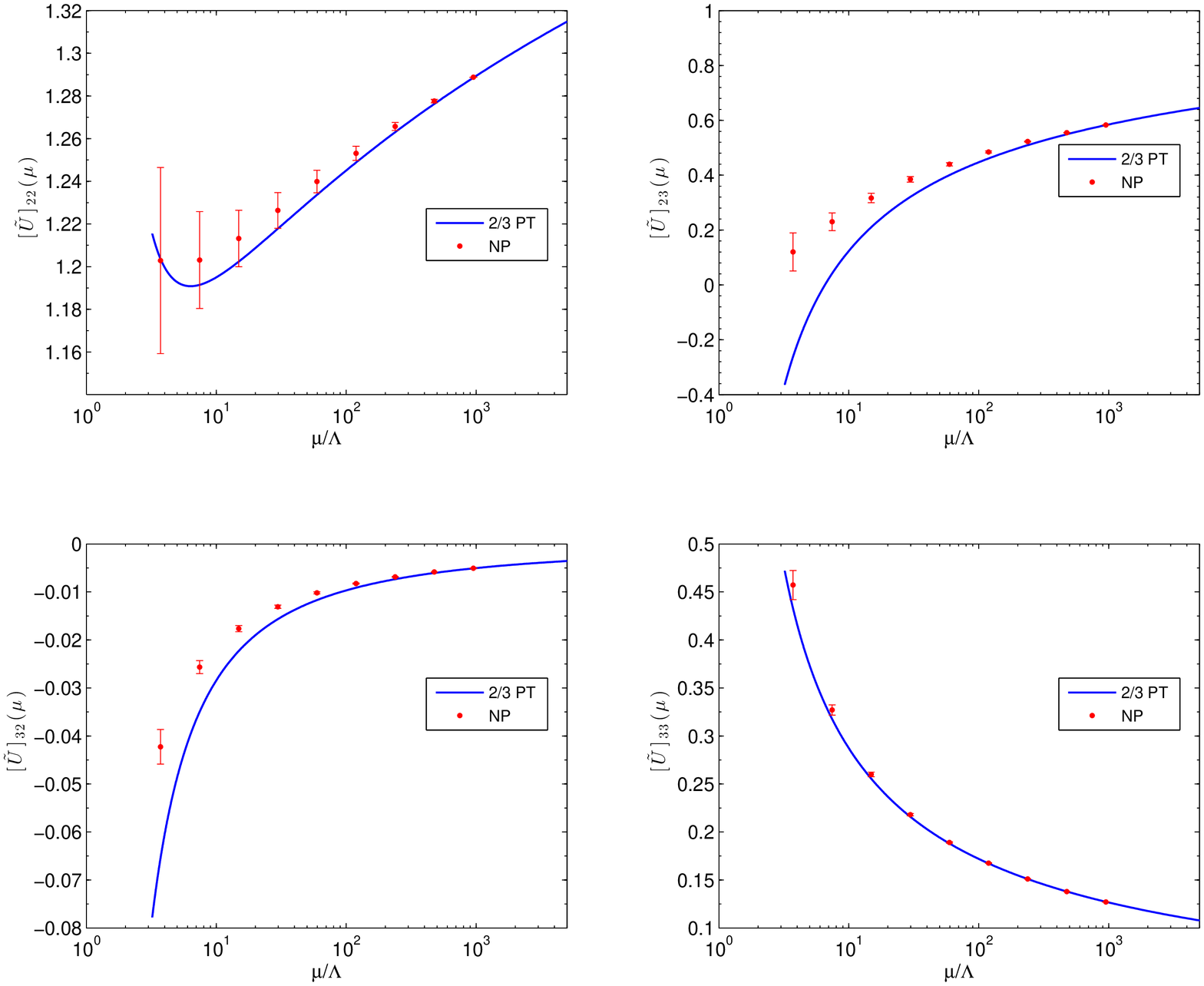}

\vspace{1mm}

\includegraphics[width=0.9\textwidth]{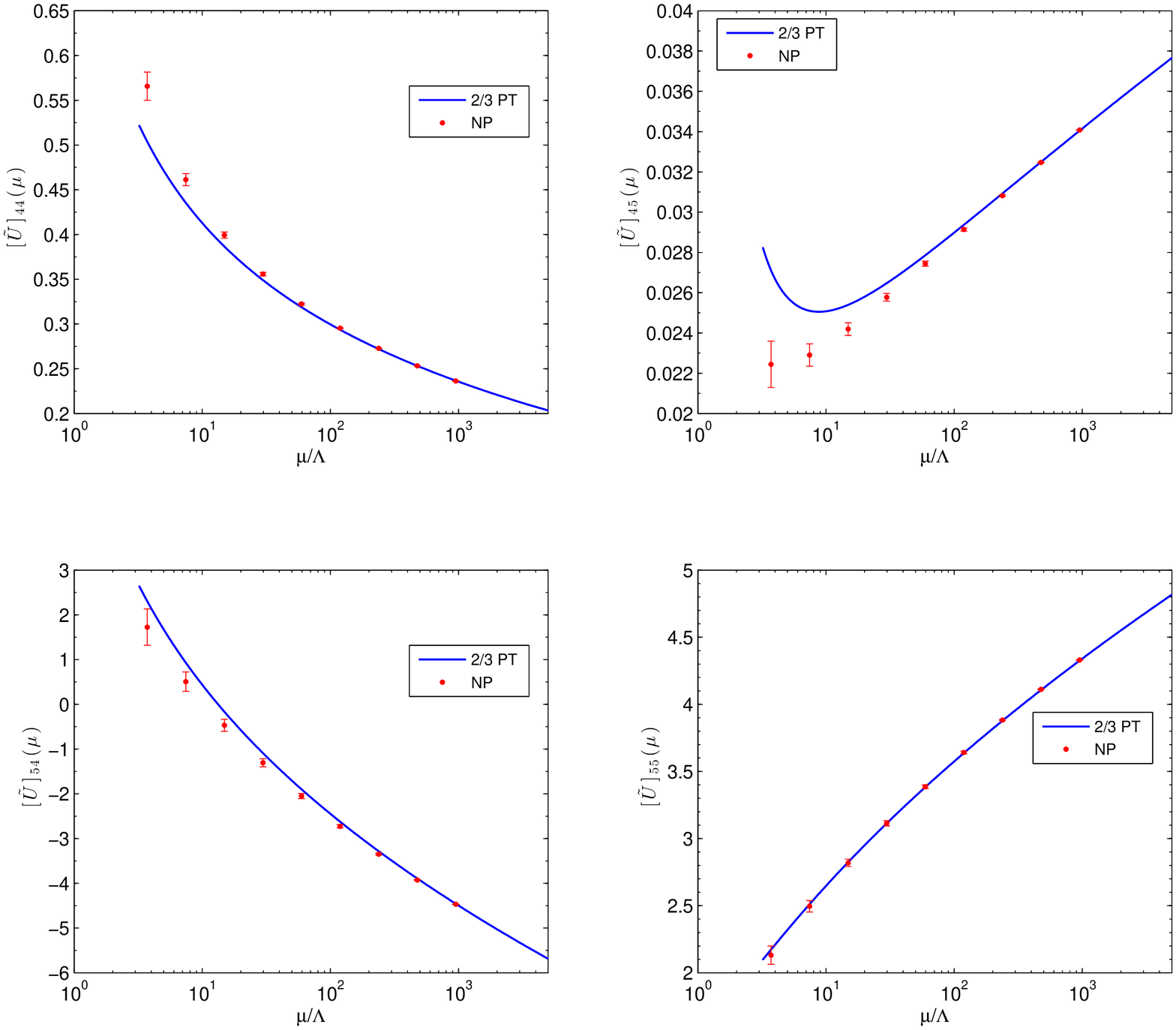}

\caption{Non-perturbative running $\tilde \bU_{(2,3)}^+(\mu)$ for the operator basis $\{ \cQ_2^+, \cQ_3^+ \}$ (top) and
$\tilde \bU_{(4,5)}^+(\mu)$ of the operator basis $\{ \cQ_4^+, \cQ_5^+ \}$ (bottom). 
Results are compared to the perturbative predictions, obtained by numerically integrating~\req{eq:rg_W}, with 
the NLO result for $\bgamma$ and the NNLO one for $\beta$, in the SF scheme.}
\label{fig:RUN+}
\end{center}
\end{figure}

\begin{figure}
\begin{center}
\vspace*{-3\baselineskip}
\includegraphics[width=0.9\textwidth]{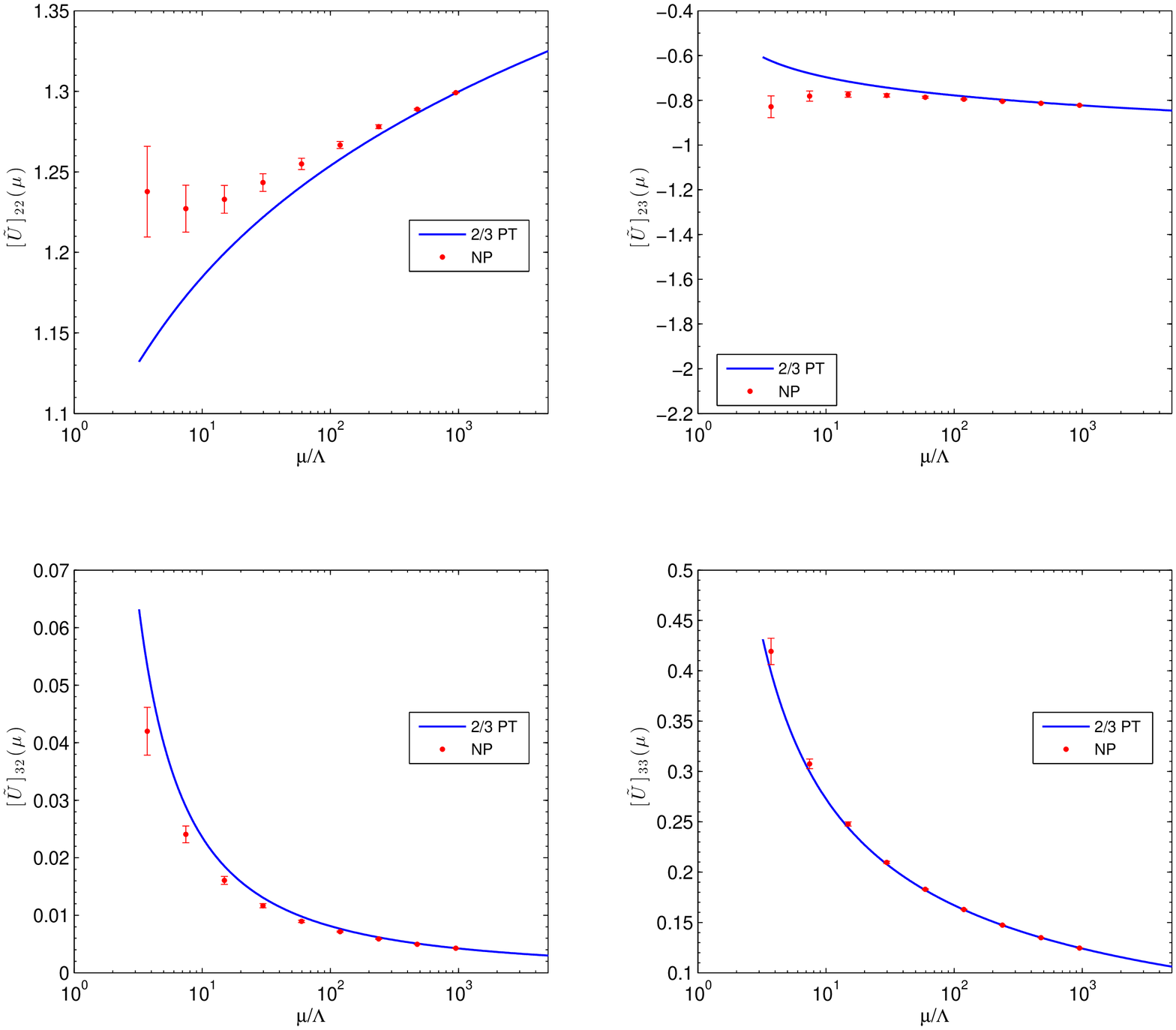}

\vspace{1mm}

\includegraphics[width=0.9\textwidth]{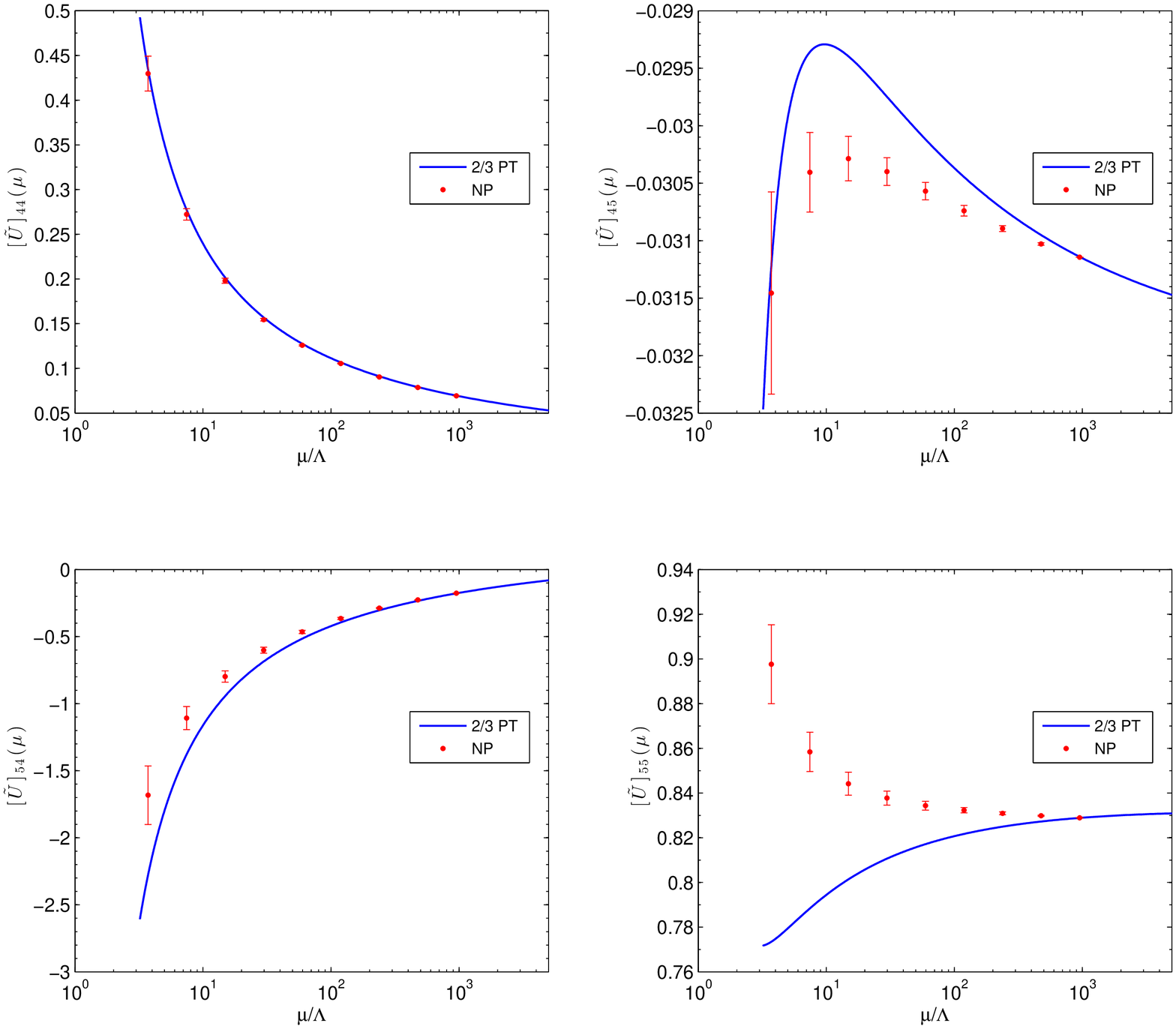}

\caption{Non-perturbative running $\tilde \bU_{(2,3)}^-(\mu)$ of the operator basis $\{ \cQ_2^-, \cQ_3^- \}$ (top) and
$\tilde \bU_{(4,5)}^-(\mu)$ of the operator basis $\{ \cQ_4^-, \cQ_5^- \}$ (bottom). 
Results are compared to the perturbative predictions, obtained by numerically integrating~\req{eq:rg_W}, with 
the NLO result for $\bgamma$ and the NNLO one for $\beta$, in the SF scheme.}
\label{fig:RUN-}
\end{center}
\end{figure}

We now play the inverse game, keeping fixed $\mu_{\rm pt}=2^8 \mu_{\rm had}$ and calculating
\begin{align}
\tilde \bU(\mu) =  \left[\frac{\gbar^2(\mu_{\rm pt})}{4\pi}\right]^{-\frac{\bgamma^{(0)}}{2b_0}}\,\bW(\mu_{\rm pt})
\left[\bU(\mu,\mu_{\rm pt})\right]^{-1} \, .
\label{eq:oprun-inv}
\end{align}
for decreasing $\mu$. The results for $\tilde \bU(\mu)$ are shown in Figs.~\ref{fig:RUN+} and \ref{fig:RUN-}. They are the first non-perturbative computation of the RG-evolution of operators which mix under renormalisation in the continuum. We stress that these results are scheme dependent.
Note that the computation thus described {\it enforces} the coincidence of our most perturbative point to the perturbative prediction, which
we {\it assume} to describe accurately the running from $\mu_{\rm pt} \sim \cO(M_{\rm W})$ to infinity. The discrepancies between perturbation theory and our results are evident at ever decreasing scales $\mu$. 
These discrepancies are sometimes dramatic; e.g. $[\tilde U^-]_{55}(\mu)$. This is related to the
discussion of Figs.~\ref{fig:SSFs+} and \ref{fig:SSFs-} above, concerning disagreements between non-perturbative and NLO
behaviour of several $\bsigma$ matrix elements. Since $\left[\bU(\mu,\mu_{\rm pt})\right]^{-1}$ in \req{eq:oprun-inv} is a product
of several $\bsigma$ matrices, these disagreements accumulate, becoming very sizeable as $\mu/\Lambda_{\rm SF}$ decreases.

Finally, we compare the perturbative (NLO-2/3PT) to the non-perturbative RG evolution $\bU(\mu,\mu_*)$ between scales $\mu$ and $\mu_*$, where $\mu_*=3.46$ GeV is kept fixed and $\mu$ is varied in the range [0.43 GeV, 110 GeV]. The comparison is described Appendix~\ref{sec:checks} and confirms the unreliability of the perturbative computation of the RG running at scales of about $3 \, \rm GeV$.

\begin{table}[t!]
\begin{scriptsize}
\begin{center}
\begin{tabular}{ccc}
\toprule
$n$ & $\tilde \bU_{(2,3)}^+(\mu_{\rm had})$ & $\tilde \bU_{(2,3)}^-(\mu_{\rm had})$ \\  
\midrule
0 & $\begin{pmatrix} 1.215505  & -0.363611 \\ -0.077786 & 0.472123 \end{pmatrix}$ & $\begin{pmatrix} 1.132141 & -0.607507 \\ 0.063161 & 0.431281 \end{pmatrix}$   \\ \midrule
1 & $\begin{pmatrix} 1.2016(190)  & -0.1649(270) \\ -0.0532(22) & 0.4562(80) \end{pmatrix}$ & $\begin{pmatrix} 1.1837(126) & -0.6972(172)  \\ 0.0484(24) & 0.4185(67) \end{pmatrix}$   \\ \midrule
2 & $\begin{pmatrix} 1.2022(283)  & -0.0773(425) \\ -0.0476(29) & 0.4580(112) \end{pmatrix}$ & $\begin{pmatrix} 1.2057(186) & -0.7419(277)  \\ 0.0452(32) & 0.4200(94) \end{pmatrix}$   \\ \midrule
3 & $\begin{pmatrix} 1.2030(336)  & -0.0212(499) \\ -0.0453(32) & 0.4595(129) \end{pmatrix}$ & $\begin{pmatrix} 1.2177(221) & -0.7693(344)  \\ 0.0440(36) & 0.4213(110) \end{pmatrix}$   \\ \midrule
4 & $\begin{pmatrix} 1.2035(369)  & 0.0212(559)  \\ -0.0441(34) & 0.4599(138) \end{pmatrix}$ & $\begin{pmatrix} 1.2250(243) & -0.7886(387)  \\ 0.0433(38) & 0.4216(118) \end{pmatrix}$   \\ \midrule
5 & $\begin{pmatrix} 1.2035(395)  & 0.0542(609)  \\ -0.0434(35) & 0.4595(144) \end{pmatrix}$ & $\begin{pmatrix} 1.2298(258) & -0.8027(422)  \\ 0.0428(40) & 0.4212(124) \end{pmatrix}$   \\ \midrule
6 & $\begin{pmatrix} 1.2033(412)  & 0.0808(644)  \\ -0.0429(35) & 0.4588(147) \end{pmatrix}$ & $\begin{pmatrix} 1.2333(268) & -0.8135(447)  \\ 0.0424(41) & 0.4206(127) \end{pmatrix}$   \\ \midrule
7 & $\begin{pmatrix} 1.2031(426)  & 0.1022(674)  \\ -0.0425(36) & 0.4580(150) \end{pmatrix}$ & $\begin{pmatrix} 1.2358(276) & -0.8220(468)  \\ 0.0422(41) & 0.4200(130) \end{pmatrix}$  \\ \midrule
8 & $\begin{pmatrix} 1.2028(436)  & 0.1202(692)  \\ -0.0423(36) & 0.4572(152) \end{pmatrix}$ & $\begin{pmatrix} 1.2377(281) & -0.8289(486)  \\ 0.0420(42) & 0.4192(131) \end{pmatrix}$  \\ \bottomrule
\end{tabular}
\end{center}
\end{scriptsize}
\caption{The matrix $\tilde \bU_{(2,3)}^\pm(\mu_{\rm had})$, corresponding to the operator bases $\{ \cQ_2^\pm, \cQ_3^\pm \}$.
It is computed for a fixed low-energy scale $\mu_{\rm had}$ and varying higher-scales $2^n\mu_{\rm had}$. For sufficiently large $n$,
the results should not depend on the higher-energy scale.}
\label{tab:mupt23}
\end{table}

\begin{table}[t!]
\begin{scriptsize}
\begin{center}
\begin{tabular}{ccc}
\toprule
$n$ & $\tilde \bU_{(4,5)}^+(\mu_{\rm had})$ & $\tilde \bU_{(4,5)}^-(\mu_{\rm had})$ \\  
\midrule
0 & $\begin{pmatrix} 0.522119 & 0.028246 \\ 2.648160 & 2.098693 \end{pmatrix}$ & $\begin{pmatrix} 0.492746 & -0.032468 \\ -2.607554 & 0.771786 \end{pmatrix}$ \\ \midrule
1 & $\begin{pmatrix} 0.5417( 73) & 0.0242( 7) \\ 2.3620(1360) & 2.1229(300) \end{pmatrix}$ & $\begin{pmatrix} 0.4531( 96) & -0.0304(5) \\ -2.2066( 850) & 0.8223( 81) \end{pmatrix}$ \\ \midrule
2 & $\begin{pmatrix} 0.5537(106) & 0.0232( 9) \\ 2.2306(2151) & 2.1222(446) \end{pmatrix}$ & $\begin{pmatrix} 0.4474(134) & -0.0305(7) \\ -2.0604(1298) & 0.8502(119) \end{pmatrix}$ \\ \midrule
3 & $\begin{pmatrix} 0.5602(126) & 0.0228(10) \\ 2.1242(2675) & 2.1205(534) \end{pmatrix}$ & $\begin{pmatrix} 0.4443(159) & -0.0307(8) \\ -1.9636(1558) & 0.8670(142) \end{pmatrix}$ \\ \midrule
4 & $\begin{pmatrix} 0.5636(136) & 0.0226(11) \\ 2.0255(3040) & 2.1214(585) \end{pmatrix}$ & $\begin{pmatrix} 0.4411(172) & -0.0309(8) \\ -1.8862(1750) & 0.8779(154) \end{pmatrix}$ \\ \midrule
5 & $\begin{pmatrix} 0.5652(143) & 0.0225(11) \\ 1.9411(3365) & 2.1237(619) \end{pmatrix}$ & $\begin{pmatrix} 0.4379(181) & -0.0311(8) \\ -1.8213(1884) & 0.8854(163) \end{pmatrix}$ \\ \midrule
6 & $\begin{pmatrix} 0.5656(150) & 0.0225(11) \\ 1.8581(3668) & 2.1266(643) \end{pmatrix}$ & $\begin{pmatrix} 0.4350(186) & -0.0312(9) \\ -1.7669(2009) & 0.8905(168) \end{pmatrix}$ \\ \midrule
7 & $\begin{pmatrix} 0.5659(154) & 0.0224(11) \\ 1.7872(3884) & 2.1292(663) \end{pmatrix}$ & $\begin{pmatrix} 0.4322(191) & -0.0314(9) \\ -1.7212(2105) & 0.8947(174) \end{pmatrix}$ \\ \midrule
8 & $\begin{pmatrix} 0.5657(158) & 0.0224(11) \\ 1.7245(4070) & 2.1317(679) \end{pmatrix}$ & $\begin{pmatrix} 0.4297(195) & -0.0315(9) \\ -1.6825(2182) & 0.8976(176) \end{pmatrix}$ \\ \bottomrule
\end{tabular}
\end{center}
\end{scriptsize}
\caption{The matrix $\tilde \bU_{(4,5)}^\pm(\mu_{\rm had})$, corresponding to the operator bases $\{ \cQ_4^\pm, \cQ_5^\pm \}$.
It is computed for a fixed low-energy scale $\mu_{\rm had}$ and varying higher-scales $2^n\mu_{\rm had}$. For sufficiently large $n$,
the results should not depend on the higher-energy scale.}
\label{tab:mupt45}
\end{table}

\subsection{Matching to hadronic observables with non-perturbatively $\Oa$ improved Wilson fermions}

Having computed the non-perturbative evolution matrices $\tilde \bU(\mu_{\rm had})$ as in \req{eq:oprun-inv}, which provide the RG-running at the low energy scale $\mu_{\rm had}$, we proceed to establish the connection between bare lattice operators and their RGI counterparts. Starting from the
definition of~\req{eq:rgi_mix}, we write the RGI operator as
\begin{eqnarray}
\label{eq:rgi_mix-2}
\hat{\boldsymbol \cQ} &\equiv& \left[\frac{\gbar^2(\mu_{\rm pt})}{4\pi}\right]^{-\frac{{\boldsymbol {\boldsymbol \gamma}}^{(0)}}{2b_0}}\bW(\mu_{\rm pt})\overline{\boldsymbol \cQ}(\mu_{\rm pt})  \\
\nonumber
&=& \left[\frac{\gbar^2(\mu_{\rm pt})}{4\pi}\right]^{-\frac{{\boldsymbol {\boldsymbol \gamma}}^{(0)}}{2b_0}}\bW(\mu_{\rm pt}) \bU(\mu_{\rm pt},\mu_{\rm had})
\lim_{g_0^2 \rightarrow 0} \Big [ {\boldsymbol \cZ}(g_0^2, a\mu_{\rm had}) \,\, {\boldsymbol \cQ} (g_0^2) \Big ] \,\,.
\end{eqnarray}
$\hat{\boldsymbol \cQ} $ is independent of any renormalisation scheme or scale; of course it is also independent of the regularisation. It is a product of 
several quantities:
\begin{itemize}
\item
The factors $[ \gbar^2(\mu_{\rm pt})/(4\pi)]^{-\frac{{\boldsymbol {\boldsymbol \gamma}}^{(0)}}{2b_0}}$ and $\bW(\mu_{\rm pt})$ depend on a high-energy scale $\mu_{\rm pt}$ and
are calculated in NLO perturbation theory. This was one of the main objectives of ref.~\cite{Papinutto:2016xpq}.
\item
The running matrix $\bU(\mu_{\rm pt},\mu_{\rm had})$ is known between the high-energy scale $\mu_{\rm pt}$ and a low-energy scale $\mu_{\rm had}$; its non-perturbative 
computation for $\NF=2$ QCD is the main objective of the present work.
\item
The product of the last two factors ${\boldsymbol \cZ}(g_0^2, a\mu_{\rm had}) \,\, {\boldsymbol \cQ} (g_0^2)$ stands for the usual lattice computation of bare hadronic quantities and their renormalisation
constants on large physical volumes and for several bare couplings, with the continuum limit taken though extrapolation.
\end{itemize}

Although the last item in the above list is beyond the scope of this paper, we have computed ${\boldsymbol \cZ}(g_0^2,a\mu_{had})$ following~\cite{DellaMorte:2005kg}, at three values of the lattice spacing, namely $\beta=6/g_0^2 = \{5.20,5.29,5.40\}$, which are in the range commonly used for simulations of $\NF=2$ QCD in physically large volumes. The results are listed in Tabs.~\ref{tab:Zhad23}, \ref{tab:Zhad45}. In order to interpolate to the target renormalized coupling $u(\mu_{\rm had})=4.61$, the data can be fitted with a polynomial.
Our numerical studies reveal that additional values of $\beta$ would
be needed to improve the quality of the interpolation to the
target value of the coupling.

\begin{table}[t!]
\begin{scriptsize}
\begin{center}
\begin{tabular}{cccccc}
\toprule
$\beta$ & $\kappa_{cr}$ & $L/a$ & $\bar{g}^2(L)$ & ${\boldsymbol \cZ}_{(23)}^+$ & ${\boldsymbol \cZ}_{(23)}^-$ \\

\midrule
\multirow{2}*{5.20} &
\multirow{2}*{0.13600}
& 4 & 3.65 & 
 $\begin{pmatrix} 0.5992(11) & 0.31835(83) \\ 
 0.08539(42) & 0.35980(88)\\ 
 \end{pmatrix}$ 
 & $\begin{pmatrix} 0.5048(11) & -0.12417(81) \\ 
 -0.08479(37) & 0.39148(77)\\ 
 \end{pmatrix}$ \\ 
& {} & 6 & 4.61 & 
 $\begin{pmatrix} 0.6026(12) & 0.34048(59) \\ 
 0.08647(33) & 0.29400(61)\\ 
 \end{pmatrix}$ 
 & $\begin{pmatrix} 0.50745(86) & -0.17402(63) \\ 
 -0.08586(34) & 0.31768(58)\\ 
 \end{pmatrix}$ \\ 

\midrule
\multirow{3}*{5.29} &
\multirow{3}*{0.13641}
& 4 & 3.39 & 
 $\begin{pmatrix} 0.6179(11) & 0.31837(69) \\ 
 0.08123(33) & 0.38268(82)\\ 
 \end{pmatrix}$ 
 & $\begin{pmatrix} 0.53117(83) & -0.12960(76) \\ 
 -0.08047(41) & 0.41335(68)\\ 
 \end{pmatrix}$ \\ 
& {} & 6 & 4.30 & 
 $\begin{pmatrix} 0.6212(11) & 0.33681(81) \\ 
 0.07975(35) & 0.31743(68)\\ 
 \end{pmatrix}$ 
 & $\begin{pmatrix} 0.53520(90) & -0.17551(80) \\ 
 -0.07941(40) & 0.34077(70)\\ 
 \end{pmatrix}$ \\ 
& {} & 8 & 5.65 & 
 $\begin{pmatrix} 0.6274(13) & 0.35466(78) \\ 
 0.08400(49) & 0.27293(68)\\ 
 \end{pmatrix}$ 
 & $\begin{pmatrix} 0.5317(10) & -0.2035(10) \\ 
 -0.08554(50) & 0.29424(62)\\ 
 \end{pmatrix}$ \\ 

\midrule
\multirow{3}*{5.40} &
\multirow{3}*{0.13669}
& 4 & 3.19 & 
 $\begin{pmatrix} 0.6367(10) & 0.31526(70) \\ 
 0.07672(32) & 0.40904(83)\\ 
 \end{pmatrix}$ 
 & $\begin{pmatrix} 0.55721(81) & -0.13146(75) \\ 
 -0.07610(28) & 0.43891(82)\\ 
 \end{pmatrix}$ \\ 
& {} & 6 & 3.86 & 
 $\begin{pmatrix} 0.63422(95) & 0.33226(72) \\ 
 0.07429(37) & 0.34047(67)\\ 
 \end{pmatrix}$ 
 & $\begin{pmatrix} 0.55768(81) & -0.17545(73) \\ 
 -0.07358(35) & 0.36360(59)\\ 
 \end{pmatrix}$ \\ 
& {} & 8 & 4.75 & 
 $\begin{pmatrix} 0.6422(13) & 0.35228(79) \\ 
 0.07738(41) & 0.29670(64)\\ 
 \end{pmatrix}$ 
 & $\begin{pmatrix} 0.55925(84) & -0.20644(70) \\ 
 -0.07761(50) & 0.31681(65)\\ 
 \end{pmatrix}$ \\ 
\midrule
\bottomrule\end{tabular}
\caption{Renormalisation constants at hadronic-scale $\beta$-values for the operator bases $\{ \cQ_2^\pm, \cQ_3^\pm \}$. }
\label{tab:Zhad23}
\end{center}
\end{scriptsize}
\end{table}
\newpage 

\begin{table}[t!]
\begin{scriptsize}
\begin{center}
\begin{tabular}{cccccc}
\toprule
$\beta$ & $\kappa_{cr}$ & $L/a$ & $\bar{g}^2(L)$ & ${\boldsymbol \cZ}_{(45)}^+$ & ${\boldsymbol \cZ}_{(45)}^-$ \\

\midrule
\multirow{2}*{5.20} &
\multirow{2}*{0.13600}
& 4 & 3.65 & 
 $\begin{pmatrix} 0.4921(11) & -0.02039(13) \\ 
 -1.1531(32) & 0.8350(19)\\ 
 \end{pmatrix}$ 
 & $\begin{pmatrix} 0.24875(92) & 0.01084(10) \\ 
 0.2681(16) & 0.5416(10)\\ 
 \end{pmatrix}$ \\ 
& {} & 6 & 4.61 & 
 $\begin{pmatrix} 0.4293(10) & -0.02340(19) \\ 
 -1.3971(38) & 0.9190(18)\\ 
 \end{pmatrix}$ 
 & $\begin{pmatrix} 0.17779(68) & 0.00886(11) \\ 
 0.2660(16) & 0.5373(11)\\ 
 \end{pmatrix}$ \\ 

\midrule
\multirow{3}*{5.29} &
\multirow{3}*{0.13641}
& 4 & 3.39 & 
 $\begin{pmatrix} 0.5133(12) & -0.01910(13) \\ 
 -1.1264(31) & 0.8385(15)\\ 
 \end{pmatrix}$ 
 & $\begin{pmatrix} 0.27459(88) & 0.009909(86) \\ 
 0.2838(16) & 0.56761(95)\\ 
 \end{pmatrix}$ \\ 
& {} & 6 & 4.30 & 
 $\begin{pmatrix} 0.4509(12) & -0.02075(23) \\ 
 -1.3442(46) & 0.9189(20)\\ 
 \end{pmatrix}$ 
 & $\begin{pmatrix} 0.20420(76) & 0.00734(13) \\ 
 0.2741(19) & 0.5621(11)\\ 
 \end{pmatrix}$ \\ 
& {} & 8 & 5.65 & 
 $\begin{pmatrix} 0.4120(11) & -0.02498(26) \\ 
 -1.5596(49) & 1.0027(26)\\ 
 \end{pmatrix}$ 
 & $\begin{pmatrix} 0.15562(58) & 0.00676(17) \\ 
 0.2607(14) & 0.5514(11)\\ 
 \end{pmatrix}$ \\ 

\midrule
\multirow{3}*{5.40} &
\multirow{3}*{0.13669}
& 4 & 3.19 & 
 $\begin{pmatrix} 0.5372(10) & -0.01782(13) \\ 
 -1.0918(31) & 0.8416(15)\\ 
 \end{pmatrix}$ 
 & $\begin{pmatrix} 0.30436(89) & 0.008962(76) \\ 
 0.2935(16) & 0.59197(98)\\ 
 \end{pmatrix}$ \\ 
& {} & 6 & 3.86 & 
 $\begin{pmatrix} 0.4717(12) & -0.01848(20) \\ 
 -1.2852(44) & 0.9099(19)\\ 
 \end{pmatrix}$ 
 & $\begin{pmatrix} 0.23038(72) & 0.006133(95) \\ 
 0.2827(18) & 0.5833(11)\\ 
 \end{pmatrix}$ \\ 
& {} & 8 & 4.75 & 
 $\begin{pmatrix} 0.43354(94) & -0.02131(20) \\ 
 -1.4867(41) & 0.9867(18)\\ 
 \end{pmatrix}$ 
 & $\begin{pmatrix} 0.18096(60) & 0.00509(12) \\ 
 0.2753(16) & 0.5788(10)\\ 
 \end{pmatrix}$ \\ 

\midrule
\bottomrule\end{tabular}
\caption{Renormalisation constants at hadronic-scale $\beta$-values for the operator bases $\{ \cQ_4^\pm, \cQ_5^\pm \}$. }
\label{tab:Zhad45}
\end{center}
\end{scriptsize}
\end{table}


\newpage

\newpage

%% file: concl.tex
\section{Conclusions}
\label{sec:concl}

In the present work we have studied the non-perturbative
RG-running of the parity-odd, dimension-six, four-fermion
operators $\cQ_2^\pm, \ldots, \cQ_5^\pm$, defined in 
Eqs.~(\ref{eq:rel_ops}) and (\ref{eq:gen_4f}). Assigning
physical flavours to the generic fermion fields
$\psi_1, \ldots, \psi_4$, the above operators describe
four-quark effective interactions for various physical processes
at low energies. Under
renormalisation, these operators mix in pairs, as discussed
in Section~\ref{sec:renorm}.
This mixing is not an artefact of the eventual loss of
symmetry due to the (lattice) regularisation; rather it is a general 
property of operators belonging to the same representations
of their symmetry groups. It follows that also the
RG-running of each operator is governed by two anomalous
dimensions, and the corresponding RG-equations are imposed
on $2 \times 2$ evolution matrices. This makes the problem
of RG-running more complicated than the cases of multiplicatively
renormalised quantities, such as the quark masses or $B_{\rm K}$.

The novelty of the present work is that, using long-established
finite-size scaling techniques and the Schr\"odinger Functional renormalisation
conditions described in Section~\ref{sec:sf}, we have computed
the non-perturbative evolution matrices  of these operators between widely varying
low- and high-energy scales 
$\mu_{\rm had} \sim \cO(\Lambda_{\rm QCD})$ and 
$\mu_{\rm pt} \sim \cO(M_W)$ for QCD with two dynamical flavours.
Our results are shown in Figs.~\ref{fig:RUN+} and \ref{fig:RUN-}
and Eqs.~(\ref{eq:U+23}) -- (\ref{eq:U-45}). The accuracy of
our results for the diagonal matrix elements ranges from 3\% to 5\%. 
The accuracy on the determination of the non-diagonal matrix elements
ranges from as high as 3\% to as poor as 60\%. Clearly there is room
for improvement.  In our next project concerning the renormalisation and
RG-running of the same operators for QCD with three dynamical flavours, 
we plan to introduce several novelties, which ought to improve the precision
of our results significantly.

Perturbation theory is to be used for the RG-running
for scales above $\mu_{\rm pt} \sim \cO(M_W)$. In our SF scheme
the perturbative results at our disposal are NNLO (3-loops)
for the Callan-Symanzik $\beta$-function and NLO (2-loops) for the 
four-fermion operator anomalous dimensions. 
In Figs.~\ref{fig:RUN+} and \ref{fig:RUN-} we see the presence of possibly relevant non-perturbative effects 
already at scales of about 3 GeV, where it is often assumed that beyond-LO perturbation theory converges
well\footnote{We have checked that other SF schemes, with
different choices of $\alpha$ and $(s_1, s_2)$ (see subsection~\ref{subsec:sf}),
display similar overall behaviour.}.
We have also performed some checks by computing the RG-evolution matrix from a generic scale to 
a scale of about 3 GeV and found some matrix elements where the NLO perturbative result 
significantly differs from the non-perturbative one (see Appendix~\ref{sec:checks}).
This should serve as a warning for other non-perturbative approaches which assume
that perturbation theory is convergent at such scales.

Finally, at a fixed hadronic scale and for three values of the bare gauge coupling,
we have computed the renormalisation constants (again in $2 \times 2$ matrix form)
of our four-fermion operators.

As a closing remark we wish to point out that the non-perturbative evolution matrices
computed in this work describe not only the RG-running of the parity-odd operators 
$\cQ_2^\pm, \ldots, \cQ_5^\pm$, but also that of their parity-even counterparts
$Q_2^\pm, \ldots, Q_5^\pm$. This is because evolution matrices are continuum 
quantities: in the continuum, each parity-odd operator combines with its parity-even
counterpart to form an operator which transforms in a given chiral representation, 
both parts having consequently the same anomalous dimension matrices.

In the case, for instance, of $\Delta S = 2$ transitions, we are
dealing with operator matrix elements between two neutral $K$-meson
states and therefore only the parity-even operators ($Q_1^+$ in the SM
and $Q_{2, \cdots, 5}^+$ for BSM) contribute. Our results for the
continuum RG-evolution, obtained for the parity-odd basis, can be used
in this case. The renormalization of the bare operators, however,
depends on the details of the lattice action. If the lattice
regularisation respects chiral symmetry (e.g. lattice QCD with
Ginsparg-Wilson fermions), then the parity-even and parity-odd parts
of a given basis of chiral operators renormalise with the same
renormalisation constants.  Consequently they also have the same
matrix-SSFs and evolution matrices. All results obtained for the
parity-odd operators $\cQ_2^\pm, \ldots, \cQ_5^\pm$ are then also
valid for the $Q_2^\pm, \ldots, Q_5^\pm$, without further ado.

Things are somewhat more complicated if the regularisation breaks chiral symmetry
(e.g. lattice QCD with Wilson fermions). Then parity-even and parity-odd operators
again have the same anomalous dimensions, as these are continuum quantities, but
the ``ratio'' of their renormalisation matrices $\Big \{ {\boldsymbol \cZ}^{-1} {\boldsymbol Z} \Big \}$
is a finite (scale-independent) matrix which is a function of the bare gauge coupling; it becomes the unit
matrix in the continuum limit. This ``ratio'' is fixed by lattice Ward identities,
as discussed for example in ref.~\cite{Donini:1999sf}. So the subtlety here is that once the 
renormalisation condition has been fixed for say, the parity-odd operator bases at a value  
$g_0^2$ of the squared gauge 
coupling, the condition for the parity-even counterparts is also
fixed through $\Big \{ {\boldsymbol \cZ}^{-1} {\boldsymbol Z} \Big \}$. Consequently,  renormalisation matrix ``ratios'' like
${\boldsymbol \cZ}\left(g_0^2,\frac{a}{2L}\right)\left[{\boldsymbol \cZ}\left(g_0^2,\frac{a}{L}\right)\right]^{-1}$
are equal to their parity-even counterparts 
${\boldsymbol Z}\left(g_0^2,\frac{a}{2L}\right)\left[{\boldsymbol Z}\left(g_0^2,\frac{a}{L}\right)\right]^{-1}$.
Thus  matrix-SSFs $\bSigma(g_0^2,a/L)$ and
evolution matrices are the same for parity-odd and parity-even cases; cf. \req{SIGMA}.
But  if we wish to use the evolution matrices of the present
work also for the parity-even operators, we must ensure that these are renormalised in the
``same" SF scheme employed for their parity-odd counterparts. This is ensured by writing
the RGI parity-even operator column (in analogy to \req{eq:rgi_mix-2}) as:
\begin{eqnarray}
\label{eq:rgi_mix-3}
\hat{\boldsymbol Q} &\equiv& \left[\frac{\gbar^2(\mu_{\rm pt})}{4\pi}\right]^{-\frac{{\boldsymbol {\boldsymbol \gamma}}^{(0)}}{2b_0}}\bW(\mu_{\rm pt})\overline{\boldsymbol Q}(\mu_{\rm pt})  \\
&=& \left[\frac{\gbar^2(\mu_{\rm pt})}{4\pi}\right]^{-\frac{{\boldsymbol {\boldsymbol \gamma}}^{(0)}}{2b_0}}\bW(\mu_{\rm pt}) \bU(\mu_{\rm pt},\mu_{\rm had})
\lim_{g_0^2 \rightarrow 0} \Big [ {\boldsymbol \cZ}(g_0^2, a\mu_{\rm had}) \,\, 
\Big \{ {\boldsymbol \cZ}^{-1} {\boldsymbol Z} \Big \}
{\boldsymbol Q}_{\rm sub} (g_0^2) \Big ] \,\,,
\nonumber
\end{eqnarray}
where ${\boldsymbol Q}_{\rm sub} \equiv ( {\bf 1} + {\boldsymbol \Delta}) {\boldsymbol Q}$ is the ``subtracted''
bare operator, as suggested by \req{eq:ren_relativistic}.
The term in square brackets of the last expression is the renormalised parity-even operator ${\boldsymbol Z} {\boldsymbol Q}$.
It is computed however in a way that ensures that the bare operator ${\boldsymbol Q} (g_0^2) $ is renormalised  in our SF scheme:
the SF renormalisation parameter ${\boldsymbol \cZ}(g_0^2, a\mu_{\rm had})$ (which removes the logarithmic divergences)
is multiplied by the scheme-independent, scale-independent ``ratio'' $\Big \{ {\boldsymbol \cZ}^{-1} {\boldsymbol Z} \Big \}$. 

Clearly, the procedure sketched above for the renormalisation of
parity-even operators is fairly cumbersome. It is also prone to
enhanced statistical uncertainties, as it involves subtracted
operators ${\boldsymbol Q}_{\rm sub}$ with non-zero ${\boldsymbol
  \Delta}$. Fortunately, there is a way to circumvent the problem: it
is well known that, using chiral (axial) transformations of the quark
fields, we can obtain continuum correlation functions of specific
parity-even composite operators in terms of bare correlation functions
of parity-odd operators of the same chiral multiplet, regularised with
twisted-mass (tmQCD) Wilson fermions~\cite{Frezzotti:2000nk}.  The
prototype example is the one expressing renormalized correlation
functions of the axial current in terms of bare twisted-mass
Wilson-fermion correlation functions of the properly renormalised
vector current. The situation is more complicated with four-fermion
operators: in ref.~\cite{Pena:2004gb} it was shown that such chiral
rotations do indeed relate parity-even to parity-odd 4-fermion
operators, but the resulting tmQCD Wilson-fermion determinant is not
real, and thus unsuitable for numerical simulations. This problem is
circumvented by working with a lattice theory with sea- and
valence-quarks regularised with different lattice
actions~\cite{Frezzotti:2004wz}.  The valence action is the so-called
Osterwalder-Seiler~\cite{Osterwalder:1977pc} variety of tmQCD, with
valence twisted-mass fermion fields suitably chosen so as to enable
the mapping of correlation functions involving parity-even operators
$\{ Q_k \}$ to those of the parity-odd basis $\{ \cQ_k \}$.  The
sea-quark action may be any tenable lattice fermion action.  While the
price to pay is the loss of unitarity at finite values of the lattice
spacing, this is, however, outweighed by the advantage of vanishing
finite subtractions ($\boldsymbol{\textcyr{\bf D}} = 0$ in
eq.~(\ref{eq:ren_relativistic})).

\input acknow.tex

\newpage
\begin{table}[h!]
\begin{tiny}
\clearpage\noindent\begin{tabular}{cccrccc}
\toprule
$\bar{g}^2(L)$ & $\beta = 6/g_0^2$ & $\kappa_{cr}$ & $L/a$ & $\left[{\boldsymbol \cZ}_{(2,3)}^+(g_0,a/L) \right]^{-1}$ & ${\boldsymbol \cZ}_{(2,3)}^+(g_0,a/2L)$ & $\bSigma_{(2,3)}^+(g_0,a/L)$\\

\midrule
\multirow{3}*{0.9793}
& 9.50000 & 0.131532 & 6 & 
 $\begin{pmatrix} 1.2133(17) & -0.3792(35) \\ 
 -0.05744(71) & 1.4505(27)\\ 
 \end{pmatrix}$ 
 & $\begin{pmatrix} 0.8410(13) & 0.2446(20) \\ 
 0.03188(52) & 0.6457(13)\\ 
 \end{pmatrix}$ 
 & $\begin{pmatrix} 1.0063(20) & 0.0358(39) \\ 
 0.00157(79) & 0.9244(26)\\ 
 \end{pmatrix}$ \\ 
& 9.73410 & 0.131305 & 8 & 
 $\begin{pmatrix} 1.2049(16) & -0.3904(39) \\ 
 -0.05411(82) & 1.4807(27)\\ 
 \end{pmatrix}$ 
 & $\begin{pmatrix} 0.8470(17) & 0.2516(42) \\ 
 0.0310(10) & 0.6326(15)\\ 
 \end{pmatrix}$ 
 & $\begin{pmatrix} 1.0069(24) & 0.0416(69) \\ 
 0.0030(14) & 0.9246(28)\\ 
 \end{pmatrix}$ \\ 
& 10.05755 & 0.131069 & 12 & 
 $\begin{pmatrix} 1.1897(14) & -0.4106(34) \\ 
 -0.05176(66) & 1.5153(25)\\ 
 \end{pmatrix}$ 
 & $\begin{pmatrix} 0.8612(30) & 0.2740(47) \\ 
 0.0332(14) & 0.6196(34)\\ 
 \end{pmatrix}$ 
 & $\begin{pmatrix} 1.0102(39) & 0.0613(84) \\ 
 0.0075(18) & 0.9253(57)\\ 
 \end{pmatrix}$ \\ 

\midrule
\multirow{3}*{1.1814}
& 8.50000 & 0.132509 & 6 & 
 $\begin{pmatrix} 1.2566(16) & -0.4655(35) \\ 
 -0.07279(84) & 1.5573(30)\\ 
 \end{pmatrix}$ 
 & $\begin{pmatrix} 0.8200(29) & 0.2682(50) \\ 
 0.0353(13) & 0.5956(23)\\ 
 \end{pmatrix}$ 
 & $\begin{pmatrix} 1.0110(38) & 0.0361(82) \\ 
 0.0010(17) & 0.9113(41)\\ 
 \end{pmatrix}$ \\ 
& 8.72230 & 0.132291 & 8 & 
 $\begin{pmatrix} 1.2436(26) & -0.4875(58) \\ 
 -0.0713(13) & 1.5927(42)\\ 
 \end{pmatrix}$ 
 & $\begin{pmatrix} 0.8332(18) & 0.2850(33) \\ 
 0.03712(84) & 0.5874(18)\\ 
 \end{pmatrix}$ 
 & $\begin{pmatrix} 1.0159(33) & 0.0475(64) \\ 
 0.0042(13) & 0.9175(38)\\ 
 \end{pmatrix}$ \\ 
& 8.99366 & 0.131975 & 12 & 
 $\begin{pmatrix} 1.2249(19) & -0.5071(46) \\ 
 -0.06697(100) & 1.6373(34)\\ 
 \end{pmatrix}$ 
 & $\begin{pmatrix} 0.8436(32) & 0.2934(44) \\ 
 0.0356(11) & 0.5700(26)\\ 
 \end{pmatrix}$ 
 & $\begin{pmatrix} 1.0137(40) & 0.0528(75) \\ 
 0.0055(15) & 0.9154(46)\\ 
 \end{pmatrix}$ \\ 

\midrule
\multirow{3}*{1.5078}
& 7.54200 & 0.133705 & 6 & 
 $\begin{pmatrix} 1.3140(20) & -0.5896(40) \\ 
 -0.09652(87) & 1.7162(33)\\ 
 \end{pmatrix}$ 
 & $\begin{pmatrix} 0.7937(27) & 0.2981(30) \\ 
 0.04239(88) & 0.5325(23)\\ 
 \end{pmatrix}$ 
 & $\begin{pmatrix} 1.0142(40) & 0.0435(59) \\ 
 0.0043(13) & 0.8890(45)\\ 
 \end{pmatrix}$ \\ 
& 7.72060 & 0.133497 & 8 & 
 $\begin{pmatrix} 1.3037(39) & -0.6279(73) \\ 
 -0.0959(17) & 1.7721(65)\\ 
 \end{pmatrix}$ 
 & $\begin{pmatrix} 0.8071(43) & 0.3162(42) \\ 
 0.0437(14) & 0.5201(34)\\ 
 \end{pmatrix}$ 
 & $\begin{pmatrix} 1.0216(59) & 0.0541(78) \\ 
 0.0070(20) & 0.8943(63)\\ 
 \end{pmatrix}$ \\ 
& 8.02599 & 0.133063 & 12 & 
 $\begin{pmatrix} 1.2785(49) & -0.660(12) \\ 
 -0.0924(25) & 1.826(11)\\ 
 \end{pmatrix}$ 
 & $\begin{pmatrix} 0.8166(40) & 0.3313(55) \\ 
 0.0439(16) & 0.5001(31)\\ 
 \end{pmatrix}$ 
 & $\begin{pmatrix} 1.0138(62) & 0.066(11) \\ 
 0.0099(23) & 0.8846(74)\\ 
 \end{pmatrix}$ \\ 

\midrule
\multirow{3}*{2.0142}
& 6.60850 & 0.135260 & 6 & 
 $\begin{pmatrix} 1.4150(27) & -0.8110(54) \\ 
 -0.1425(12) & 2.0090(49)\\ 
 \end{pmatrix}$ 
 & $\begin{pmatrix} 0.7514(19) & 0.3335(15) \\ 
 0.05167(57) & 0.4401(13)\\ 
 \end{pmatrix}$ 
 & $\begin{pmatrix} 1.0157(31) & 0.0605(40) \\ 
 0.01041(96) & 0.8421(32)\\ 
 \end{pmatrix}$ \\ 
& 6.82170 & 0.134891 & 8 & 
 $\begin{pmatrix} 1.3812(53) & -0.814(11) \\ 
 -0.1295(26) & 2.043(10)\\ 
 \end{pmatrix}$ 
 & $\begin{pmatrix} 0.7719(23) & 0.3554(27) \\ 
 0.05420(98) & 0.4320(17)\\ 
 \end{pmatrix}$ 
 & $\begin{pmatrix} 1.0199(52) & 0.0974(74) \\ 
 0.0189(17) & 0.8385(51)\\ 
 \end{pmatrix}$ \\ 
& 7.09300 & 0.134432 & 12 & 
 $\begin{pmatrix} 1.3519(40) & -0.8891(89) \\ 
 -0.1335(22) & 2.1328(77)\\ 
 \end{pmatrix}$ 
 & $\begin{pmatrix} 0.7844(26) & 0.3691(25) \\ 
 0.05261(86) & 0.4150(18)\\ 
 \end{pmatrix}$ 
 & $\begin{pmatrix} 1.0113(43) & 0.0897(67) \\ 
 0.0157(14) & 0.8385(47)\\ 
 \end{pmatrix}$ \\ 

\midrule
\multirow{3}*{2.4792}
& 6.13300 & 0.136110 & 6 & 
 $\begin{pmatrix} 1.4969(45) & -1.0111(81) \\ 
 -0.1878(20) & 2.2820(81)\\ 
 \end{pmatrix}$ 
 & $\begin{pmatrix} 0.7280(50) & 0.3519(39) \\ 
 0.0587(16) & 0.3780(35)\\ 
 \end{pmatrix}$ 
 & $\begin{pmatrix} 1.0236(79) & 0.0672(88) \\ 
 0.0169(25) & 0.8031(86)\\ 
 \end{pmatrix}$ \\ 
& 6.32290 & 0.135767 & 8 & 
 $\begin{pmatrix} 1.4557(36) & -1.0256(76) \\ 
 -0.1743(20) & 2.3371(71)\\ 
 \end{pmatrix}$ 
 & $\begin{pmatrix} 0.7340(53) & 0.3602(44) \\ 
 0.0558(17) & 0.3621(39)\\ 
 \end{pmatrix}$ 
 & $\begin{pmatrix} 1.0063(86) & 0.089(12) \\ 
 0.0183(29) & 0.7893(99)\\ 
 \end{pmatrix}$ \\ 
& 6.63164 & 0.135227 & 12 & 
 $\begin{pmatrix} 1.4076(47) & -1.078(10) \\ 
 -0.1677(28) & 2.401(10)\\ 
 \end{pmatrix}$ 
 & $\begin{pmatrix} 0.7640(65) & 0.3826(32) \\ 
 0.0569(15) & 0.3608(41)\\ 
 \end{pmatrix}$ 
 & $\begin{pmatrix} 1.0111(96) & 0.0945(94) \\ 
 0.0195(23) & 0.8051(98)\\ 
 \end{pmatrix}$ \\ 

\midrule
\multirow{3}*{3.3340}
& 5.62150 & 0.136665 & 6 & 
 $\begin{pmatrix} 1.6592(74) & -1.411(13) \\ 
 -0.2842(37) & 2.863(15)\\ 
 \end{pmatrix}$ 
 & $\begin{pmatrix} 0.6776(84) & 0.3716(53) \\ 
 0.0714(29) & 0.2796(41)\\ 
 \end{pmatrix}$ 
 & $\begin{pmatrix} 1.019(14) & 0.1080(98) \\ 
 0.0392(41) & 0.700(10)\\ 
 \end{pmatrix}$ \\ 
& 5.80970 & 0.136608 & 8 & 
 $\begin{pmatrix} 1.5838(67) & -1.415(16) \\ 
 -0.2655(46) & 2.888(16)\\ 
 \end{pmatrix}$ 
 & $\begin{pmatrix} 0.7001(70) & 0.3851(43) \\ 
 0.0738(19) & 0.2822(31)\\ 
 \end{pmatrix}$ 
 & $\begin{pmatrix} 1.006(11) & 0.1215(95) \\ 
 0.0419(28) & 0.7107(94)\\ 
 \end{pmatrix}$ \\ 
& 6.11816 & 0.136139 & 12 & 
 $\begin{pmatrix} 1.5063(82) & -1.424(20) \\ 
 -0.2385(57) & 2.907(19)\\ 
 \end{pmatrix}$ 
 & $\begin{pmatrix} 0.7347(86) & 0.4113(46) \\ 
 0.0755(26) & 0.2776(29)\\ 
 \end{pmatrix}$ 
 & $\begin{pmatrix} 1.009(14) & 0.149(11) \\ 
 0.0474(38) & 0.6999(72)\\ 
 \end{pmatrix}$ \\ 

\midrule
\bottomrule\end{tabular}
\noindent\begin{tabular}{cccrccc}
\toprule
$\bar{g}^2(L)$ & $\beta = 6/g_0^2$ & $\kappa_{cr}$ & $L/a$ & $\left[{\boldsymbol \cZ}_{(2,3)}^-(g_0,a/L) \right]^{-1}$ & ${\boldsymbol \cZ}_{(2,3)}^-(g_0,a/2L)$ & $\bSigma_{(2,3)}^-(g_0,a/L)$\\

\midrule
\multirow{3}*{0.9793}
& 9.50000 & 0.131532 & 6 & 
 $\begin{pmatrix} 1.2280(14) & 0.1801(22) \\ 
 0.05161(71) & 1.4101(23)\\ 
 \end{pmatrix}$ 
 & $\begin{pmatrix} 0.8246(12) & -0.1423(16) \\ 
 -0.02854(53) & 0.6572(14)\\ 
 \end{pmatrix}$ 
 & $\begin{pmatrix} 1.0052(19) & -0.0522(28) \\ 
 -0.00113(82) & 0.9217(25)\\ 
 \end{pmatrix}$ \\ 
& 9.73410 & 0.131305 & 8 & 
 $\begin{pmatrix} 1.2186(14) & 0.2053(27) \\ 
 0.04869(82) & 1.4443(23)\\ 
 \end{pmatrix}$ 
 & $\begin{pmatrix} 0.8302(14) & -0.1575(27) \\ 
 -0.0277(10) & 0.6420(16)\\ 
 \end{pmatrix}$ 
 & $\begin{pmatrix} 1.0040(21) & -0.0568(45) \\ 
 -0.0026(14) & 0.9215(28)\\ 
 \end{pmatrix}$ \\ 
& 10.05755 & 0.131069 & 12 & 
 $\begin{pmatrix} 1.2029(14) & 0.2361(27) \\ 
 0.04650(80) & 1.4813(25)\\ 
 \end{pmatrix}$ 
 & $\begin{pmatrix} 0.8418(32) & -0.1770(34) \\ 
 -0.0302(13) & 0.6284(34)\\ 
 \end{pmatrix}$ 
 & $\begin{pmatrix} 1.0044(42) & -0.0633(57) \\ 
 -0.0070(18) & 0.9239(56)\\ 
 \end{pmatrix}$ \\ 

\midrule
\multirow{3}*{1.1814}
& 8.50000 & 0.132509 & 6 & 
 $\begin{pmatrix} 1.2763(15) & 0.2281(25) \\ 
 0.06654(82) & 1.5057(24)\\ 
 \end{pmatrix}$ 
 & $\begin{pmatrix} 0.8000(23) & -0.1615(34) \\ 
 -0.0323(13) & 0.6079(25)\\ 
 \end{pmatrix}$ 
 & $\begin{pmatrix} 1.0102(34) & -0.0607(59) \\ 
 -0.0008(18) & 0.9078(45)\\ 
 \end{pmatrix}$ \\ 
& 8.72230 & 0.132291 & 8 & 
 $\begin{pmatrix} 1.2627(26) & 0.2609(42) \\ 
 0.0644(13) & 1.5455(46)\\ 
 \end{pmatrix}$ 
 & $\begin{pmatrix} 0.8110(16) & -0.1758(23) \\ 
 -0.03383(87) & 0.5988(18)\\ 
 \end{pmatrix}$ 
 & $\begin{pmatrix} 1.0127(28) & -0.0600(45) \\ 
 -0.0042(14) & 0.9166(38)\\ 
 \end{pmatrix}$ \\ 
& 8.99366 & 0.131975 & 12 & 
 $\begin{pmatrix} 1.2437(17) & 0.2982(31) \\ 
 0.06091(98) & 1.5937(30)\\ 
 \end{pmatrix}$ 
 & $\begin{pmatrix} 0.8219(26) & -0.1912(22) \\ 
 -0.03184(97) & 0.5797(28)\\ 
 \end{pmatrix}$ 
 & $\begin{pmatrix} 1.0107(35) & -0.0597(44) \\ 
 -0.0043(14) & 0.9145(48)\\ 
 \end{pmatrix}$ \\ 

\midrule
\multirow{3}*{1.5078}
& 7.54200 & 0.133705 & 6 & 
 $\begin{pmatrix} 1.3425(17) & 0.2977(28) \\ 
 0.08950(82) & 1.6457(28)\\ 
 \end{pmatrix}$ 
 & $\begin{pmatrix} 0.7646(17) & -0.1822(17) \\ 
 -0.03943(92) & 0.5461(17)\\ 
 \end{pmatrix}$ 
 & $\begin{pmatrix} 1.0102(25) & -0.0723(36) \\ 
 -0.0040(13) & 0.8869(30)\\ 
 \end{pmatrix}$ \\ 
& 7.72060 & 0.133497 & 8 & 
 $\begin{pmatrix} 1.3295(34) & 0.3403(68) \\ 
 0.0886(21) & 1.7028(62)\\ 
 \end{pmatrix}$ 
 & $\begin{pmatrix} 0.7765(30) & -0.2011(34) \\ 
 -0.0405(14) & 0.5319(30)\\ 
 \end{pmatrix}$ 
 & $\begin{pmatrix} 1.0143(49) & -0.0787(71) \\ 
 -0.0067(21) & 0.8915(64)\\ 
 \end{pmatrix}$ \\ 
& 8.02599 & 0.133063 & 12 & 
 $\begin{pmatrix} 1.3040(46) & 0.3987(88) \\ 
 0.0858(26) & 1.7652(88)\\ 
 \end{pmatrix}$ 
 & $\begin{pmatrix} 0.7873(35) & -0.2226(47) \\ 
 -0.0406(17) & 0.5114(36)\\ 
 \end{pmatrix}$ 
 & $\begin{pmatrix} 1.0076(57) & -0.0792(98) \\ 
 -0.0089(27) & 0.8864(79)\\ 
 \end{pmatrix}$ \\ 

\midrule
\multirow{3}*{2.0142}
& 6.60850 & 0.135260 & 6 & 
 $\begin{pmatrix} 1.4601(24) & 0.4226(35) \\ 
 0.1357(12) & 1.8967(42)\\ 
 \end{pmatrix}$ 
 & $\begin{pmatrix} 0.7107(13) & -0.2099(13) \\ 
 -0.04891(53) & 0.4548(13)\\ 
 \end{pmatrix}$ 
 & $\begin{pmatrix} 1.0092(24) & -0.0978(33) \\ 
 -0.00972(94) & 0.8419(29)\\ 
 \end{pmatrix}$ \\ 
& 6.82170 & 0.134891 & 8 & 
 $\begin{pmatrix} 1.4195(56) & 0.4610(75) \\ 
 0.1223(30) & 1.9437(87)\\ 
 \end{pmatrix}$ 
 & $\begin{pmatrix} 0.7288(21) & -0.2327(21) \\ 
 -0.0515(12) & 0.4451(16)\\ 
 \end{pmatrix}$ 
 & $\begin{pmatrix} 1.0060(46) & -0.1166(59) \\ 
 -0.0187(21) & 0.8414(49)\\ 
 \end{pmatrix}$ \\ 
& 7.09300 & 0.134432 & 12 & 
 $\begin{pmatrix} 1.3903(33) & 0.5490(66) \\ 
 0.1259(24) & 2.0377(67)\\ 
 \end{pmatrix}$ 
 & $\begin{pmatrix} 0.7451(24) & -0.2508(24) \\ 
 -0.0502(13) & 0.4277(25)\\ 
 \end{pmatrix}$ 
 & $\begin{pmatrix} 1.0042(38) & -0.1019(61) \\ 
 -0.0160(19) & 0.8436(53)\\ 
 \end{pmatrix}$ \\ 

\midrule
\multirow{3}*{2.4792}
& 6.13300 & 0.136110 & 6 & 
 $\begin{pmatrix} 1.5594(39) & 0.5362(61) \\ 
 0.1815(21) & 2.1251(69)\\ 
 \end{pmatrix}$ 
 & $\begin{pmatrix} 0.6741(37) & -0.2273(29) \\ 
 -0.0565(18) & 0.3925(32)\\ 
 \end{pmatrix}$ 
 & $\begin{pmatrix} 1.0097(68) & -0.1217(65) \\ 
 -0.0169(28) & 0.8037(75)\\ 
 \end{pmatrix}$ \\ 
& 6.32290 & 0.135767 & 8 & 
 $\begin{pmatrix} 1.5131(30) & 0.6039(49) \\ 
 0.1678(21) & 2.2016(54)\\ 
 \end{pmatrix}$ 
 & $\begin{pmatrix} 0.6854(38) & -0.2428(24) \\ 
 -0.0545(17) & 0.3758(34)\\ 
 \end{pmatrix}$ 
 & $\begin{pmatrix} 0.9962(57) & -0.1208(70) \\ 
 -0.0194(26) & 0.7946(73)\\ 
 \end{pmatrix}$ \\ 
& 6.63164 & 0.135227 & 12 & 
 $\begin{pmatrix} 1.4544(40) & 0.6853(76) \\ 
 0.1586(30) & 2.2771(86)\\ 
 \end{pmatrix}$ 
 & $\begin{pmatrix} 0.7177(41) & -0.2635(17) \\ 
 -0.0553(16) & 0.3752(31)\\ 
 \end{pmatrix}$ 
 & $\begin{pmatrix} 1.0019(64) & -0.1080(61) \\ 
 -0.0208(26) & 0.8163(78)\\ 
 \end{pmatrix}$ \\ 

\midrule
\multirow{3}*{3.3340}
& 5.62150 & 0.136665 & 6 & 
 $\begin{pmatrix} 1.7596(65) & 0.7674(91) \\ 
 0.2862(43) & 2.597(12)\\ 
 \end{pmatrix}$ 
 & $\begin{pmatrix} 0.6020(53) & -0.2474(28) \\ 
 -0.0741(28) & 0.2977(42)\\ 
 \end{pmatrix}$ 
 & $\begin{pmatrix} 0.9882(98) & -0.1803(75) \\ 
 -0.0450(42) & 0.716(11)\\ 
 \end{pmatrix}$ \\ 
& 5.80970 & 0.136608 & 8 & 
 $\begin{pmatrix} 1.6746(64) & 0.850(14) \\ 
 0.2609(51) & 2.664(15)\\ 
 \end{pmatrix}$ 
 & $\begin{pmatrix} 0.6190(48) & -0.2516(46) \\ 
 -0.0754(16) & 0.2997(28)\\ 
 \end{pmatrix}$ 
 & $\begin{pmatrix} 0.9710(78) & -0.145(12) \\ 
 -0.0479(26) & 0.7344(78)\\ 
 \end{pmatrix}$ \\ 
& 6.11816 & 0.136139 & 12 & 
 $\begin{pmatrix} 1.5789(79) & 0.924(16) \\ 
 0.2346(74) & 2.710(18)\\ 
 \end{pmatrix}$ 
 & $\begin{pmatrix} 0.6588(47) & -0.2865(40) \\ 
 -0.0755(18) & 0.2924(23)\\ 
 \end{pmatrix}$ 
 & $\begin{pmatrix} 0.9733(74) & -0.168(10) \\ 
 -0.0504(32) & 0.7229(72)\\ 
 \end{pmatrix}$ \\ 
\midrule
\bottomrule
\end{tabular}
\clearpage\end{tiny}
\caption{Renormalisation matrices and lattice matrix-SSFs for the operator bases $\{ \cQ_2^\pm, \cQ_3^\pm \}$.}
\label{tab:Z23}
\end{table}


\begin{table}
\begin{tiny}
\clearpage\noindent\begin{tabular}{cccrccc}
\toprule
$\bar{g}^2(L)$ & $\beta = 6/g_0^2$ & $\kappa_{cr}$ & $L/a$ & $\left[{\boldsymbol \cZ}_{(4,5)}^+(g_0,a/L) \right]^{-1}$ & ${\boldsymbol \cZ}_{(4,5)}^+(g_0,a/2L)$ & $\bSigma_{(4,5)}^+(g_0,a/L)$\\

\midrule
\multirow{3}*{0.9793}
& 9.50000 & 0.131532 & 6 & 
 $\begin{pmatrix} 1.3013(20) & 0.00768(15) \\ 
 0.8563(75) & 1.0872(15)\\ 
 \end{pmatrix}$ 
 & $\begin{pmatrix} 0.7328(14) & -0.00488(16) \\ 
 -0.7433(64) & 0.9690(19)\\ 
 \end{pmatrix}$ 
 & $\begin{pmatrix} 0.9495(23) & 0.00031(20) \\ 
 -0.138(10) & 1.0479(25)\\ 
 \end{pmatrix}$ \\ 
& 9.73410 & 0.131305 & 8 & 
 $\begin{pmatrix} 1.3213(18) & 0.00683(18) \\ 
 0.8960(86) & 1.0678(16)\\ 
 \end{pmatrix}$ 
 & $\begin{pmatrix} 0.7228(17) & -0.00473(36) \\ 
 -0.789(14) & 0.9851(20)\\ 
 \end{pmatrix}$ 
 & $\begin{pmatrix} 0.9508(26) & -0.00011(41) \\ 
 -0.160(20) & 1.0465(27)\\ 
 \end{pmatrix}$ \\ 
& 10.05755 & 0.131069 & 12 & 
 $\begin{pmatrix} 1.3389(19) & 0.00600(18) \\ 
 0.9538(82) & 1.0363(15)\\ 
 \end{pmatrix}$ 
 & $\begin{pmatrix} 0.7180(25) & -0.00528(31) \\ 
 -0.884(11) & 1.0188(35)\\ 
 \end{pmatrix}$ 
 & $\begin{pmatrix} 0.9562(38) & -0.00115(35) \\ 
 -0.211(17) & 1.0505(39)\\ 
 \end{pmatrix}$ \\ 

\midrule
\multirow{3}*{1.1814}
& 8.50000 & 0.132509 & 6 & 
 $\begin{pmatrix} 1.3671(20) & 0.00986(18) \\ 
 1.0429(87) & 1.1019(15)\\ 
 \end{pmatrix}$ 
 & $\begin{pmatrix} 0.6919(27) & -0.00566(47) \\ 
 -0.845(18) & 0.9711(45)\\ 
 \end{pmatrix}$ 
 & $\begin{pmatrix} 0.9400(41) & 0.00059(53) \\ 
 -0.142(27) & 1.0618(50)\\ 
 \end{pmatrix}$ \\ 
& 8.72230 & 0.132291 & 8 & 
 $\begin{pmatrix} 1.3865(33) & 0.00919(29) \\ 
 1.103(12) & 1.0754(24)\\ 
 \end{pmatrix}$ 
 & $\begin{pmatrix} 0.6906(14) & -0.00607(26) \\ 
 -0.9209(87) & 1.0022(20)\\ 
 \end{pmatrix}$ 
 & $\begin{pmatrix} 0.9509(30) & -0.00019(36) \\ 
 -0.171(18) & 1.0694(35)\\ 
 \end{pmatrix}$ \\ 
& 8.99366 & 0.131975 & 12 & 
 $\begin{pmatrix} 1.4106(22) & 0.00796(22) \\ 
 1.1609(94) & 1.0381(17)\\ 
 \end{pmatrix}$ 
 & $\begin{pmatrix} 0.6780(32) & -0.00536(40) \\ 
 -0.984(12) & 1.0382(43)\\ 
 \end{pmatrix}$ 
 & $\begin{pmatrix} 0.9501(47) & -0.00017(40) \\ 
 -0.182(20) & 1.0700(48)\\ 
 \end{pmatrix}$ \\ 

\midrule
\multirow{3}*{1.5078}
& 7.54200 & 0.133705 & 6 & 
 $\begin{pmatrix} 1.4600(23) & 0.01299(19) \\ 
 1.2985(86) & 1.1144(18)\\ 
 \end{pmatrix}$ 
 & $\begin{pmatrix} 0.6422(24) & -0.00750(31) \\ 
 -0.998(10) & 0.9825(37)\\ 
 \end{pmatrix}$ 
 & $\begin{pmatrix} 0.9278(37) & -0.00002(38) \\ 
 -0.180(18) & 1.0819(43)\\ 
 \end{pmatrix}$ \\ 
& 7.72060 & 0.133497 & 8 & 
 $\begin{pmatrix} 1.4907(51) & 0.01228(49) \\ 
 1.381(20) & 1.0858(37)\\ 
 \end{pmatrix}$ 
 & $\begin{pmatrix} 0.6360(43) & -0.00776(54) \\ 
 -1.081(15) & 1.0163(55)\\ 
 \end{pmatrix}$ 
 & $\begin{pmatrix} 0.9372(68) & -0.00064(61) \\ 
 -0.208(29) & 1.0904(70)\\ 
 \end{pmatrix}$ \\ 
& 8.02599 & 0.133063 & 12 & 
 $\begin{pmatrix} 1.5172(63) & 0.01140(51) \\ 
 1.471(23) & 1.0442(45)\\ 
 \end{pmatrix}$ 
 & $\begin{pmatrix} 0.6214(33) & -0.00777(69) \\ 
 -1.175(18) & 1.0523(64)\\ 
 \end{pmatrix}$ 
 & $\begin{pmatrix} 0.9308(66) & -0.00102(78) \\ 
 -0.236(36) & 1.0849(80)\\ 
 \end{pmatrix}$ \\ 

\midrule
\multirow{3}*{2.0142}
& 6.60850 & 0.135260 & 6 & 
 $\begin{pmatrix} 1.6235(31) & 0.01888(23) \\ 
 1.7229(97) & 1.1352(20)\\ 
 \end{pmatrix}$ 
 & $\begin{pmatrix} 0.5644(17) & -0.01041(21) \\ 
 -1.2111(59) & 1.0006(32)\\ 
 \end{pmatrix}$ 
 & $\begin{pmatrix} 0.8983(32) & -0.00114(27) \\ 
 -0.242(12) & 1.1131(42)\\ 
 \end{pmatrix}$ \\ 
& 6.82170 & 0.134891 & 8 & 
 $\begin{pmatrix} 1.6431(61) & 0.01625(58) \\ 
 1.743(23) & 1.0896(38)\\ 
 \end{pmatrix}$ 
 & $\begin{pmatrix} 0.5624(22) & -0.01143(44) \\ 
 -1.324(11) & 1.0439(36)\\ 
 \end{pmatrix}$ 
 & $\begin{pmatrix} 0.9044(48) & -0.00330(56) \\ 
 -0.355(27) & 1.1162(53)\\ 
 \end{pmatrix}$ \\ 
& 7.09300 & 0.134432 & 12 & 
 $\begin{pmatrix} 1.6778(56) & 0.01590(50) \\ 
 1.890(18) & 1.0359(35)\\ 
 \end{pmatrix}$ 
 & $\begin{pmatrix} 0.5494(19) & -0.01069(34) \\ 
 -1.4154(83) & 1.0946(36)\\ 
 \end{pmatrix}$ 
 & $\begin{pmatrix} 0.9016(40) & -0.00233(44) \\ 
 -0.306(22) & 1.1116(51)\\ 
 \end{pmatrix}$ \\ 

\midrule
\multirow{3}*{2.4792}
& 6.13300 & 0.136110 & 6 & 
 $\begin{pmatrix} 1.7653(57) & 0.02424(41) \\ 
 2.075(17) & 1.1435(34)\\ 
 \end{pmatrix}$ 
 & $\begin{pmatrix} 0.5112(34) & -0.01334(62) \\ 
 -1.380(16) & 1.0334(63)\\ 
 \end{pmatrix}$ 
 & $\begin{pmatrix} 0.8747(60) & -0.00287(73) \\ 
 -0.294(27) & 1.1484(75)\\ 
 \end{pmatrix}$ \\ 
& 6.32290 & 0.135767 & 8 & 
 $\begin{pmatrix} 1.7950(40) & 0.02187(39) \\ 
 2.133(13) & 1.0929(23)\\ 
 \end{pmatrix}$ 
 & $\begin{pmatrix} 0.4950(39) & -0.01251(59) \\ 
 -1.452(15) & 1.0690(87)\\ 
 \end{pmatrix}$ 
 & $\begin{pmatrix} 0.8621(73) & -0.00284(67) \\ 
 -0.326(28) & 1.1366(96)\\ 
 \end{pmatrix}$ \\ 
& 6.63164 & 0.135227 & 12 & 
 $\begin{pmatrix} 1.8152(59) & 0.01965(57) \\ 
 2.224(18) & 1.0303(33)\\ 
 \end{pmatrix}$ 
 & $\begin{pmatrix} 0.5009(24) & -0.01228(34) \\ 
 -1.5776(83) & 1.1426(48)\\ 
 \end{pmatrix}$ 
 & $\begin{pmatrix} 0.8820(52) & -0.00283(43) \\ 
 -0.323(21) & 1.1463(64)\\ 
 \end{pmatrix}$ \\ 

\midrule
\multirow{3}*{3.3340}
& 5.62150 & 0.136665 & 6 & 
 $\begin{pmatrix} 2.0537(100) & 0.03521(64) \\ 
 2.720(21) & 1.1637(50)\\ 
 \end{pmatrix}$ 
 & $\begin{pmatrix} 0.4202(72) & -0.0213(15) \\ 
 -1.662(33) & 1.079(15)\\ 
 \end{pmatrix}$ 
 & $\begin{pmatrix} 0.805(12) & -0.0099(15) \\ 
 -0.479(40) & 1.196(17)\\ 
 \end{pmatrix}$ \\ 
& 5.80970 & 0.136608 & 8 & 
 $\begin{pmatrix} 2.0571(90) & 0.03204(84) \\ 
 2.758(27) & 1.0912(47)\\ 
 \end{pmatrix}$ 
 & $\begin{pmatrix} 0.4272(61) & -0.0214(11) \\ 
 -1.770(29) & 1.153(14)\\ 
 \end{pmatrix}$ 
 & $\begin{pmatrix} 0.819(11) & -0.0097(11) \\ 
 -0.458(36) & 1.202(15)\\ 
 \end{pmatrix}$ \\ 
& 6.11816 & 0.136139 & 12 & 
 $\begin{pmatrix} 2.052(14) & 0.0275(12) \\ 
 2.771(37) & 1.0145(70)\\ 
 \end{pmatrix}$ 
 & $\begin{pmatrix} 0.4279(39) & -0.0219(13) \\ 
 -1.950(28) & 1.241(14)\\ 
 \end{pmatrix}$ 
 & $\begin{pmatrix} 0.8171(75) & -0.0106(13) \\ 
 -0.563(41) & 1.206(16)\\ 
 \end{pmatrix}$ \\ 

\midrule
\bottomrule\end{tabular}
\noindent\begin{tabular}{cccrccc}
\toprule
$\bar{g}^2(L)$ & $\beta = 6/g_0^2$ & $\kappa_{cr}$ & $L/a$ & $\left[{\boldsymbol \cZ}_{(4,5)}^-(g_0,a/L) \right]^{-1}$ & ${\boldsymbol \cZ}_{(4,5)}^-(g_0,a/2L)$ & $\bSigma_{(4,5)}^-(g_0,a/L)$\\

\midrule
\multirow{3}*{0.9793}
& 9.50000 & 0.131532 & 6 & 
 $\begin{pmatrix} 1.5644(33) & -0.00363(17) \\ 
 -0.5050(81) & 1.1866(11)\\ 
 \end{pmatrix}$ 
 & $\begin{pmatrix} 0.5723(14) & -0.00084(12) \\ 
 0.2759(46) & 0.83999(93)\\ 
 \end{pmatrix}$ 
 & $\begin{pmatrix} 0.8957(29) & -0.00308(18) \\ 
 0.0073(97) & 0.9956(15)\\ 
 \end{pmatrix}$ \\ 
& 9.73410 & 0.131305 & 8 & 
 $\begin{pmatrix} 1.6118(35) & -0.00099(22) \\ 
 -0.5070(92) & 1.1835(11)\\ 
 \end{pmatrix}$ 
 & $\begin{pmatrix} 0.5548(19) & -0.00178(28) \\ 
 0.280(11) & 0.8424(21)\\ 
 \end{pmatrix}$ 
 & $\begin{pmatrix} 0.8952(37) & -0.00267(36) \\ 
 0.023(19) & 0.9967(26)\\ 
 \end{pmatrix}$ \\ 
& 10.05755 & 0.131069 & 12 & 
 $\begin{pmatrix} 1.6685(36) & 0.00201(20) \\ 
 -0.5117(99) & 1.17541(97)\\ 
 \end{pmatrix}$ 
 & $\begin{pmatrix} 0.5338(26) & -0.00232(23) \\ 
 0.2988(79) & 0.8526(16)\\ 
 \end{pmatrix}$ 
 & $\begin{pmatrix} 0.8920(47) & -0.00166(30) \\ 
 0.062(15) & 1.0027(21)\\ 
 \end{pmatrix}$ \\ 

\midrule
\multirow{3}*{1.1814}
& 8.50000 & 0.132509 & 6 & 
 $\begin{pmatrix} 1.7063(38) & -0.00486(20) \\ 
 -0.6226(89) & 1.2260(11)\\ 
 \end{pmatrix}$ 
 & $\begin{pmatrix} 0.5133(25) & -0.00119(31) \\ 
 0.291(11) & 0.8139(28)\\ 
 \end{pmatrix}$ 
 & $\begin{pmatrix} 0.8766(49) & -0.00395(40) \\ 
 -0.009(19) & 0.9964(36)\\ 
 \end{pmatrix}$ \\ 
& 8.72230 & 0.132291 & 8 & 
 $\begin{pmatrix} 1.7642(57) & -0.00233(32) \\ 
 -0.643(15) & 1.2165(17)\\ 
 \end{pmatrix}$ 
 & $\begin{pmatrix} 0.4994(21) & -0.00162(21) \\ 
 0.3023(76) & 0.8242(11)\\ 
 \end{pmatrix}$ 
 & $\begin{pmatrix} 0.8821(48) & -0.00313(29) \\ 
 0.004(16) & 1.0019(21)\\ 
 \end{pmatrix}$ \\ 
& 8.99366 & 0.131975 & 12 & 
 $\begin{pmatrix} 1.8399(44) & 0.00194(25) \\ 
 -0.636(12) & 1.2090(10)\\ 
 \end{pmatrix}$ 
 & $\begin{pmatrix} 0.4773(24) & -0.00316(25) \\ 
 0.2918(91) & 0.8278(24)\\ 
 \end{pmatrix}$ 
 & $\begin{pmatrix} 0.8801(49) & -0.00288(33) \\ 
 0.010(18) & 1.0014(30)\\ 
 \end{pmatrix}$ \\ 

\midrule
\multirow{3}*{1.5078}
& 7.54200 & 0.133705 & 6 & 
 $\begin{pmatrix} 1.9239(42) & -0.00667(20) \\ 
 -0.7804(88) & 1.2815(13)\\ 
 \end{pmatrix}$ 
 & $\begin{pmatrix} 0.4388(22) & -0.00085(24) \\ 
 0.3045(66) & 0.7780(16)\\ 
 \end{pmatrix}$ 
 & $\begin{pmatrix} 0.8450(45) & -0.00400(33) \\ 
 -0.021(13) & 0.9950(22)\\ 
 \end{pmatrix}$ \\ 
& 7.72060 & 0.133497 & 8 & 
 $\begin{pmatrix} 2.0113(90) & -0.00333(55) \\ 
 -0.809(22) & 1.2730(25)\\ 
 \end{pmatrix}$ 
 & $\begin{pmatrix} 0.4203(35) & -0.00157(36) \\ 
 0.3177(100) & 0.7884(38)\\ 
 \end{pmatrix}$ 
 & $\begin{pmatrix} 0.8470(77) & -0.00340(53) \\ 
 -0.000(25) & 1.0026(53)\\ 
 \end{pmatrix}$ \\ 
& 8.02599 & 0.133063 & 12 & 
 $\begin{pmatrix} 2.112(14) & 0.00116(67) \\ 
 -0.844(29) & 1.2565(27)\\ 
 \end{pmatrix}$ 
 & $\begin{pmatrix} 0.3959(38) & -0.00270(40) \\ 
 0.323(12) & 0.7938(21)\\ 
 \end{pmatrix}$ 
 & $\begin{pmatrix} 0.8380(97) & -0.00291(53) \\ 
 0.013(30) & 0.9978(32)\\ 
 \end{pmatrix}$ \\ 

\midrule
\multirow{3}*{2.0142}
& 6.60850 & 0.135260 & 6 & 
 $\begin{pmatrix} 2.3354(67) & -0.01089(30) \\ 
 -1.059(11) & 1.3788(19)\\ 
 \end{pmatrix}$ 
 & $\begin{pmatrix} 0.3342(11) & -0.00052(14) \\ 
 0.3116(33) & 0.7219(12)\\ 
 \end{pmatrix}$ 
 & $\begin{pmatrix} 0.7808(38) & -0.00436(22) \\ 
 -0.037(10) & 0.9920(22)\\ 
 \end{pmatrix}$ \\ 
& 6.82170 & 0.134891 & 8 & 
 $\begin{pmatrix} 2.405(15) & -0.00394(70) \\ 
 -1.025(27) & 1.3541(36)\\ 
 \end{pmatrix}$ 
 & $\begin{pmatrix} 0.3200(15) & -0.00089(22) \\ 
 0.3311(50) & 0.7392(20)\\ 
 \end{pmatrix}$ 
 & $\begin{pmatrix} 0.7702(60) & -0.00249(38) \\ 
 0.039(19) & 0.9996(39)\\ 
 \end{pmatrix}$ \\ 
& 7.09300 & 0.134432 & 12 & 
 $\begin{pmatrix} 2.570(11) & 0.00073(73) \\ 
 -1.111(22) & 1.3266(27)\\ 
 \end{pmatrix}$ 
 & $\begin{pmatrix} 0.3004(15) & -0.00244(21) \\ 
 0.3275(48) & 0.7489(18)\\ 
 \end{pmatrix}$ 
 & $\begin{pmatrix} 0.7751(49) & -0.00303(34) \\ 
 0.009(18) & 0.9939(35)\\ 
 \end{pmatrix}$ \\ 

\midrule
\multirow{3}*{2.4792}
& 6.13300 & 0.136110 & 6 & 
 $\begin{pmatrix} 2.737(11) & -0.01545(61) \\ 
 -1.300(18) & 1.4600(33)\\ 
 \end{pmatrix}$ 
 & $\begin{pmatrix} 0.2654(21) & -0.00006(36) \\ 
 0.2997(81) & 0.6818(36)\\ 
 \end{pmatrix}$ 
 & $\begin{pmatrix} 0.7264(65) & -0.00421(60) \\ 
 -0.067(23) & 0.9909(55)\\ 
 \end{pmatrix}$ \\ 
& 6.32290 & 0.135767 & 8 & 
 $\begin{pmatrix} 2.8542(96) & -0.00675(53) \\ 
 -1.288(16) & 1.4303(23)\\ 
 \end{pmatrix}$ 
 & $\begin{pmatrix} 0.2483(34) & -0.00164(38) \\ 
 0.2972(87) & 0.6861(34)\\ 
 \end{pmatrix}$ 
 & $\begin{pmatrix} 0.711(10) & -0.00401(58) \\ 
 -0.035(26) & 0.9792(51)\\ 
 \end{pmatrix}$ \\ 
& 6.63164 & 0.135227 & 12 & 
 $\begin{pmatrix} 2.992(14) & 0.00089(77) \\ 
 -1.343(20) & 1.3787(29)\\ 
 \end{pmatrix}$ 
 & $\begin{pmatrix} 0.2424(28) & -0.00267(43) \\ 
 0.3056(54) & 0.7129(36)\\ 
 \end{pmatrix}$ 
 & $\begin{pmatrix} 0.7286(98) & -0.00346(62) \\ 
 -0.043(19) & 0.9830(56)\\ 
 \end{pmatrix}$ \\ 

\midrule
\multirow{3}*{3.3340}
& 5.62150 & 0.136665 & 6 & 
 $\begin{pmatrix} 3.640(20) & -0.0259(11) \\ 
 -1.759(25) & 1.6207(53)\\ 
 \end{pmatrix}$ 
 & $\begin{pmatrix} 0.1619(27) & 0.00166(52) \\ 
 0.2764(65) & 0.6050(52)\\ 
 \end{pmatrix}$ 
 & $\begin{pmatrix} 0.586(10) & -0.00149(84) \\ 
 -0.058(24) & 0.9736(88)\\ 
 \end{pmatrix}$ \\ 
& 5.80970 & 0.136608 & 8 & 
 $\begin{pmatrix} 3.750(24) & -0.0150(13) \\ 
 -1.758(36) & 1.5532(45)\\ 
 \end{pmatrix}$ 
 & $\begin{pmatrix} 0.1617(23) & 0.00119(34) \\ 
 0.2714(49) & 0.6194(59)\\ 
 \end{pmatrix}$ 
 & $\begin{pmatrix} 0.6044(91) & -0.00055(55) \\ 
 -0.073(23) & 0.9581(96)\\ 
 \end{pmatrix}$ \\ 
& 6.11816 & 0.136139 & 12 & 
 $\begin{pmatrix} 3.836(26) & -0.0011(17) \\ 
 -1.736(43) & 1.4802(54)\\ 
 \end{pmatrix}$ 
 & $\begin{pmatrix} 0.1535(15) & 0.00048(57) \\ 
 0.2928(68) & 0.6551(37)\\ 
 \end{pmatrix}$ 
 & $\begin{pmatrix} 0.5883(80) & 0.00053(92) \\ 
 -0.013(31) & 0.9694(66)\\ 
 \end{pmatrix}$ \\ 

\midrule
\bottomrule
\end{tabular}
\clearpage\end{tiny}
\caption{Renormalisation matrices and lattice matrix-SSFs for the operator bases $\{ \cQ_4^\pm, \cQ_5^\pm \}$.}
\label{tab:Z45}
\end{table}

\newpage

\begin{figure}
\begin{center}
\vspace*{-5\baselineskip}
\begin{minipage}[t]{0.45\textwidth}
\includegraphics[width=7.5cm]{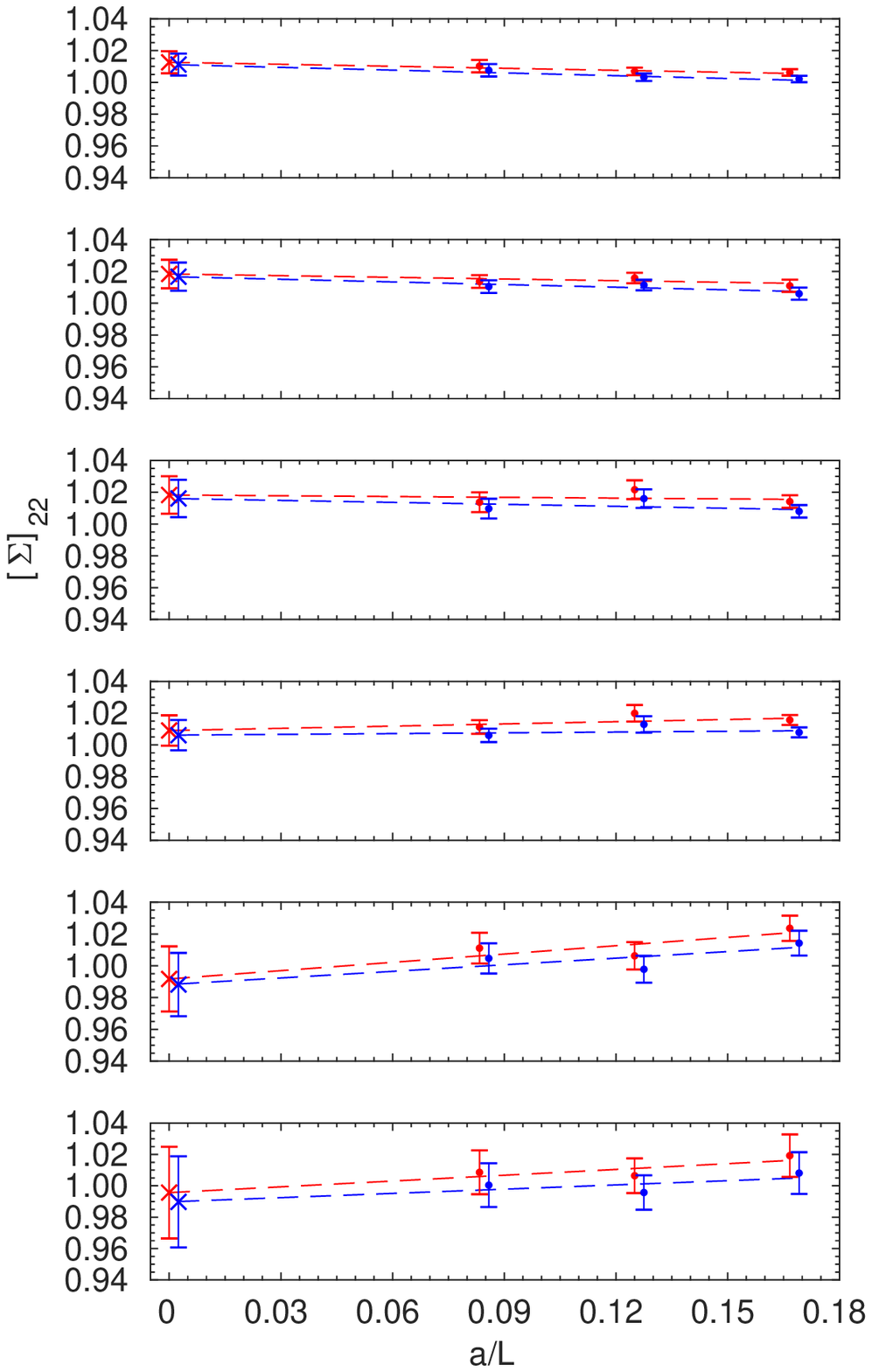}
\end{minipage}
\hspace{1.2cm}
\begin{minipage}[t]{0.45\textwidth}
\includegraphics[width=7.5cm]{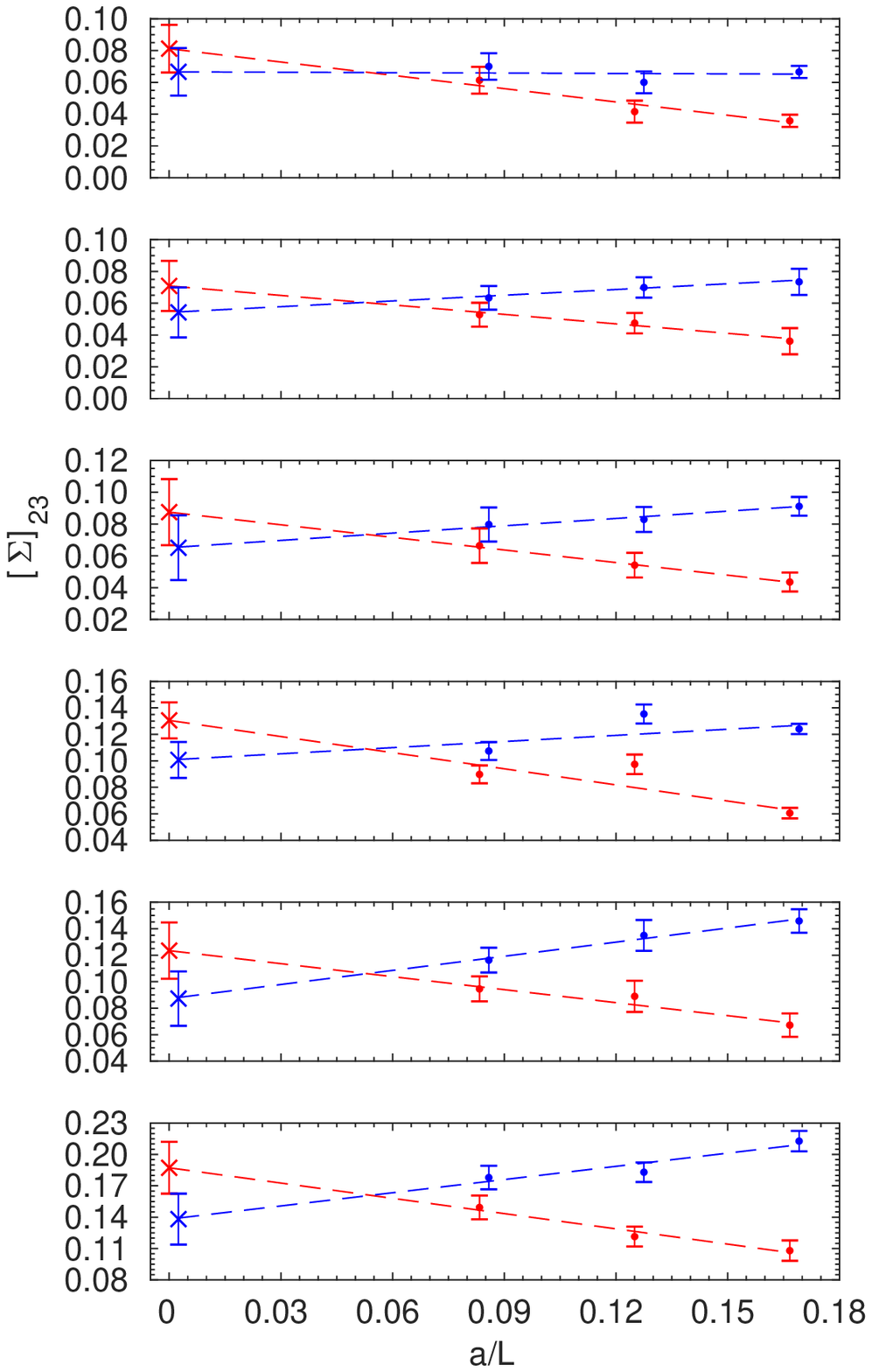}
\end{minipage}

\vspace{1mm}

\begin{minipage}[t]{0.45\textwidth}
\includegraphics[width=7.5cm]{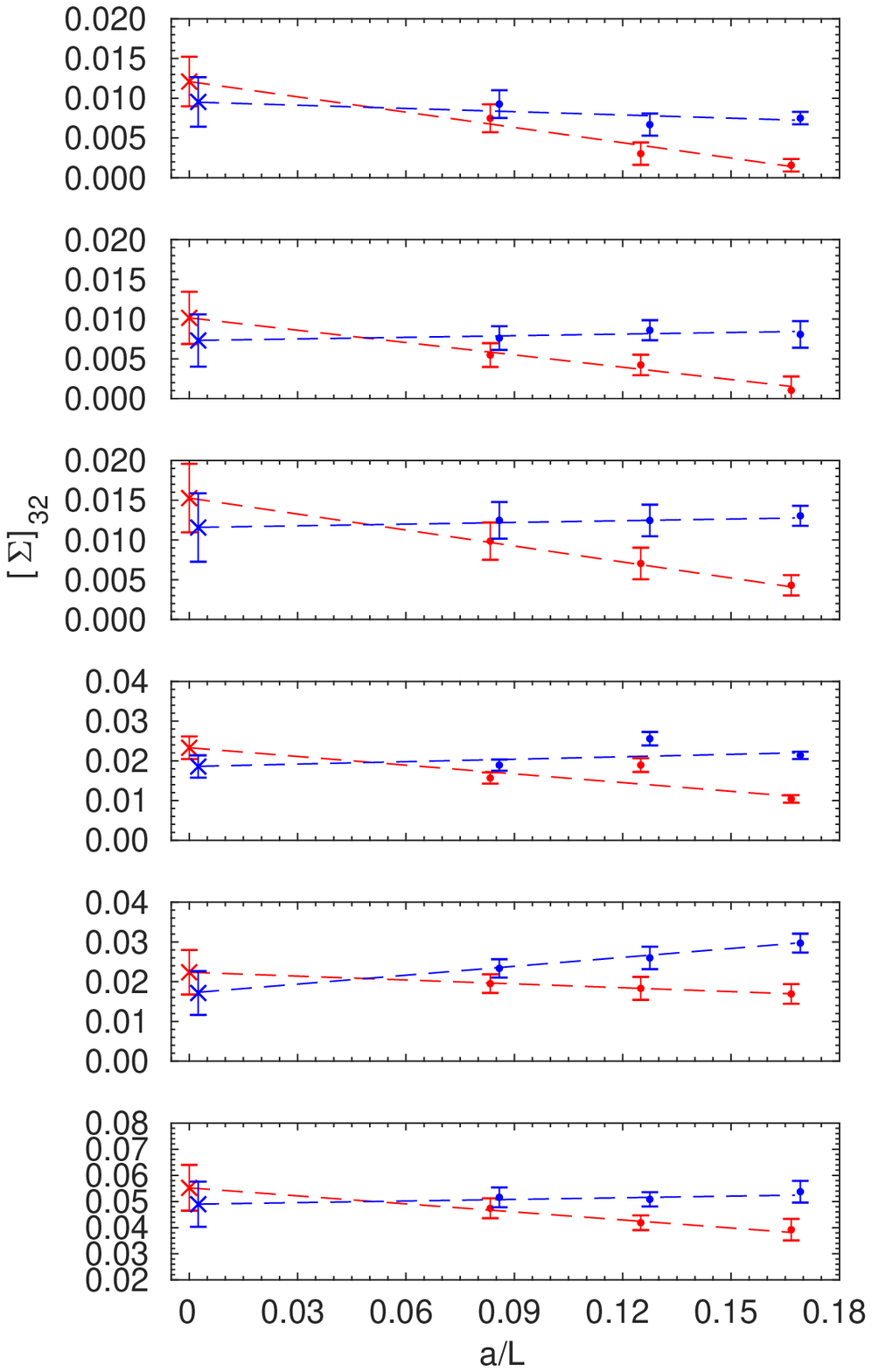}
\end{minipage}
\hspace{1cm}
\begin{minipage}[t]{0.45\textwidth}
\includegraphics[width=7.5cm]{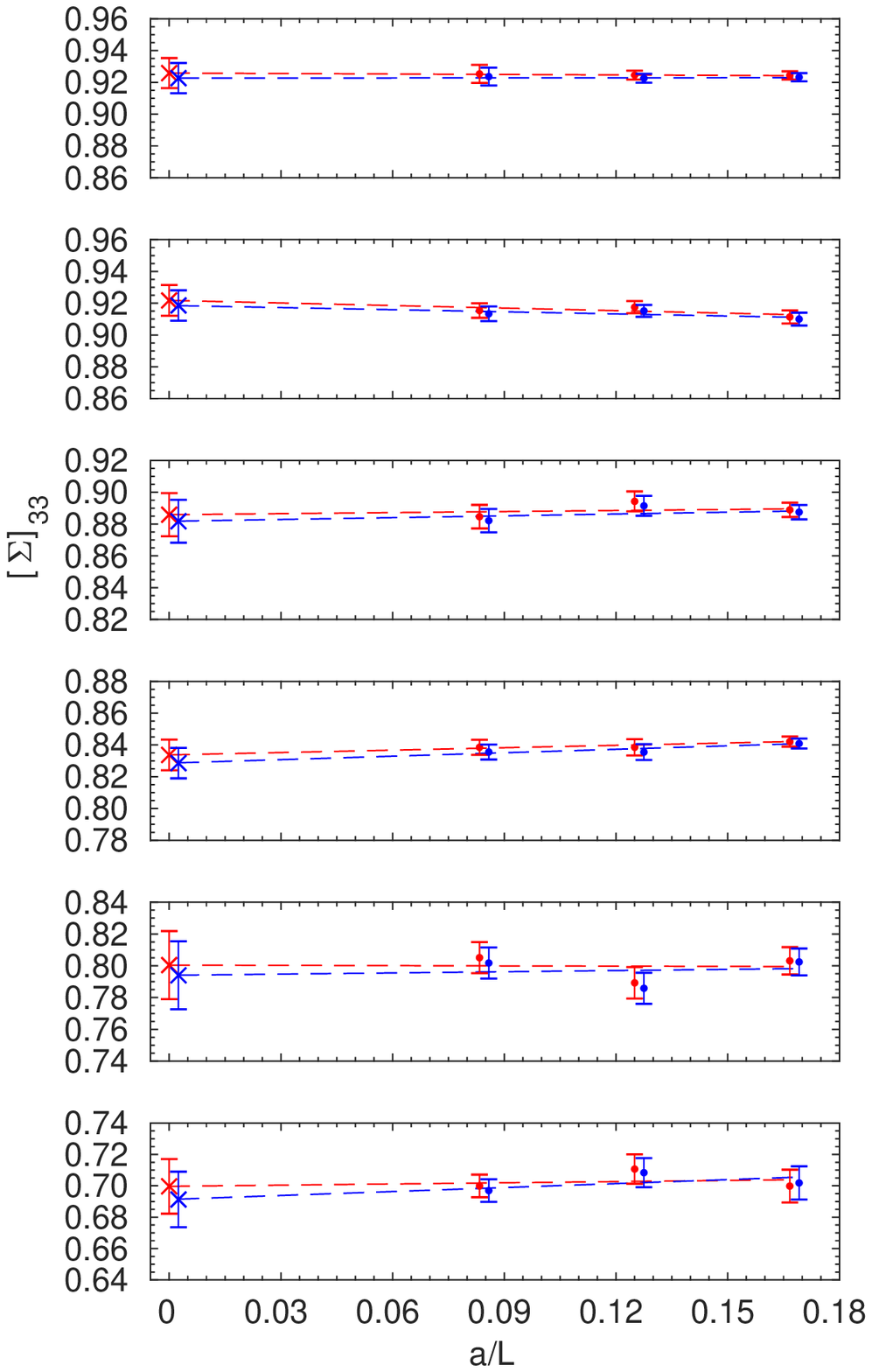}
\end{minipage}
\caption{Continuum limit extrapolation of $\bSigma^+(u,a/L)$ in red
  and $\tilde{\bSigma}^+(u,a/L)$ in blue, the operator basis $\{
  \cQ_2^+, \cQ_3^+ \}$. The values of the renormalised coupling, $u=0.9793, 1.1814, 1.5078, 2.0142, 2.4792, 3.3340$, grow from top to bottom for each element of the matrix-SSFs.}
\label{fig:cont23+}
\end{center}
\end{figure}

\newpage
\begin{figure}
\begin{center}
\vspace*{-5\baselineskip}
\begin{minipage}[t]{0.45\textwidth}
\includegraphics[width=7.5cm]{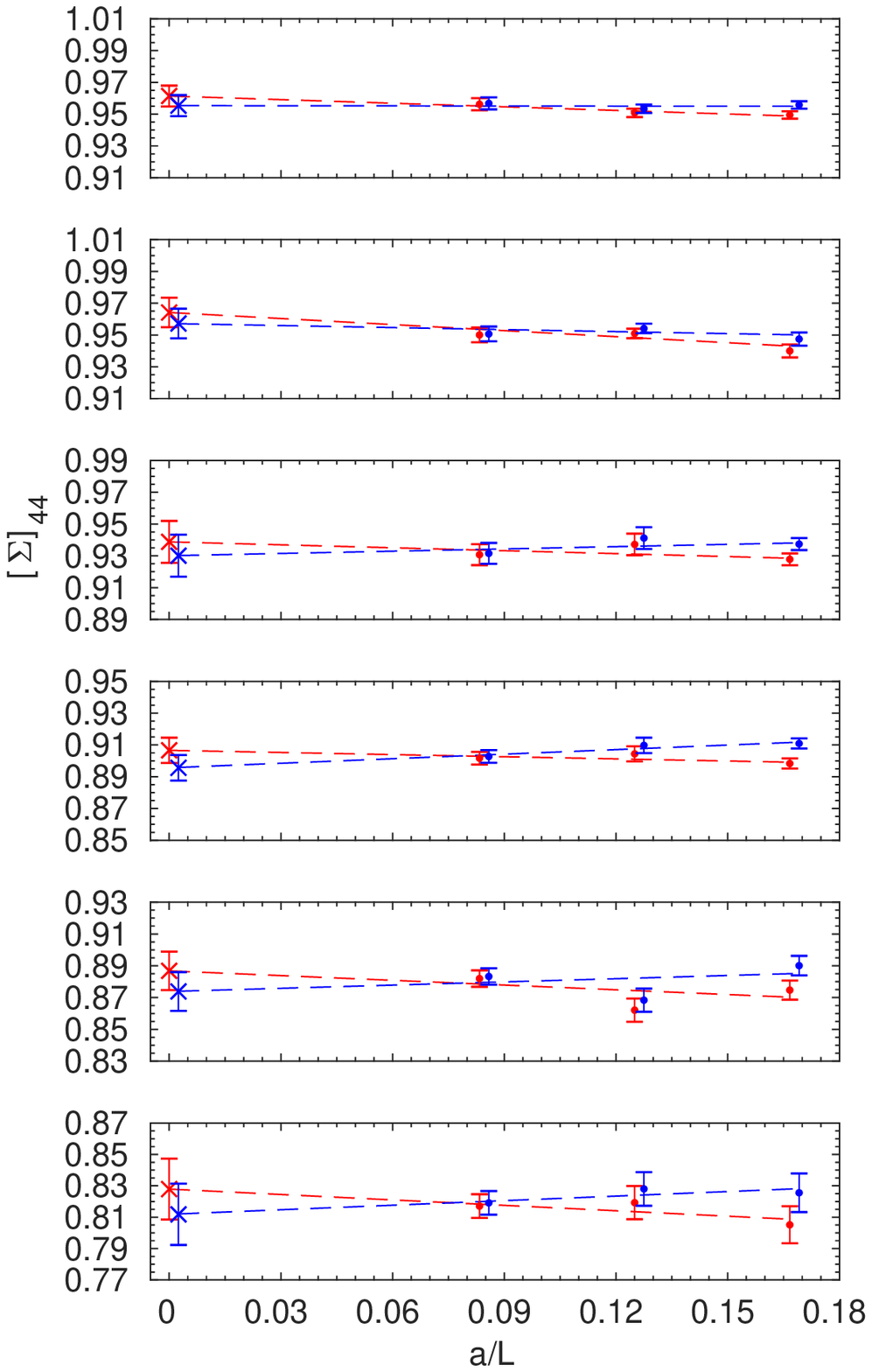}
\end{minipage}
\hspace{1.2cm}
\begin{minipage}[t]{0.45\textwidth}
\includegraphics[width=7.5cm]{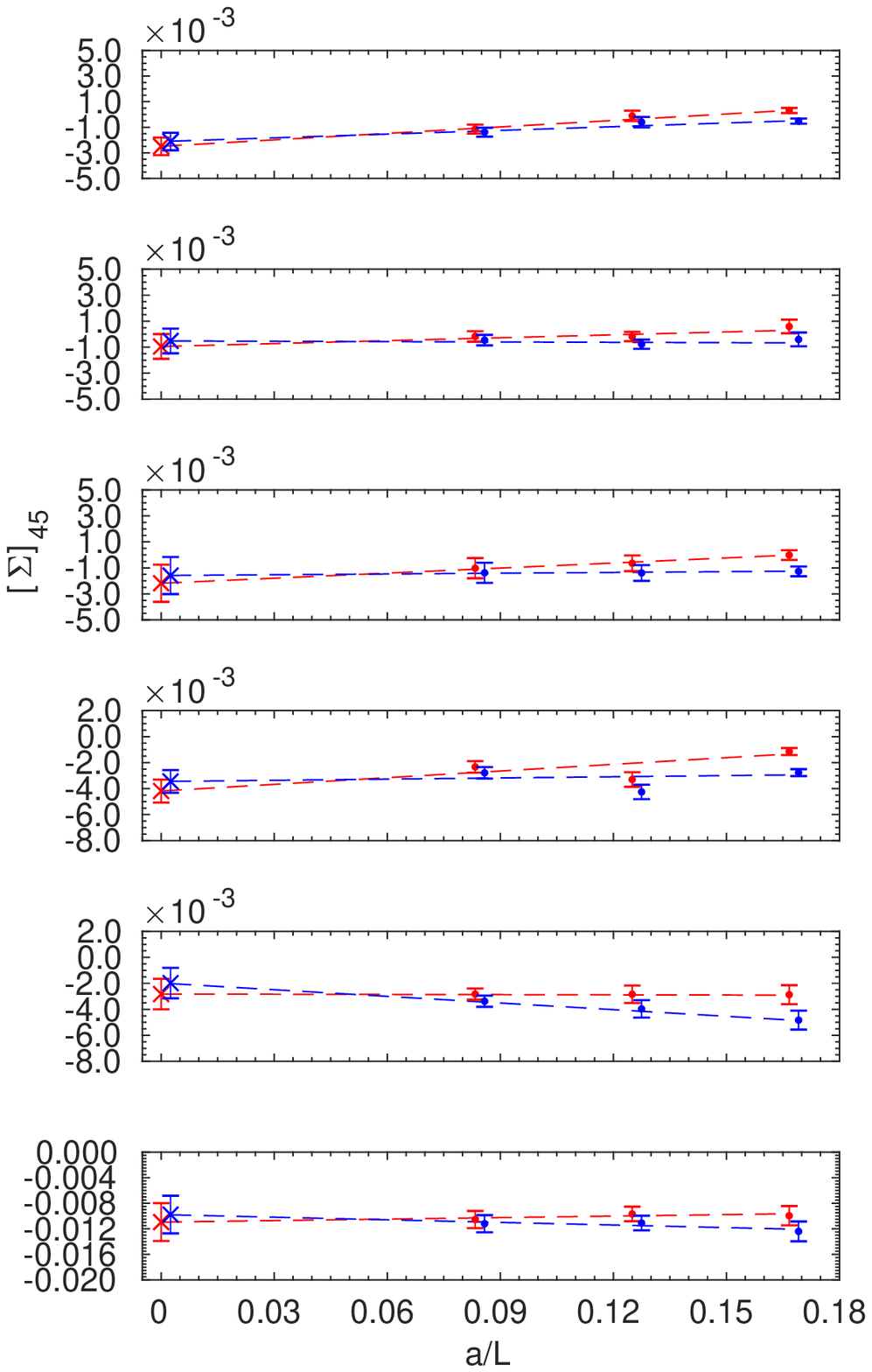}
\end{minipage}

\vspace{1mm}

\begin{minipage}[t]{0.45\textwidth}
\includegraphics[width=7.5cm]{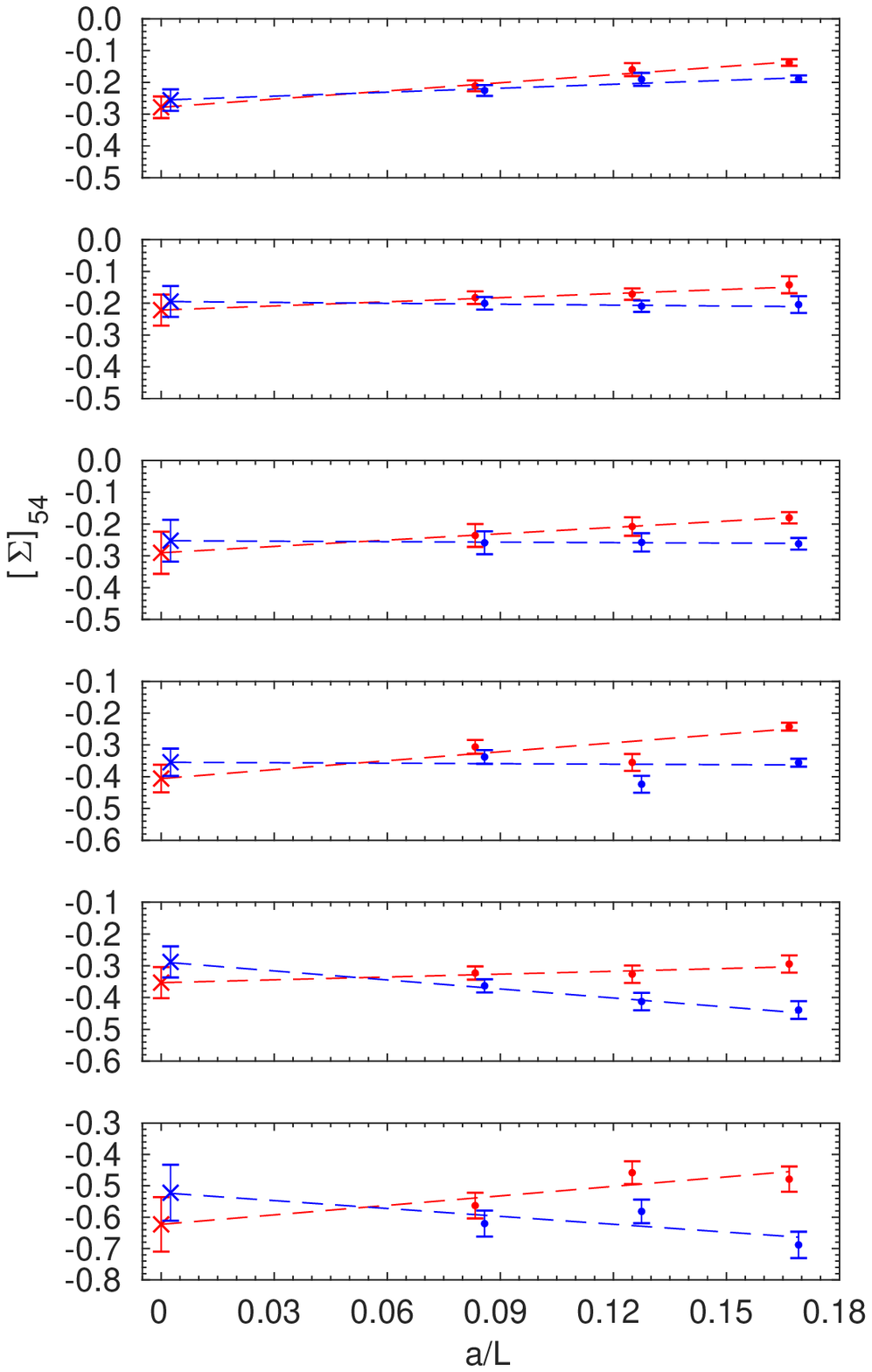}
\end{minipage}
\hspace{1cm}
\begin{minipage}[t]{0.45\textwidth}
\includegraphics[width=7.5cm]{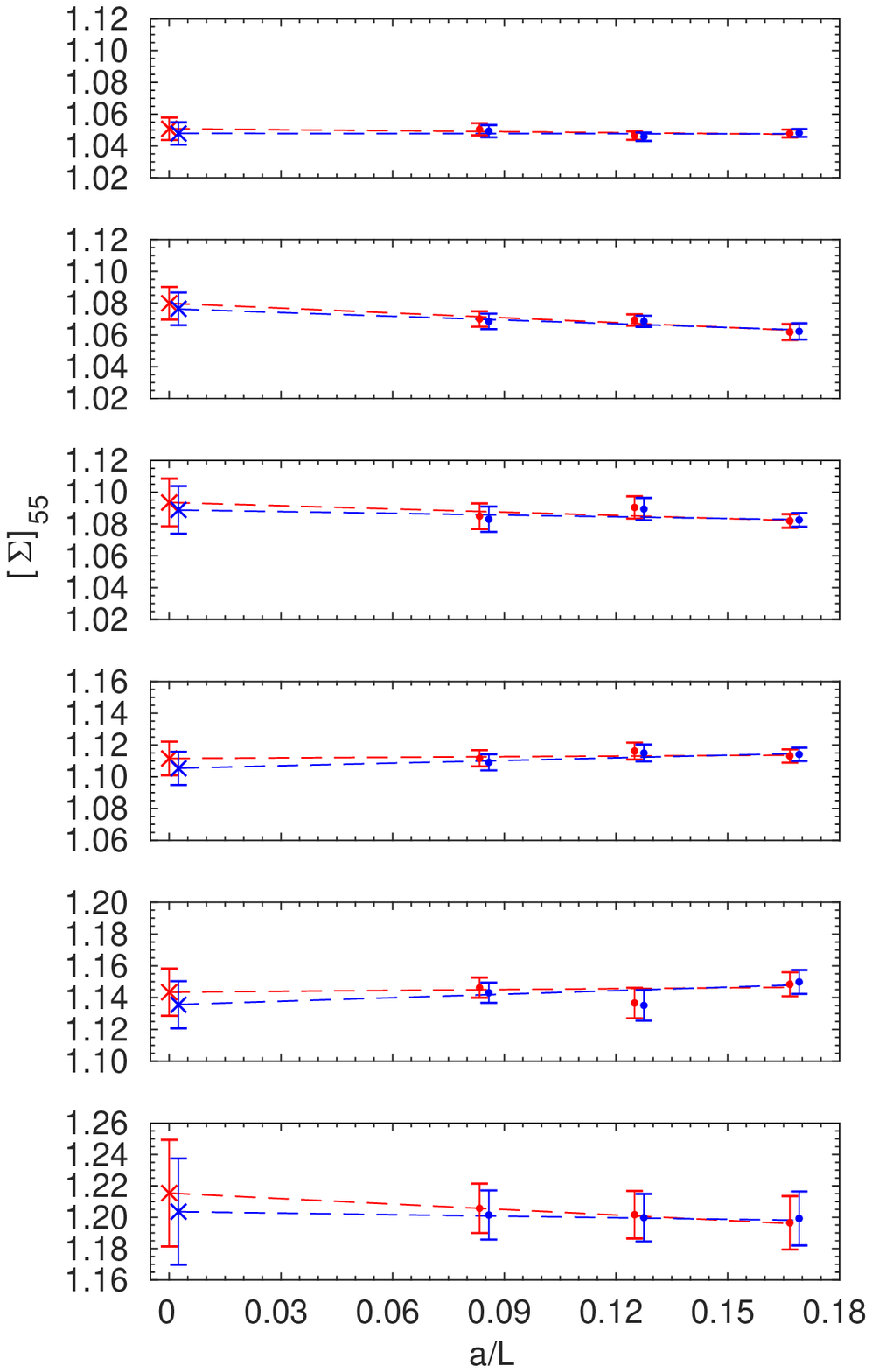}
\end{minipage}
\caption{Continuum limit extrapolation of $\bSigma^+(u,a/L)$ in red and $\tilde{\bSigma}^+(u,a/L)$ in blue, the operator basis $\{ \cQ_4^+, \cQ_5^+ \}$. The values of the renormalised coupling, $u=0.9793, 1.1814, 1.5078, 2.0142, 2.4792, 3.3340$, grow from top to bottom for each element of the matrix-SSFs.}
\label{fig:cont45+}
\end{center}
\end{figure}

\newpage
\begin{figure}
\begin{center}
\vspace*{-5\baselineskip}
\begin{minipage}[t]{0.45\textwidth}
\includegraphics[width=7.5cm]{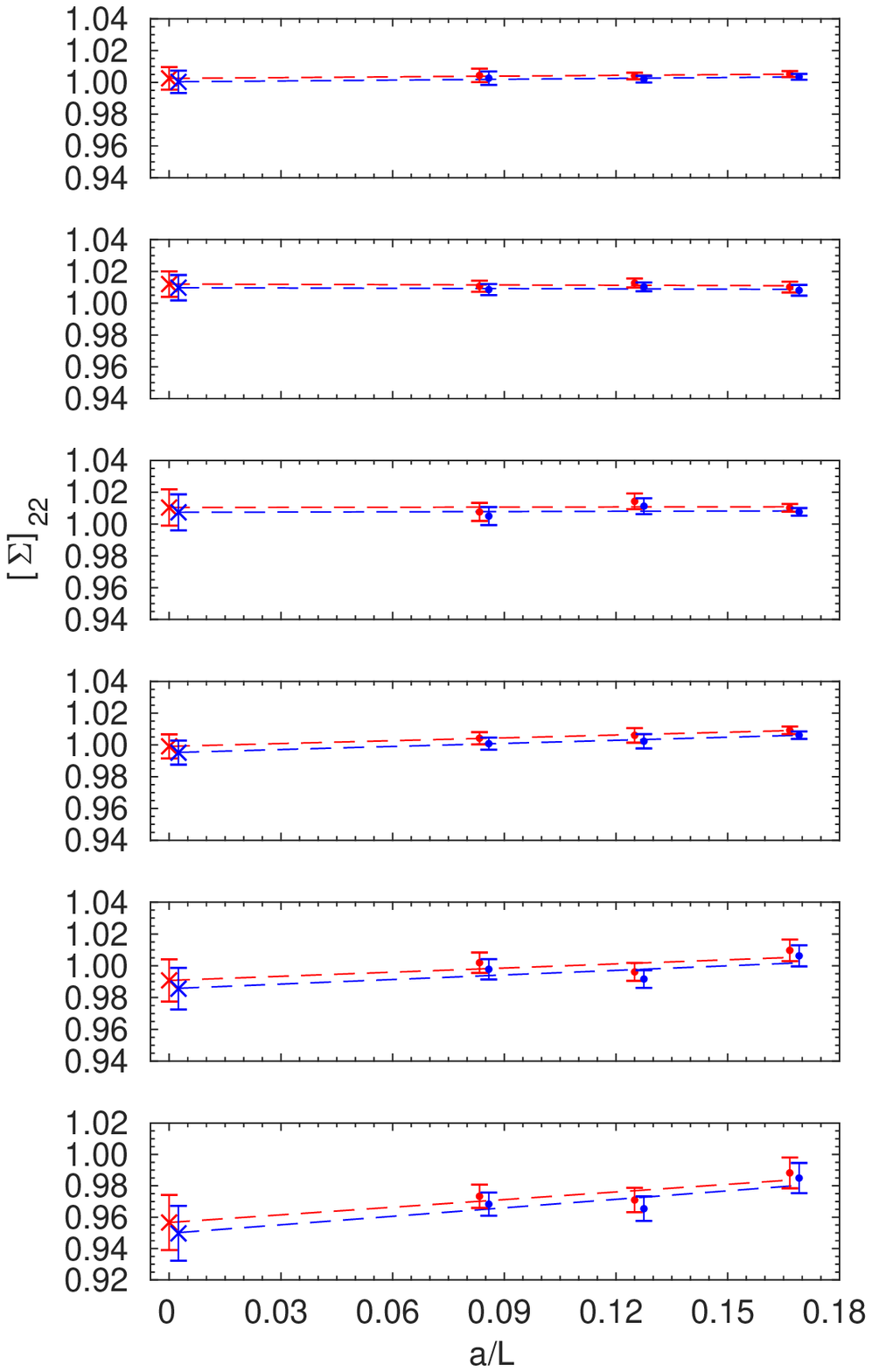}
\end{minipage}
\hspace{1.2cm}
\begin{minipage}[t]{0.45\textwidth}
\includegraphics[width=7.5cm]{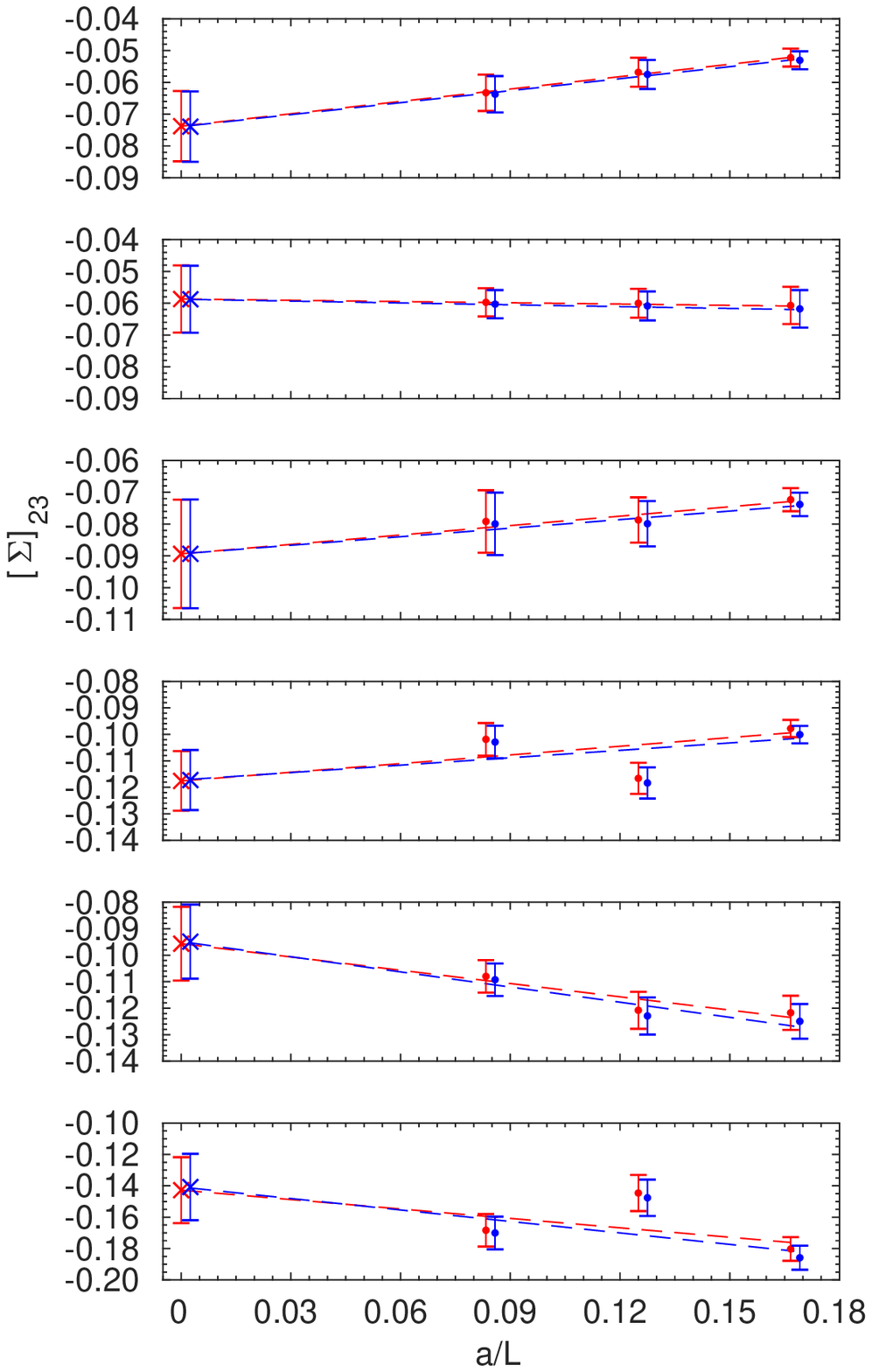}
\end{minipage}

\vspace{1mm}

\begin{minipage}[t]{0.45\textwidth}
\includegraphics[width=7.5cm]{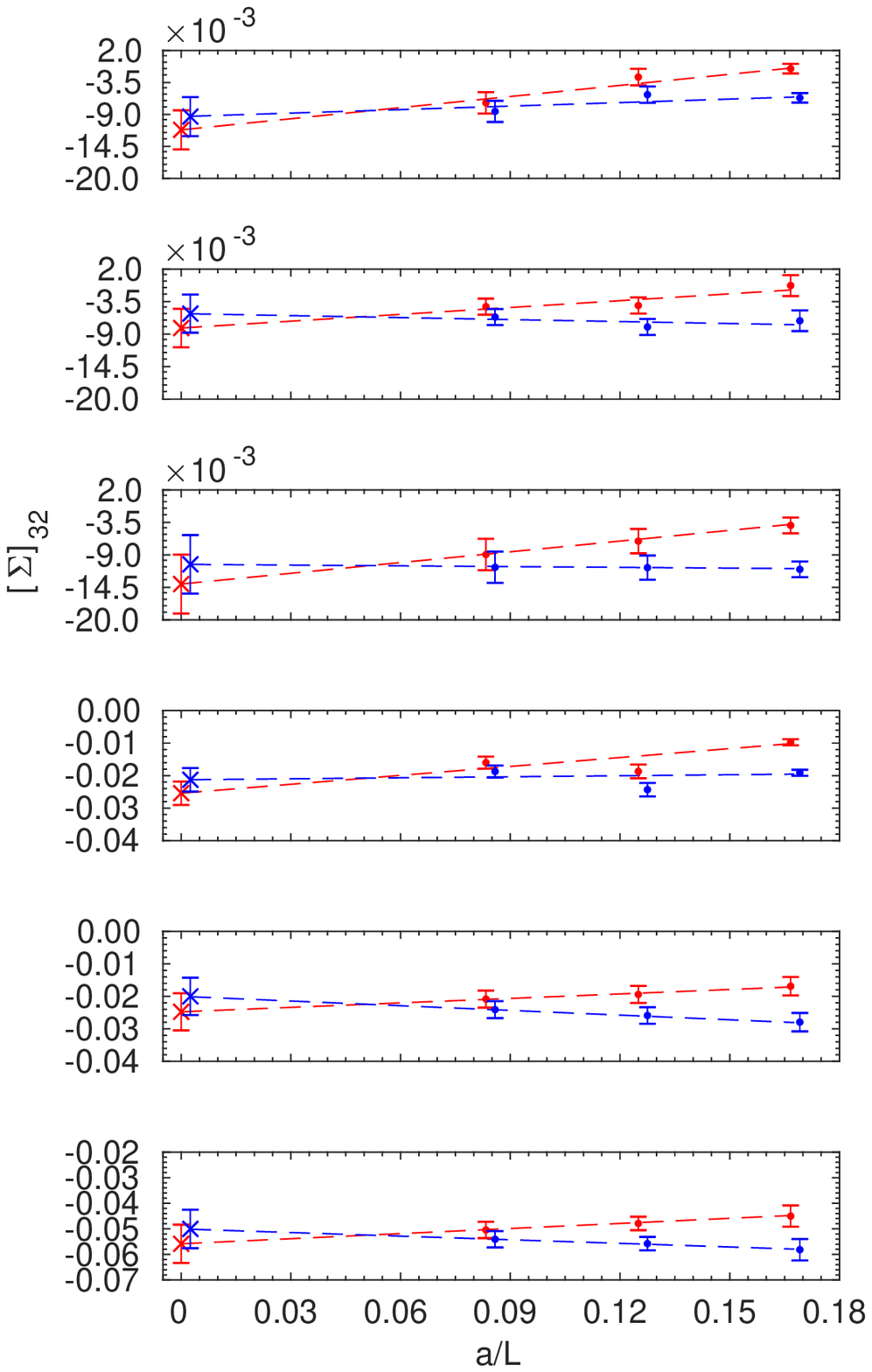}
\end{minipage}
\hspace{1cm}
\begin{minipage}[t]{0.45\textwidth}
\includegraphics[width=7.5cm]{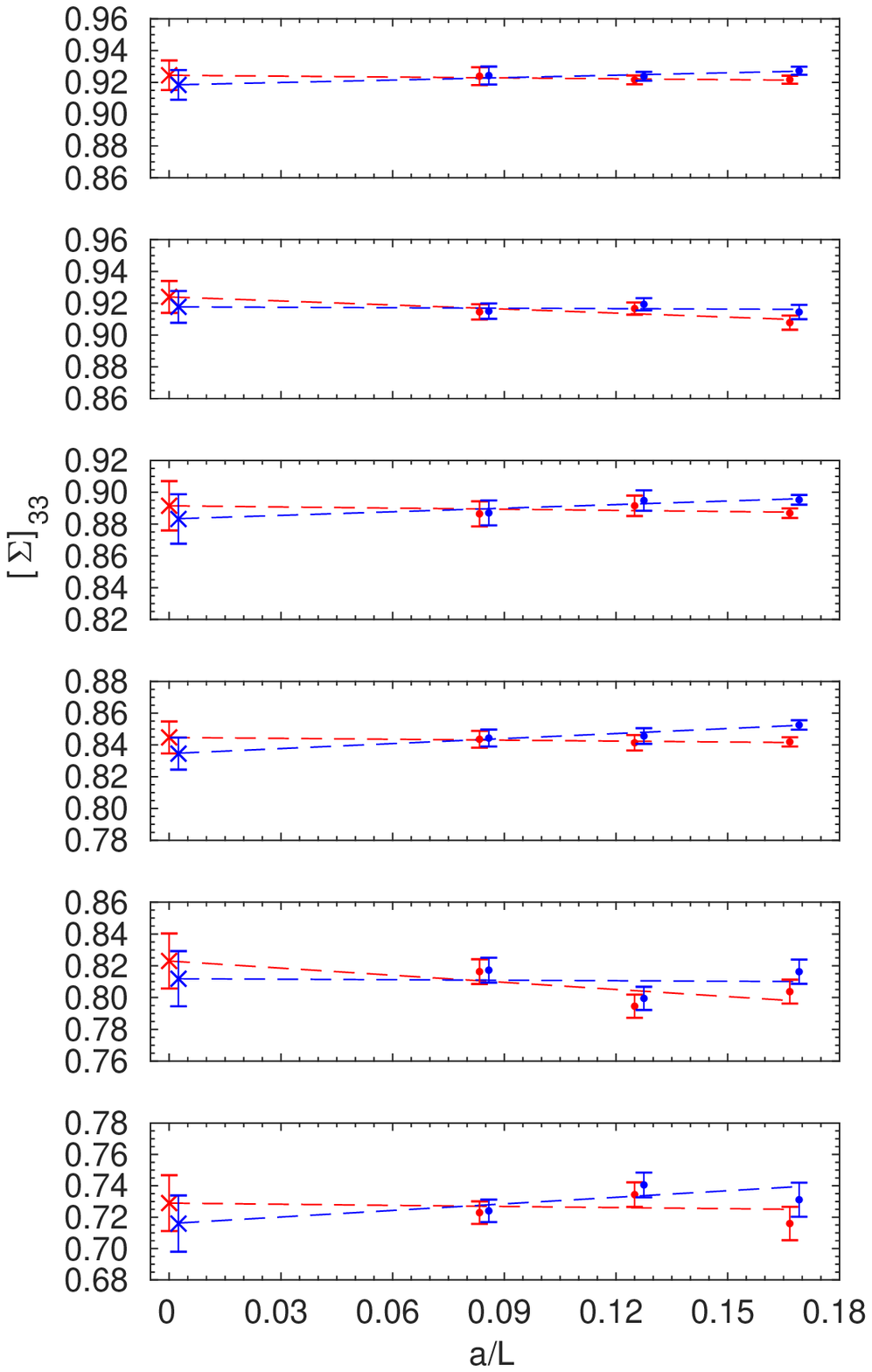}
\end{minipage}
\caption{Continuum limit extrapolation of $\bSigma^-(u,a/L)$ in red and $\tilde{\bSigma}^-(u,a/L)$ in blue, the operator basis $\{ \cQ_2^-, \cQ_3^- \}$. The values of the renormalised coupling, $u=0.9793, 1.1814, 1.5078, 2.0142, 2.4792, 3.3340$, grow from top to bottom for each element of the matrix-SSFs.}\label{fig:cont23-}
\label{fig:cont23-}
\end{center}
\end{figure}

\newpage
\begin{figure}
\begin{center}
\vspace*{-5\baselineskip}
\begin{minipage}[t]{0.45\textwidth}
\includegraphics[width=7.5cm]{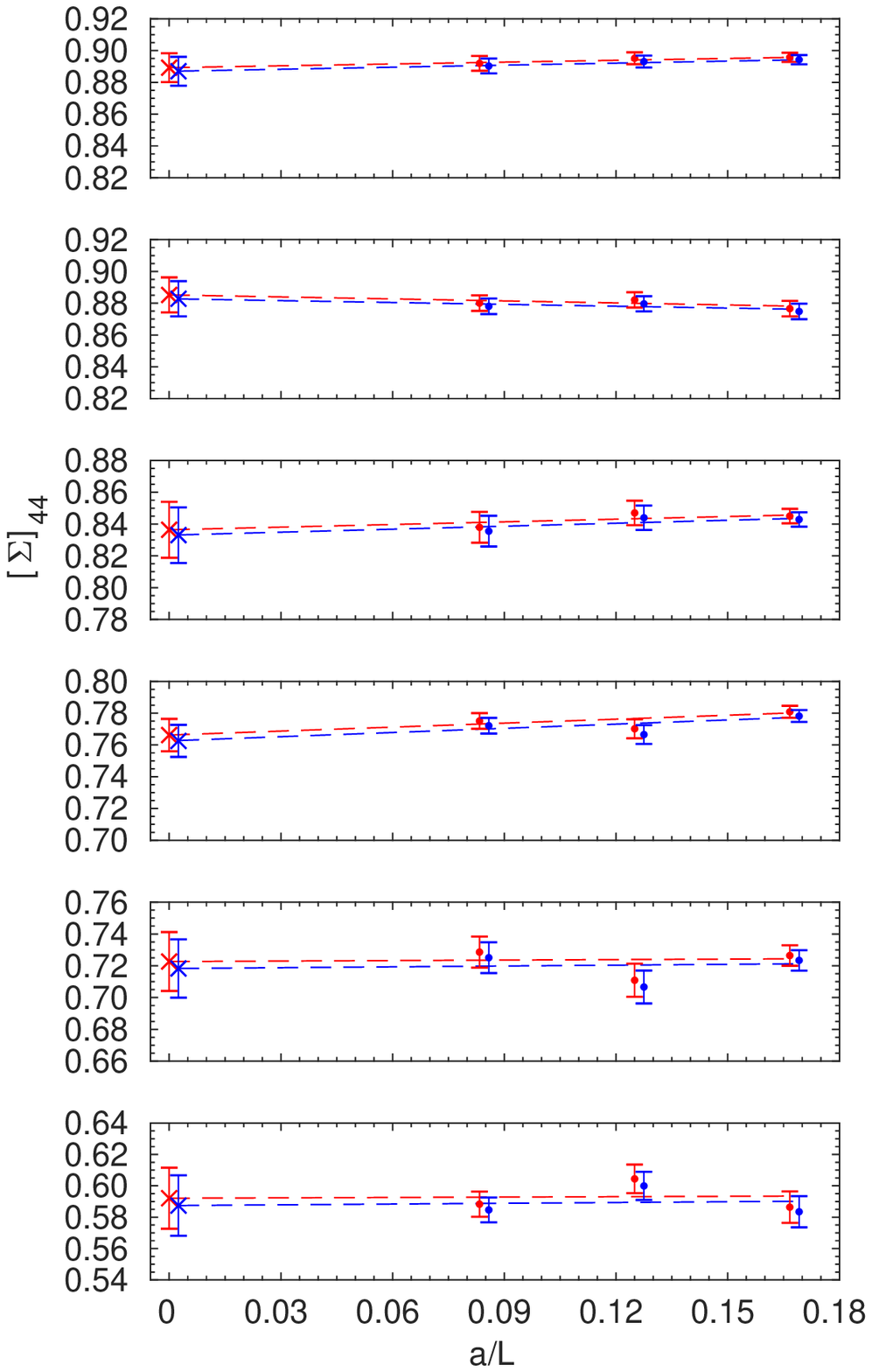}
\end{minipage}
\hspace{1.2cm}
\begin{minipage}[t]{0.45\textwidth}
\includegraphics[width=7.5cm]{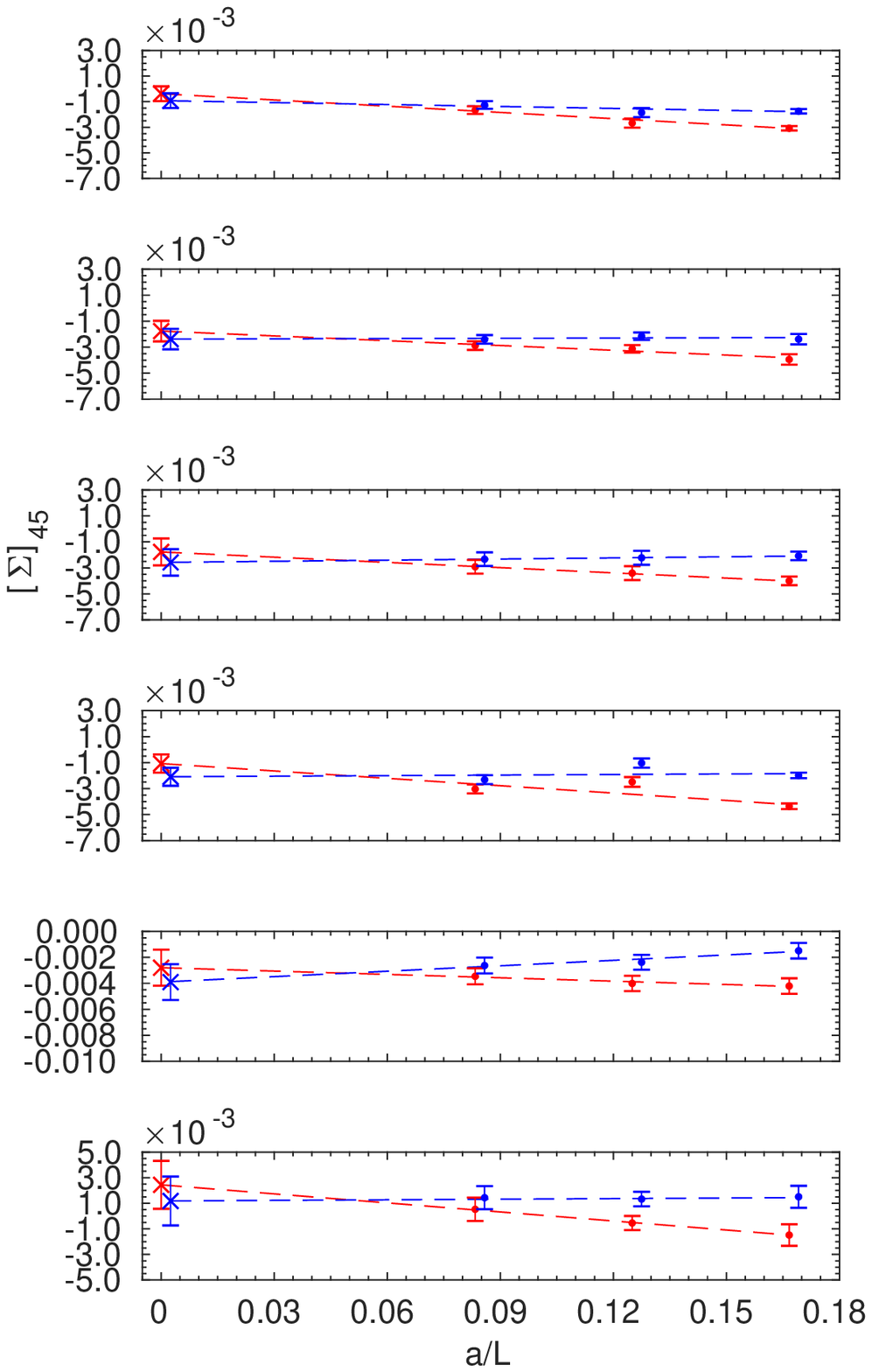}
\end{minipage}

\vspace{1mm}

\begin{minipage}[t]{0.45\textwidth}
\includegraphics[width=7.5cm]{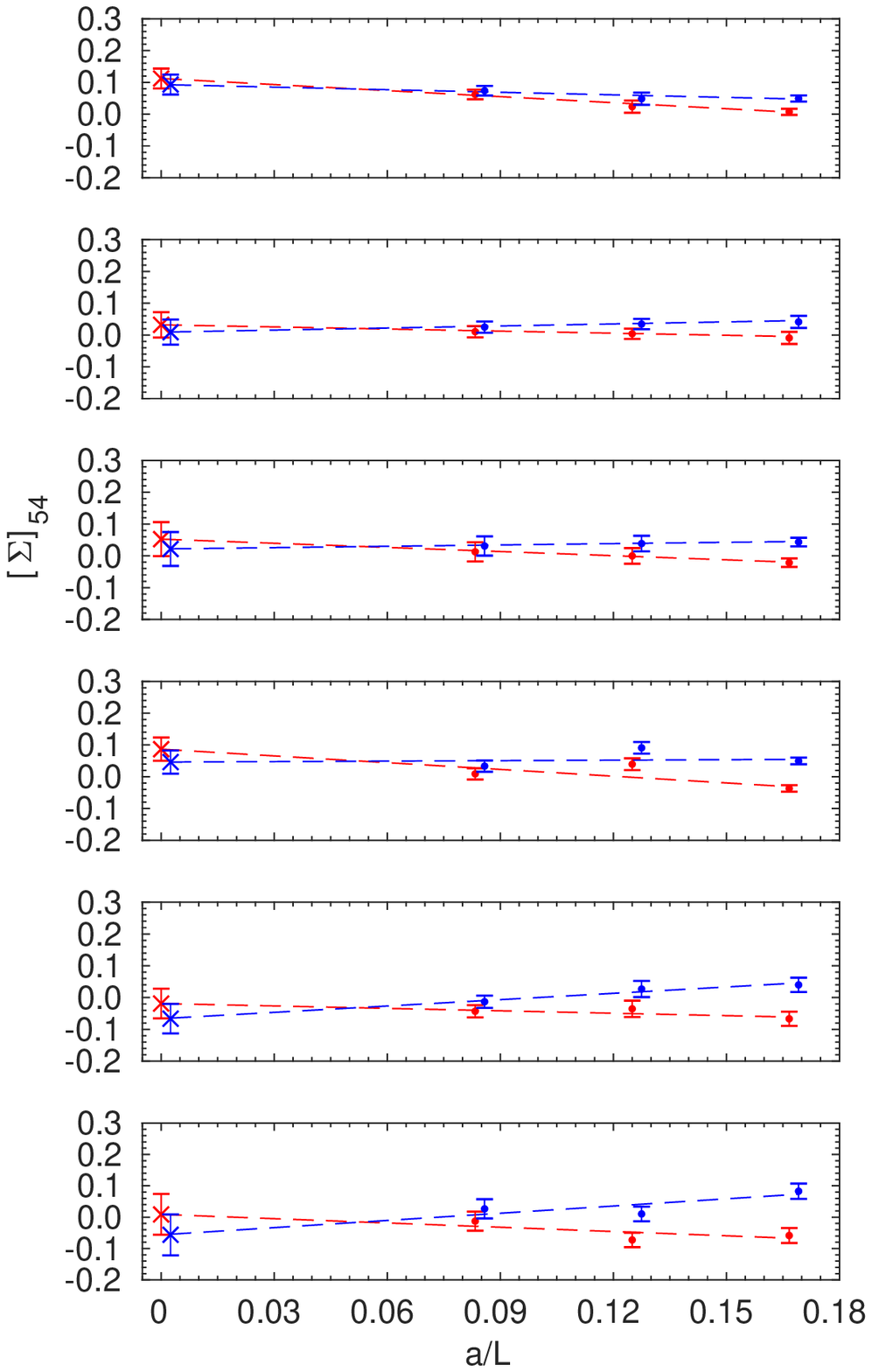}
\end{minipage}
\hspace{1cm}
\begin{minipage}[t]{0.45\textwidth}
\includegraphics[width=7.5cm]{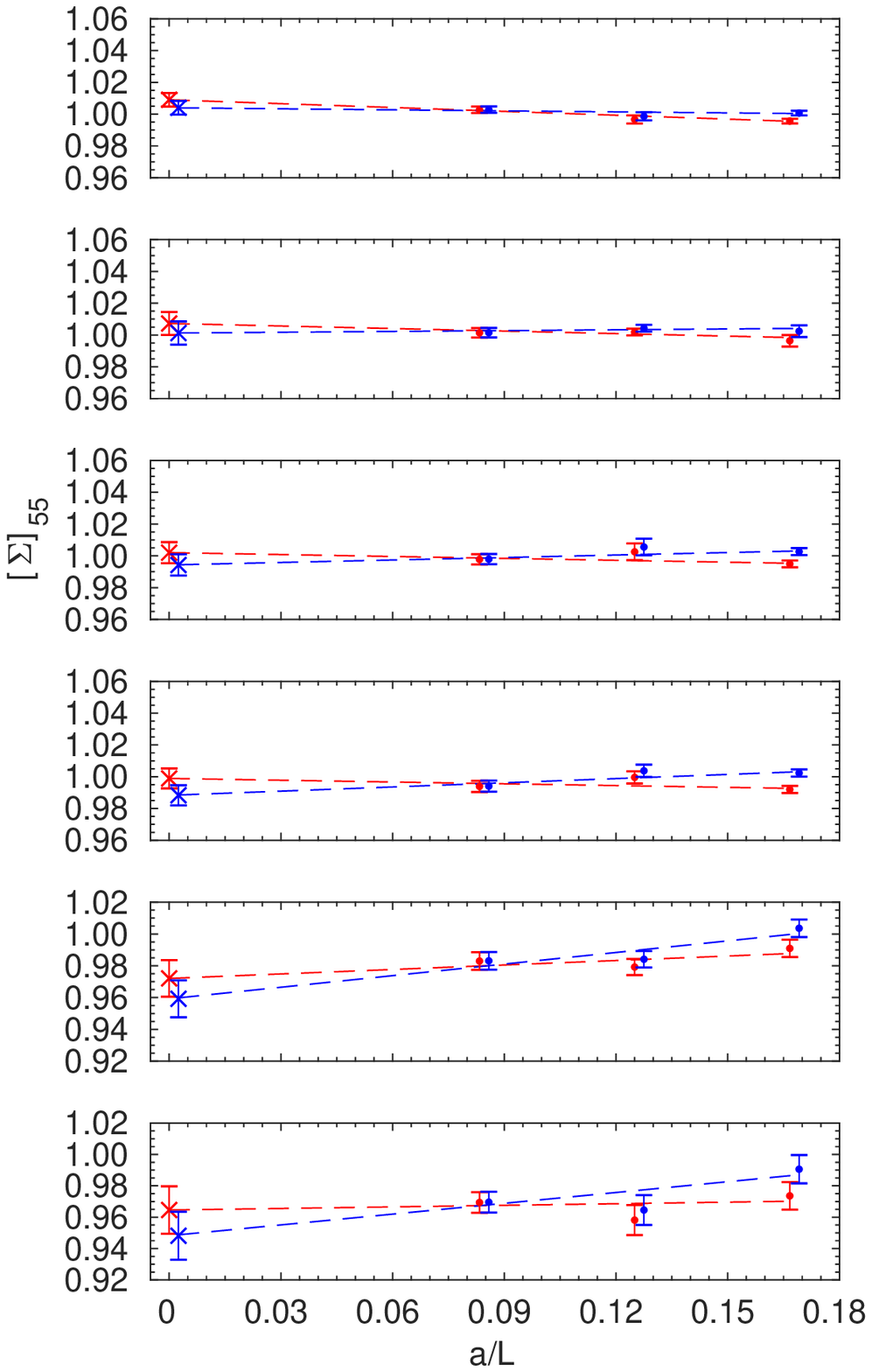}
\end{minipage}

\caption{Continuum limit extrapolation of $\bSigma^-(u,a/L)$ in red and $\tilde{\bSigma}^-(u,a/L)$ in blue, the operator basis $\{ \cQ_4^-, \cQ_5^- \}$. The values of the renormalised coupling, $u=0.9793, 1.1814, 1.5078, 2.0142, 2.4792, 3.3340$, grow from top to bottom for each element of the matrix-SSFs.}
\label{fig:cont45-}
\end{center}
\end{figure}

%% file: acknow.tex
\section*{Acknowledgments}

The present work is an extension of previous efforts dedicated to the
SF renormalisation and running of the $B_{\rm K}$-parameter in the
Standard Model. We are indebted to our collaborators at the time,
namely M.~Guagnelli, J. Heitger, F.~Palombi, and S.~Sint, for their
early contributions. We owe a lot to their participation in the
defining phase of this project.
G.H., C.P. and D.P. acknowledge support by the Spanish MINECO
grant FPA2015-68541-P (MINECO/FEDER), the MINECO's Centro de Excelencia Severo Ochoa
Programme under grant SEV-2016-0597 and the Ram\'on y Cajal Programme
RYC-2012-10819.
M.P. acknowledges partial support by the MIUR-PRIN grant 2010YJ2NYW and by
the INFN SUMA project.

%% file: app_check.tex
\section{Non-perturbative vs perturbative behaviour of the RG evolution}
\label{sec:checks}

\begin{figure}
\begin{center}
\vspace*{-3\baselineskip}
\includegraphics[width=0.9\textwidth]{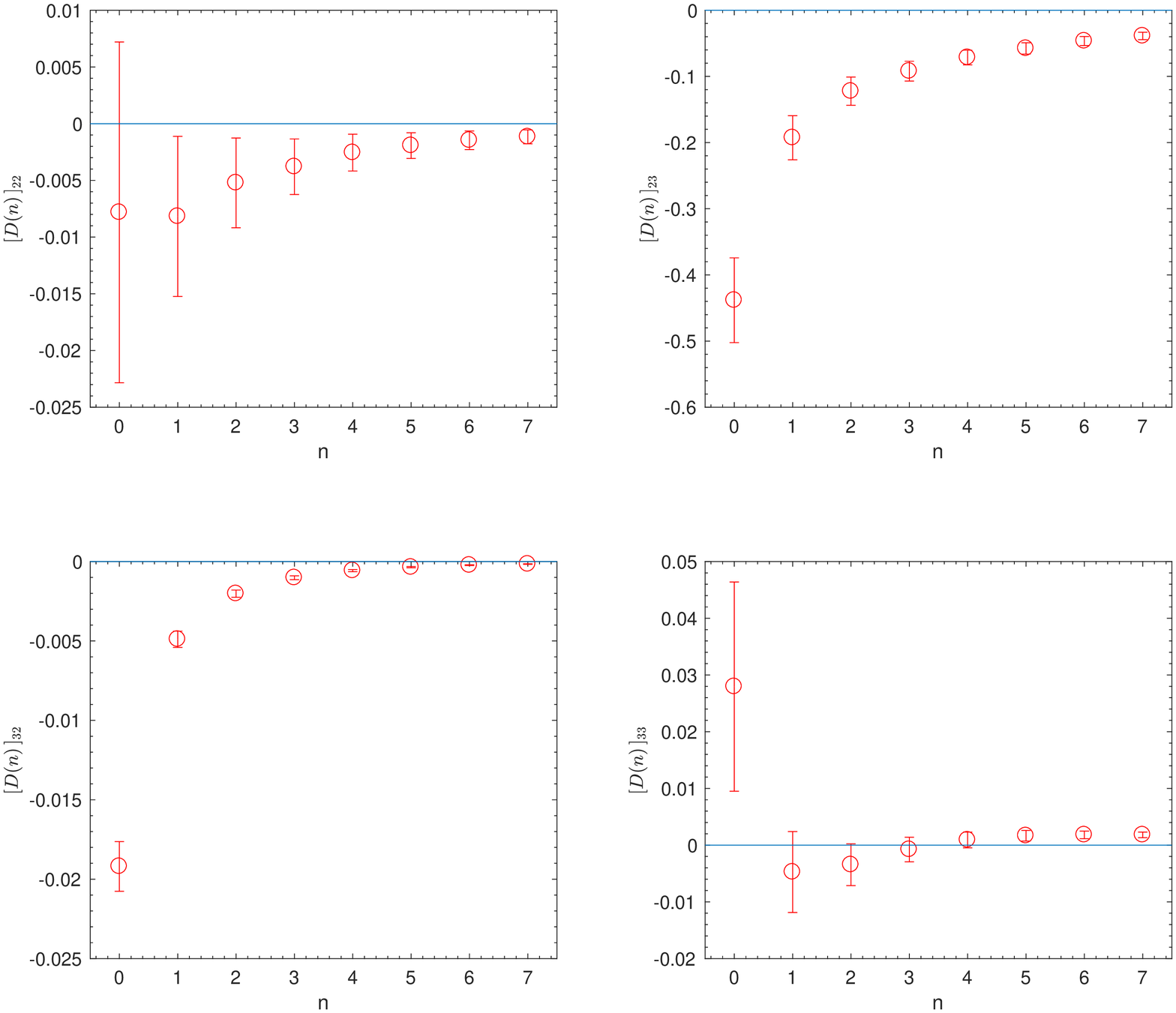}

\vspace{1mm}

\includegraphics[width=0.9\textwidth]{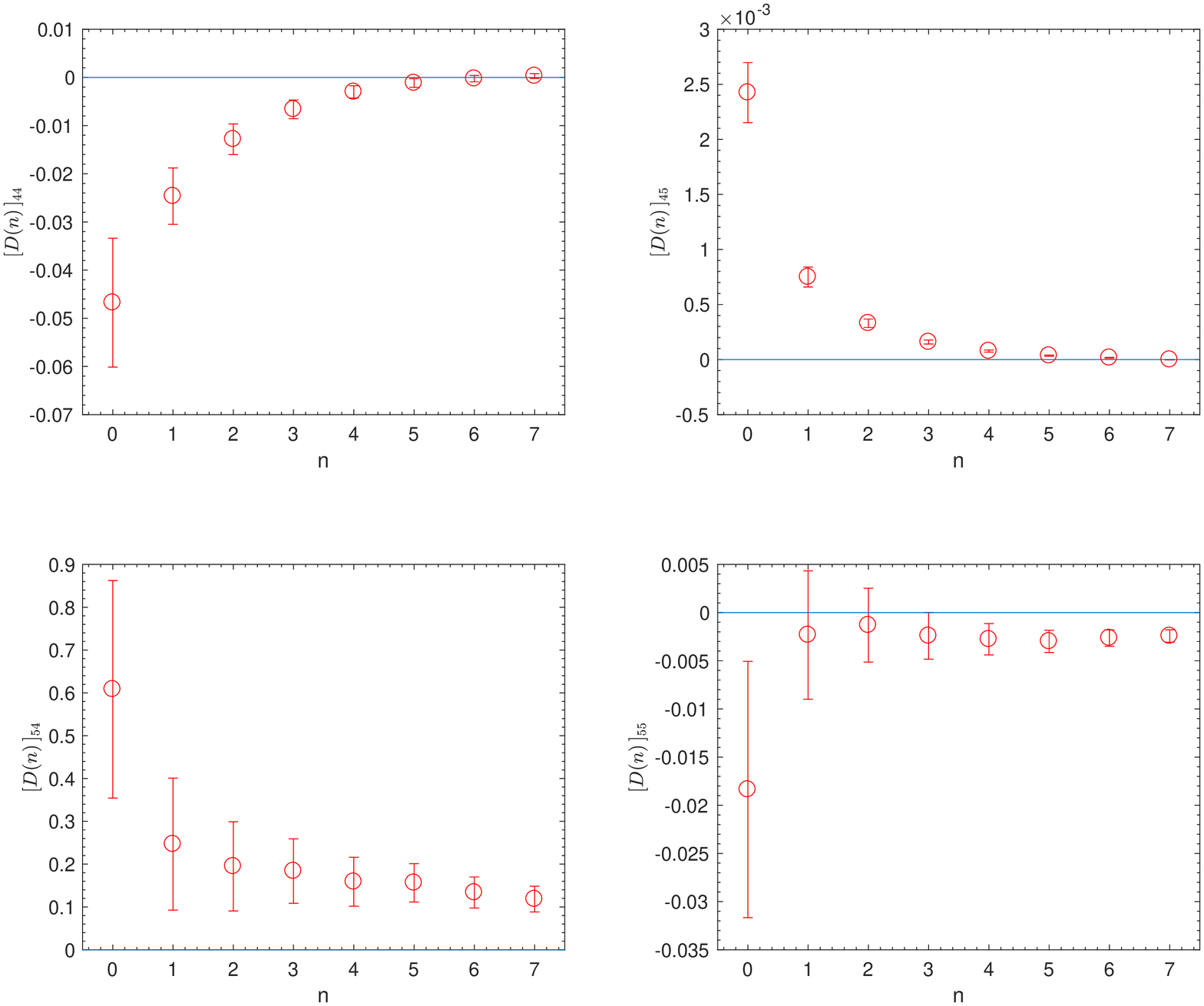}

\caption{The quantity $\boldsymbol D (n)$ \req{eq:D_systematics} for the operator bases $\{ \cQ_2^+, \cQ_3^+ \}$ (top) and
$\{ \cQ_4^+, \cQ_5^+ \}$ (bottom).}
\label{fig:plot_DELTA_sys+}
\end{center}
\end{figure}

\begin{figure}
\begin{center}
\vspace*{-3\baselineskip}
\includegraphics[width=0.9\textwidth]{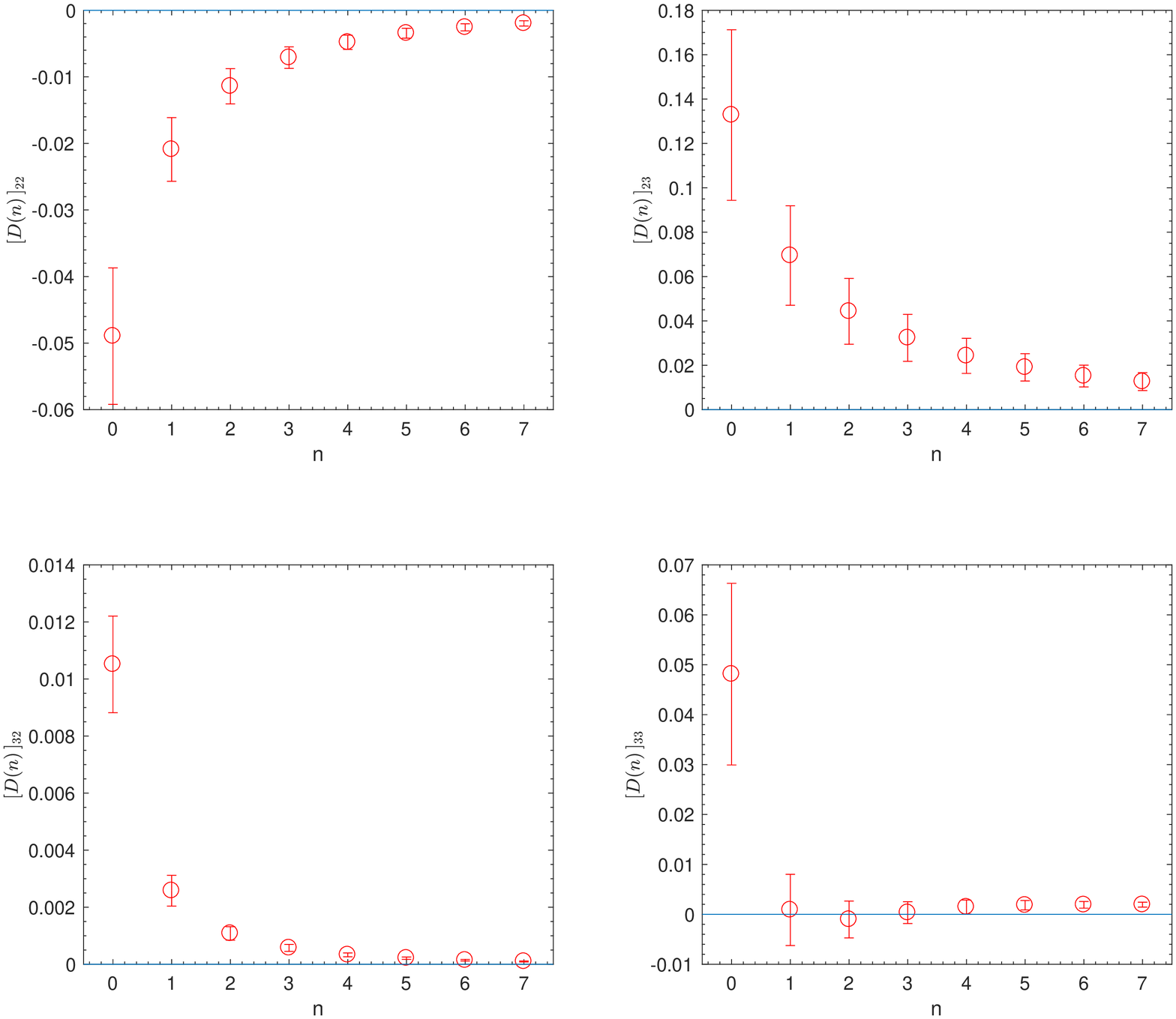}

\vspace{1mm}

\includegraphics[width=0.9\textwidth]{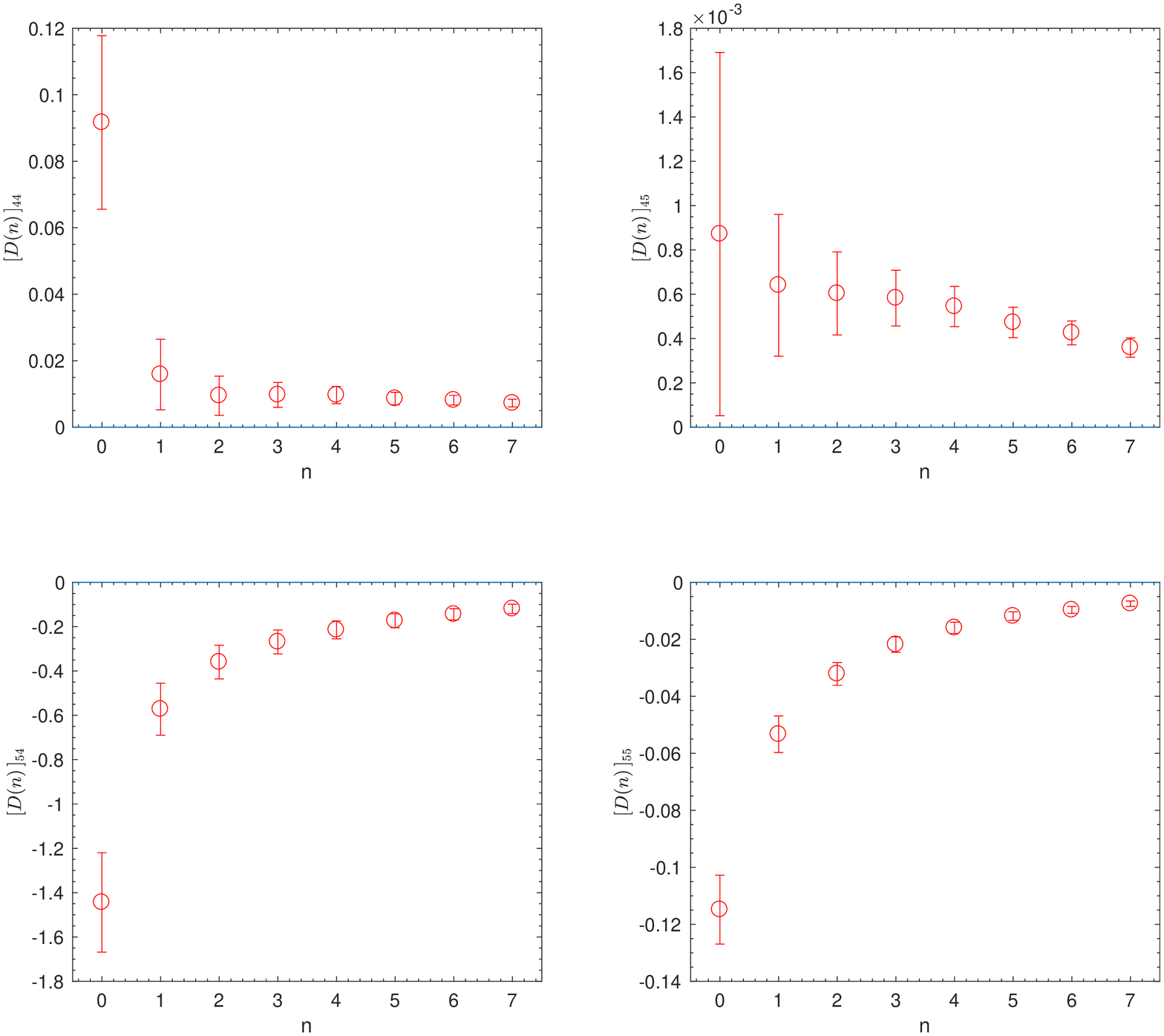}

\caption{The quantity $\boldsymbol D (n)$ \req{eq:D_systematics} for the operator bases $\{ \cQ_2^-, \cQ_3^- \}$ (top) and
$\{ \cQ_4^-, \cQ_5^- \}$ (bottom).}
\label{fig:plot_DELTA_sys-}
\end{center}
\end{figure}

In analogy to Appendix C of ref.~\cite{Boyle:2017skn} we construct the quantity:  
\begin{eqnarray}
\boldsymbol D(n)& \equiv &[\tilde \bU(2^n \mu_{\rm had}) \bU(\mu_{\rm had},2^n\mu_{\rm had})^{-1}][\tilde \bU(2^{n+1}\mu_{\rm had}) \bU(\mu_{\rm had},2^{n+1}\mu_{\rm had})^{-1}]^{-1}- {\bf 1} \nonumber \\
&=& [\tilde \bU(2^n \mu_{\rm had})]\bsigma(u_{n+1})[\tilde \bU(2^{n+1}\mu_{\rm had})]^{-1} -  {\bf 1} \nonumber \\
&=& [\tilde \bU(2^{n+1} \mu_{\rm had})][\bU(2^n\mu_{\rm had},2^{n+1}\mu_{\rm had})^{-1}\bsigma(u_{n+1})][\tilde \bU(2^{n+1}\mu_{\rm had})]^{-1} - {\bf 1}
\label{eq:D_systematics}
\end{eqnarray}
Once again $\tilde \bU(2^n \mu_{\rm had})$, $\tilde \bU(2^{n+1}\mu_{\rm had})$, and $\bU(2^{n}\mu_{\rm had},2^{n+1}\mu_{\rm had})$ are perturbative quantities known in NLO-2/3PT , while $\bsigma(u_{n+1})$ is a single non-perturbative
matrix-SSF. In the last line of \req{eq:D_systematics}, the product $[\bU(2^n\mu_{\rm had},2^{n+1}\mu_{\rm had})^{-1} \bsigma(u_{n\
+1})]$ is the ratio of the non-perturbative over the perturbative RG evolution between scales $2^n\mu_{\rm had}$ and $2^{n+1}\mu_{\rm had}$. 
If perturbation theory were reliable at these high scales, $\boldsymbol D(n)$ would vanish at large $n$. The results for the $\boldsymbol D(n)$ matrix elements are shown in Figs.~\ref{fig:plot_DELTA_sys+}, \ref{fig:plot_DELTA_sys-}. At the largest $n$ values some of them are compatible with 0 while others are not. The latter case signals that due to large anomalous dimensions, NLO-2/3PT performs poorly even at scales as high as $2^n \mu_{\rm had}$ and $2^{n+1} \mu_{\rm had}$.

\begin{figure}
\begin{center}
\vspace*{-3\baselineskip}
\includegraphics[width=0.88\textwidth]{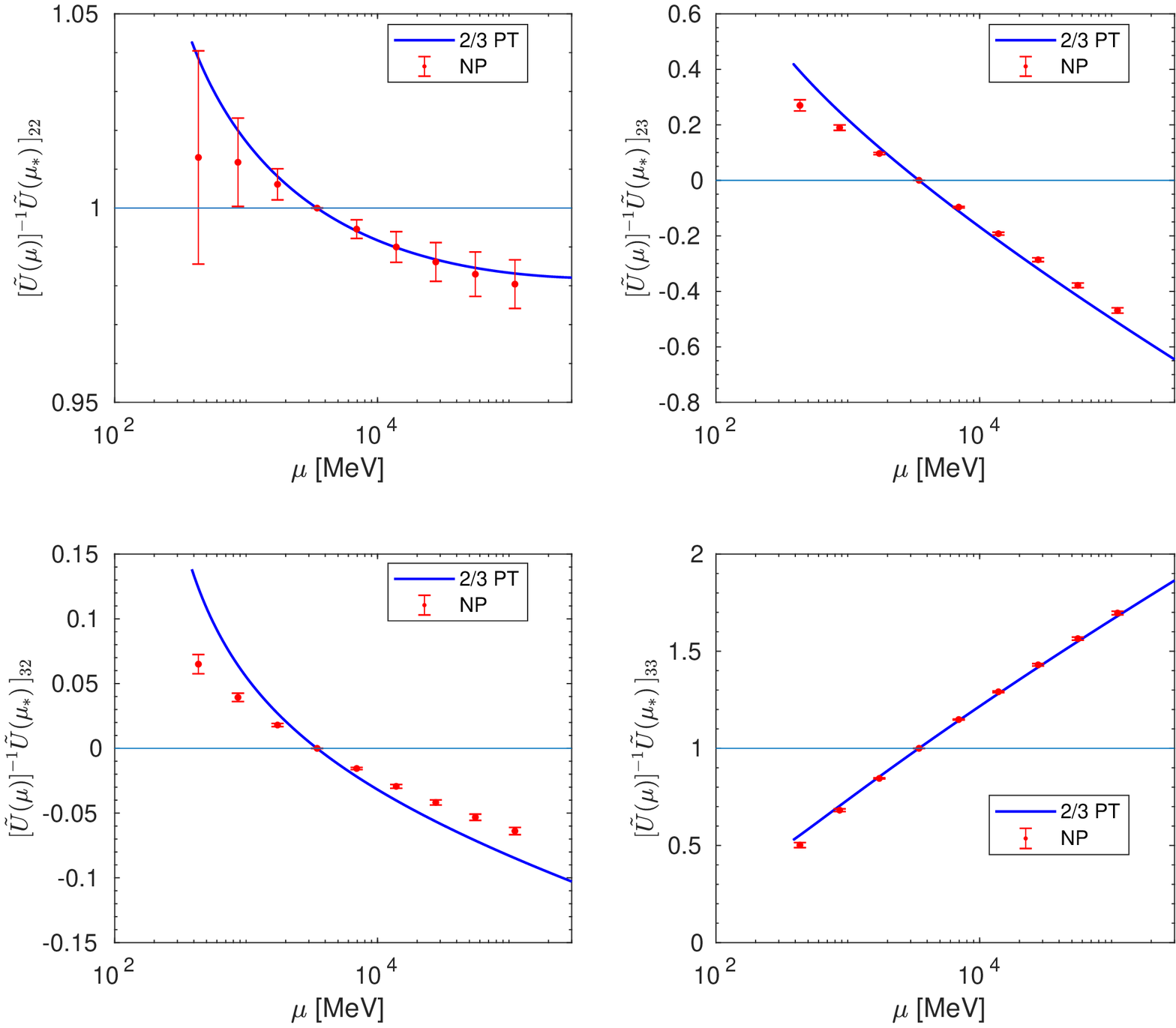}

\vspace{1mm}

\includegraphics[width=0.88\textwidth]{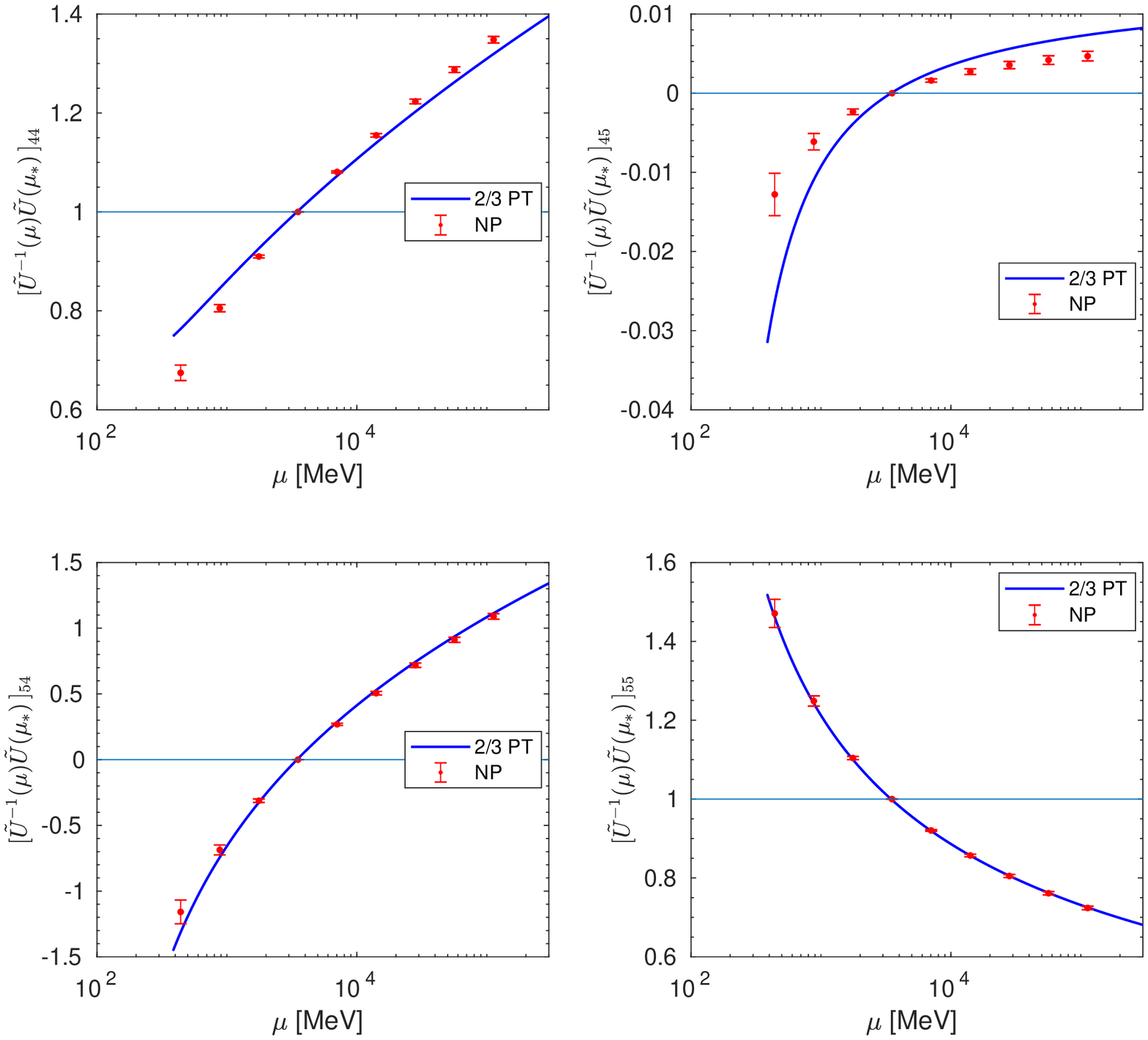}

\caption{Non-perturbative evolution factor $\bU^+(\mu,\mu_*)=[\tilde
    \bU^+(\mu)]^{-1} \tilde \bU^+(\mu_*)$, where $\mu_*=3.46\, \rm GeV$, for the operator bases $\{ \cQ_2^+, \cQ_3^+ \}$ (top) and for  $\{ \cQ_4^+, \cQ_5^+ \}$ (bottom). 
Results are compared to the perturbative prediction, obtained by numerically integrating~\req{eq:rg_W}, with 
$\bgamma$ (at NLO) and $\beta$ (at NNLO) in the SF scheme.}
\label{fig:evol_mustar+}
\end{center}
\end{figure}

\begin{figure}
\begin{center}
\vspace*{-3\baselineskip}
\includegraphics[width=0.88\textwidth]{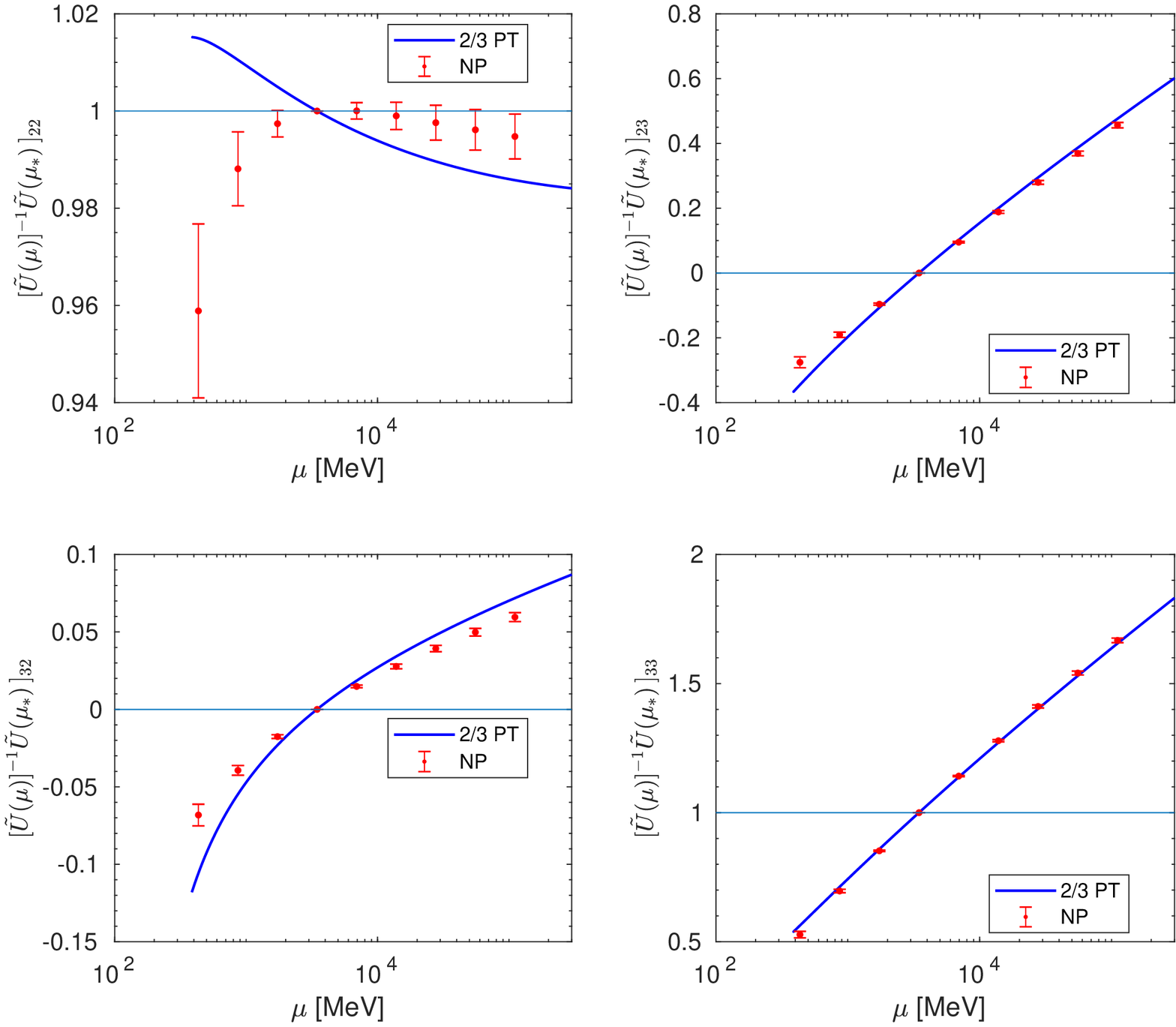}

\vspace{1mm}

\includegraphics[width=0.88\textwidth]{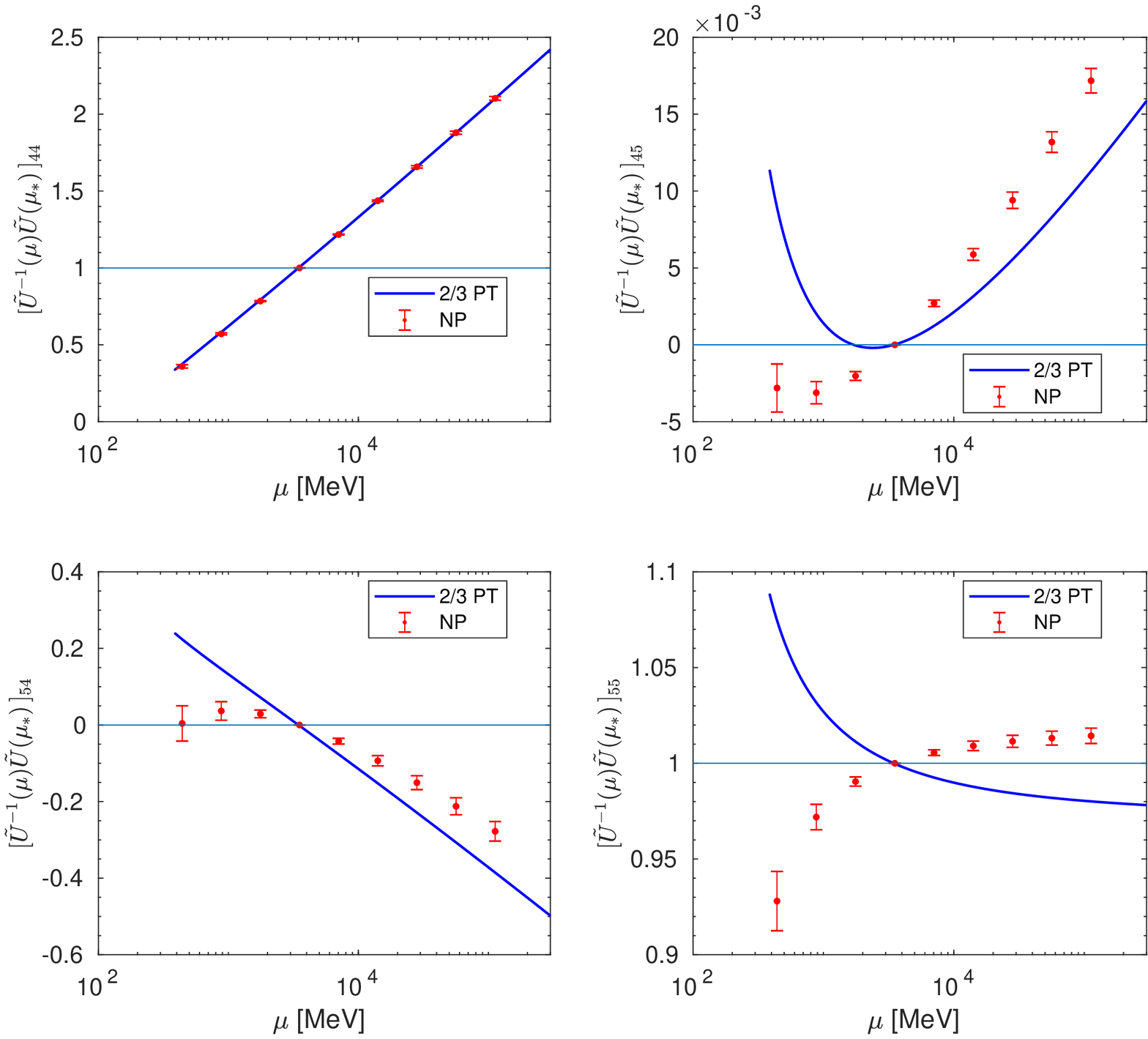}

\caption{Non-perturbative evolution factor $\bU^-(\mu,\mu_*)=[\tilde \bU^-(\mu)]^{-1} \tilde \bU^-(\mu_*)$, where $\mu_*=3.46\, \rm GeV$, for the operator bases $\{ \cQ_2^-, \cQ_3^- \}$ (top) and for  $\{ \cQ_4^-, \cQ_5^- \}$ (bottom). 
Results are compared to the perturbative prediction, obtained by numerically integrating~\req{eq:rg_W}, with 
$\bgamma$ (at NLO) and $\beta$ (at NNLO) in the SF scheme.}
\label{fig:evol_mustar-}
\end{center}
\end{figure}

Moreover, in Appendix C of ref.~\cite{Boyle:2017skn} the non-perturbative RG evolution $\bU(\mu,\mu_*)$ between scales $\mu$ and $\mu_*$ has been compared to the result from NLO-2/3PT. In ref.~\cite{Boyle:2017skn}, $\mu$ is kept fixed to 2 GeV while $\mu_*$ is varied in the range [2 GeV, 3 GeV]. 
We perform a similar study by fixing the reference scale $\mu_*=3.46\, \rm GeV=2^3\mu_{\rm had}$, corresponding to the squared
coupling $u_3$. This is the scale closest to the interval [2 GeV, 3 GeV] of ref.~\cite{Boyle:2017skn},  for which we have directly computed  the matrix-SSFs
non-perturbatively.
The scale $\mu$ is varied in the range [0.43 GeV, 110 GeV]. We compute $\bU(\mu,\mu_*)$ in the following way: 
 \begin{eqnarray}
 \bU(\mu,\mu_*) & = & [\tilde \bU(\mu)]^{-1}\tilde \bU(\mu_*)  \nonumber \\
 & = & \left[\bU(\mu,\mu_{\rm pt})\right] \, [\bW(\mu_{\rm pt})]^{-1}  \, \left[\frac{\gbar^2(\mu_{\rm pt})}{4\pi}\right]^{\frac{\bgamma^{(0)}}{2b_0}}\, \left[\frac{\gbar^2(\mu_{\rm pt})}{4\pi}\right]^{-\frac{\bgamma^{(0)}}{2b_0}}\, \bW(\mu_{\rm pt})\left[\bU(\mu_*,\mu_{\rm pt})\right]^{-1} \nonumber \\
& = & \bU(\mu,\mu_{\rm pt})\bU(\mu_*,\mu_{\rm pt})^{-1}\,.
 \end{eqnarray}
$\bU(\mu,\mu_*)$ can be evaluated in a purely non-pertubative way for integer 
$n_1 = {\rm log}_2(\mu_{\rm pt}/\mu) $ and $n_2 = {\rm log}_2(\mu_{\rm pt}/\mu_*)$.
The results are presented Figs.~\ref{fig:evol_mustar+}, \ref{fig:evol_mustar-}. Very dominant non-perturbative effects are clearly visible for the elements $(2,2)$, $(4,5)$, $(5,4)$ and $(5,5)$ of the operators $\{ \cQ_2^-, \cQ_3^- \}$ and $\{ \cQ_4^-, \cQ_5^- \}$.
Given the large deviation from the NLO-2/3PT running already seen in Fig.~\ref{fig:RUN-}, these results are not surprising and simply confirm the non-reliability of the NLO-2/3PT computation of the RG running at scales around $3 \, \rm GeV$. Notice that the scale interval where this comparison has been performed in ref.~\cite{Boyle:2017skn} is completely contained in our plots between the third and the fourth point which correspond to scales of 1.73 GeV and 3.46 GeV.
We remind the reader that a direct comparison between our results and those of ref.~\cite{Boyle:2017skn} is meaningless, the crucial differences being, among many others, the renormalisation scheme and the $\NF$-value. 
\newpage

\newpage

\newpage

\newpage

%% file: app_cutoff.tex
\section{One-loop cutoff effects in the step scaling function}
\label{sec:cutoff}

\begin{figure}[t!]
\begin{center}
\includegraphics[width=0.8\textwidth]{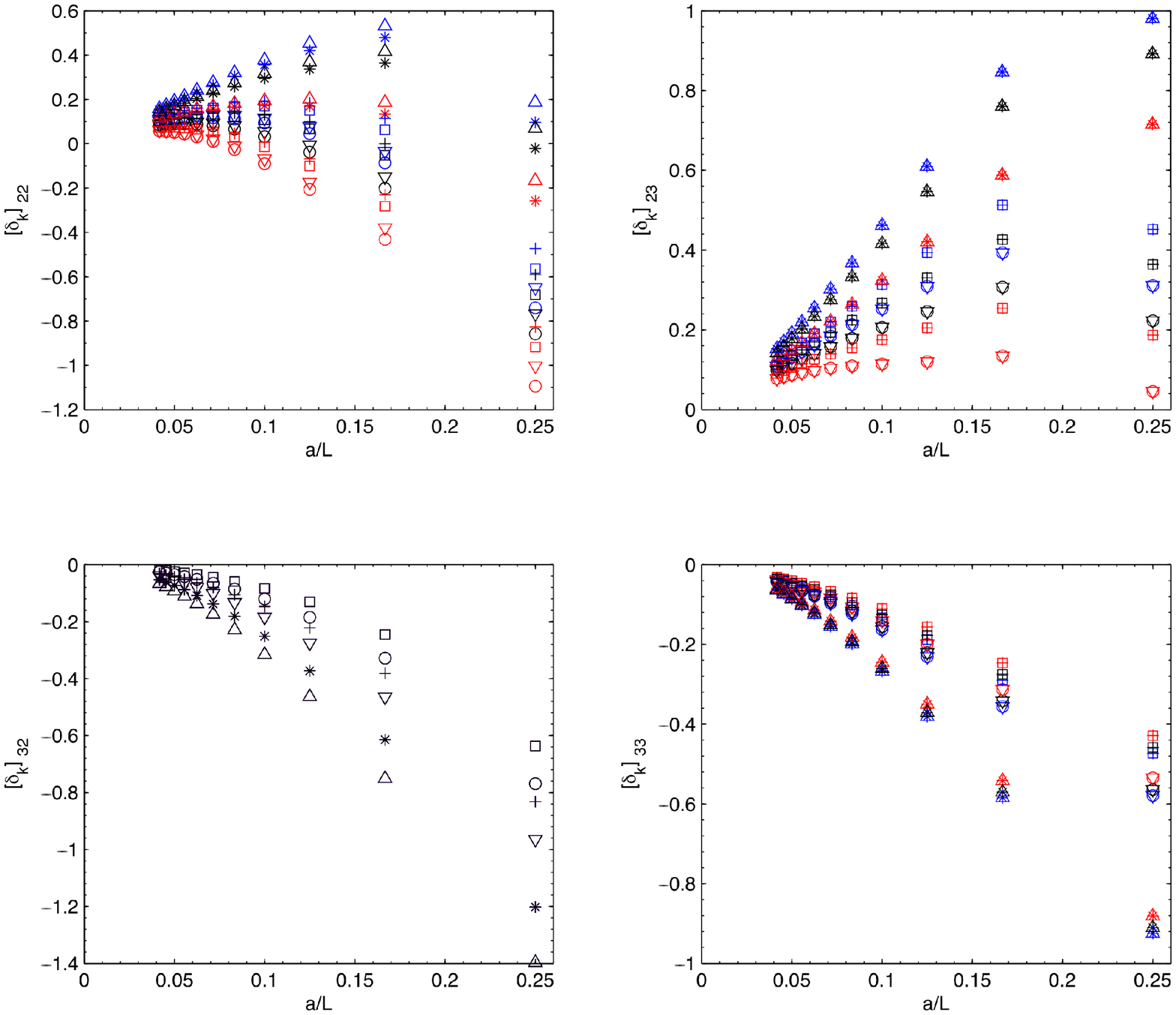}
\includegraphics[width=0.8\textwidth]{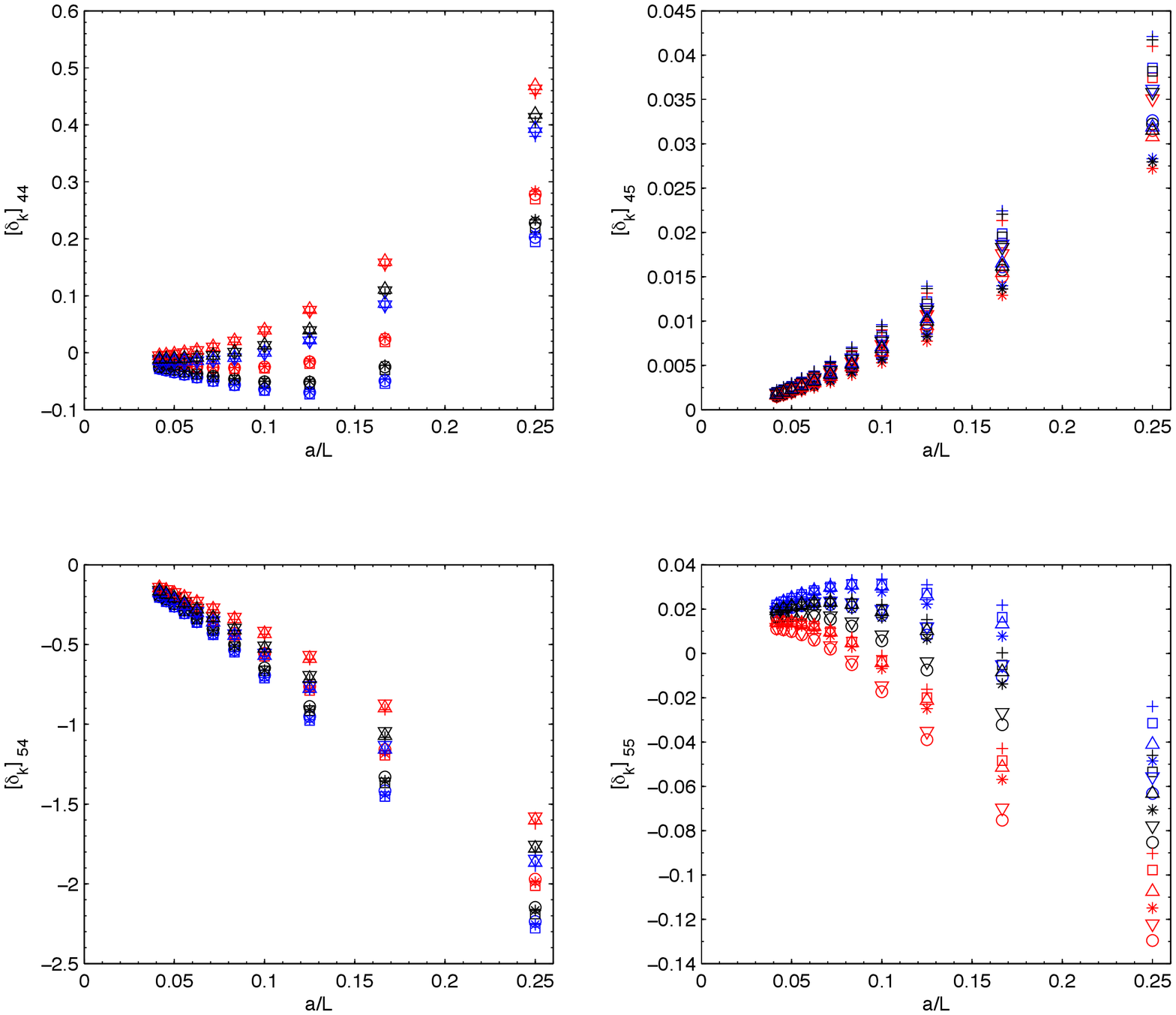}
\end{center}
\caption{Matrix elements of $\bdelta_k(L/a)$ with $c_{\rm sw}=1$ for the operator bases $\{ \cQ_2^+, \cQ_3^+ \}$ (top) and $\{ \cQ_4^+, \cQ_5^+ \}$ (bottom). Different colours distinguish the various choices of $\alpha$ and different symbols the various choices of $(s_1, s_2)$.}
\label{fig:cutoffcsw1p}
\end{figure}

\begin{figure}[t!]
\begin{center}
\includegraphics[width=0.8\textwidth]{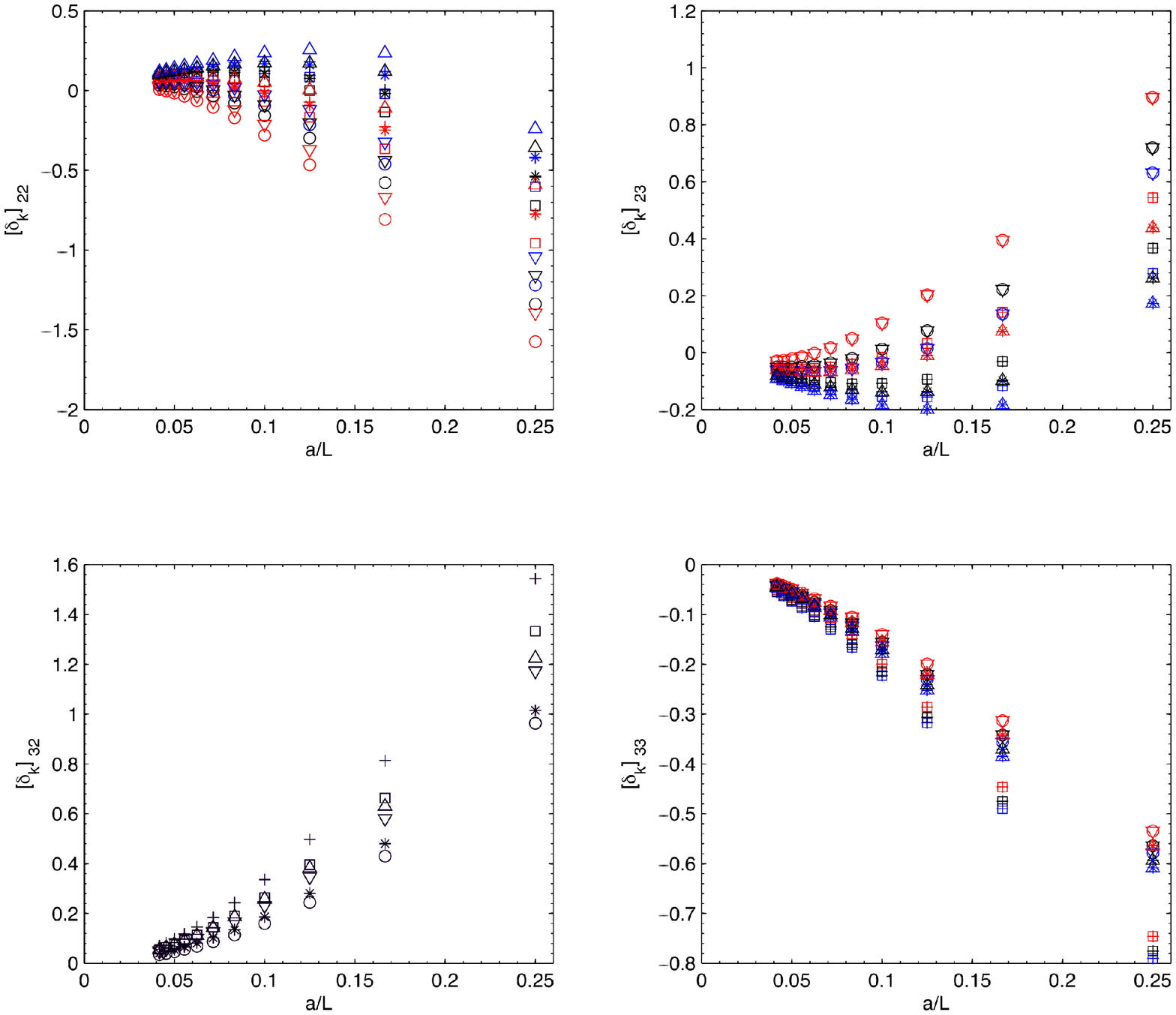}
\includegraphics[width=0.8\textwidth]{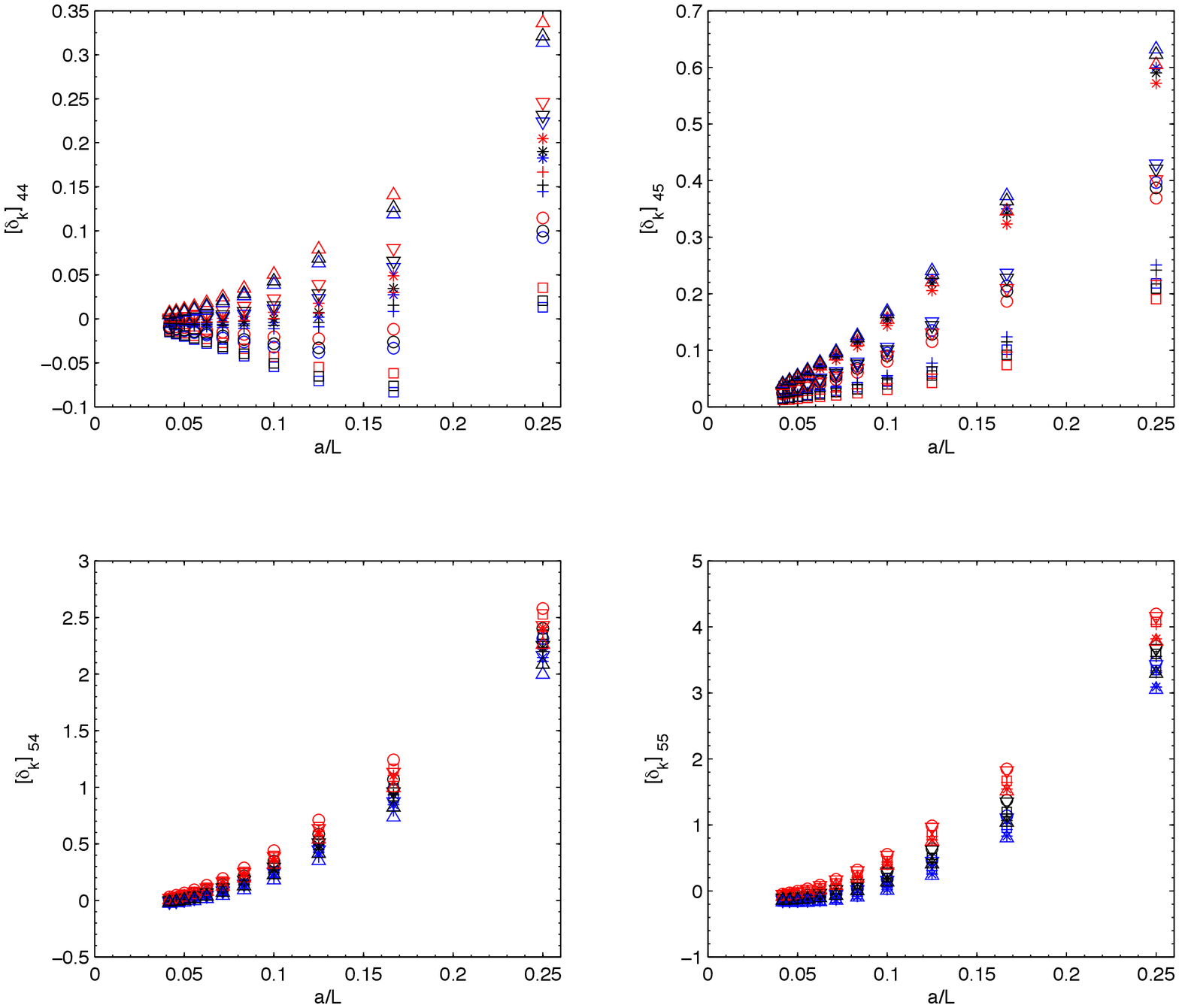}
\end{center}
\caption{Matrix elements of $\bdelta_k(L/a)$ with $c_{\rm sw}=1$ for the operator bases $\{ \cQ_2^-, \cQ_3^- \}$ (top) and $\{ \cQ_4^-, \cQ_5^- \}$ (bottom). Different colours distinguish the various choices of $\alpha$ and different symbols the various choices of $(s_1, s_2)$.}
\label{fig:cutoffcsw1m}
\end{figure}

\begin{figure}[t!]
\begin{center}
\includegraphics[width=0.8\textwidth]{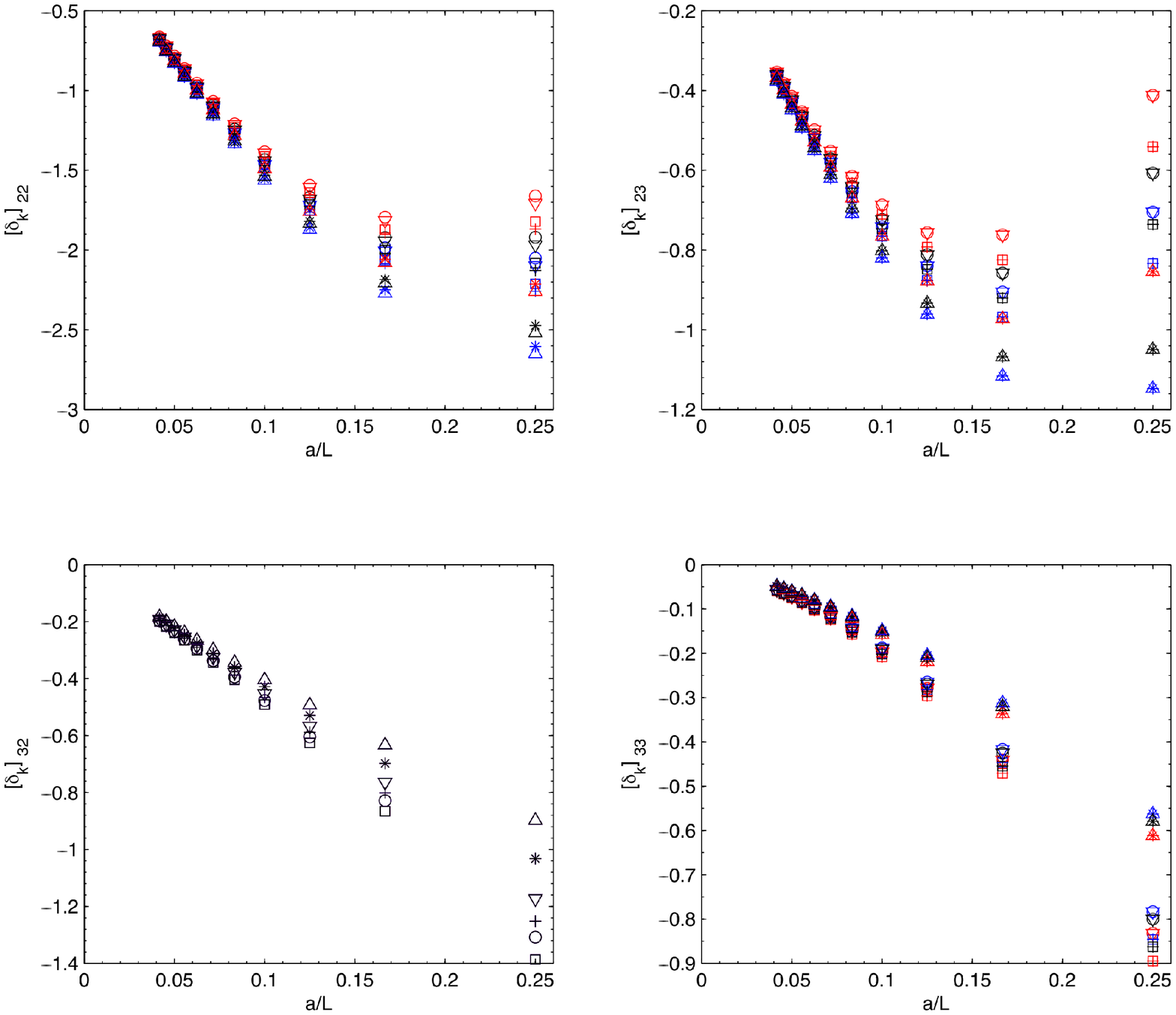}
\includegraphics[width=0.8\textwidth]{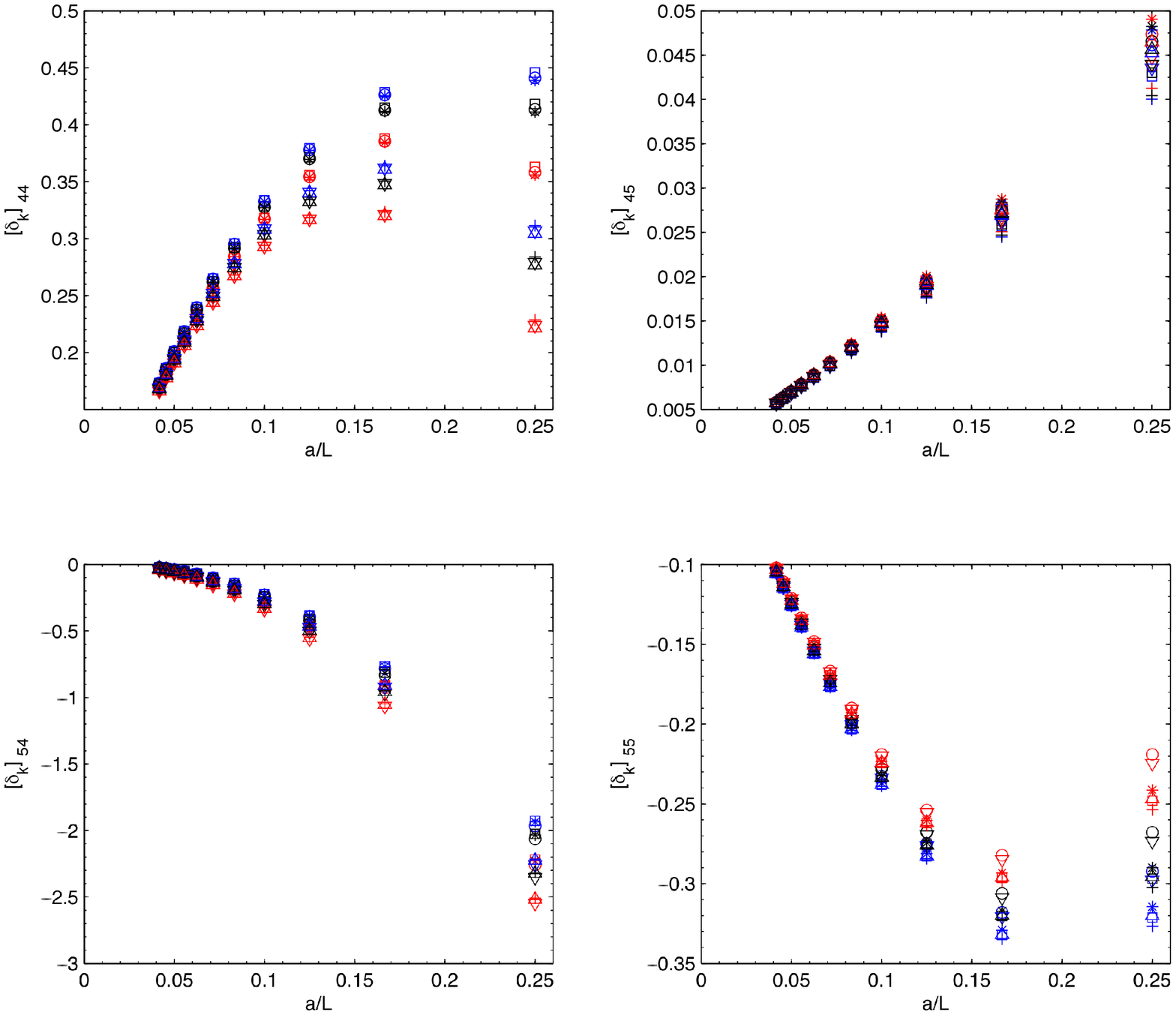}
\end{center}
\caption{Matrix elements of $\bdelta_k(L/a)$ with $c_{\rm sw}=0$ for the operator bases $\{ \cQ_2^+, \cQ_3^+ \}$ (top) and $\{ \cQ_4^+, \cQ_5^+ \}$ (bottom). Different colours distinguish the various choices of $\alpha$ and different symbols the various choices of $(s_1, s_2)$.}
\label{fig:cutoffcsw0p}
\end{figure}

\begin{figure}[t!]
\begin{center}
\includegraphics[width=0.8\textwidth]{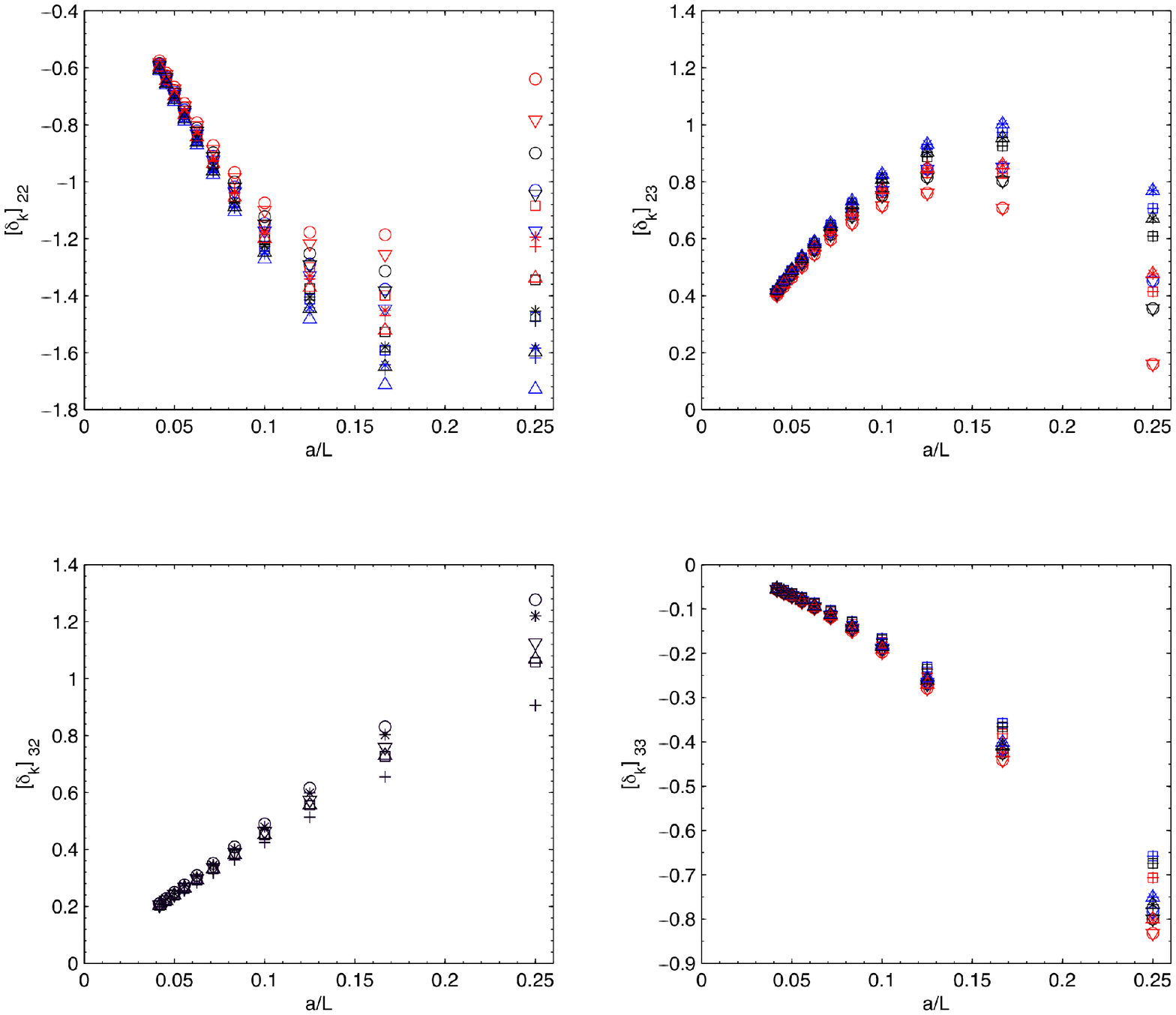}
\includegraphics[width=0.8\textwidth]{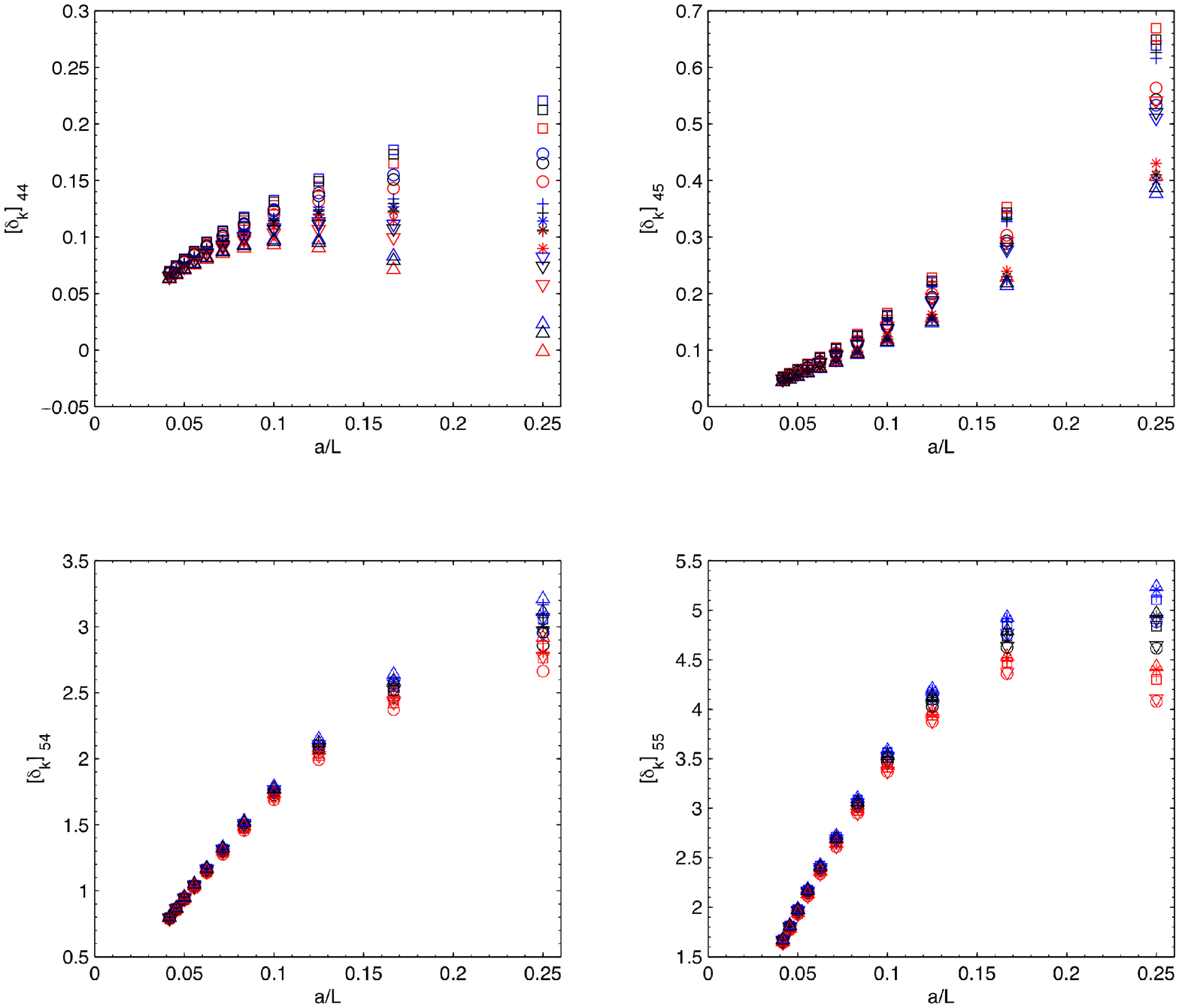}
\end{center}
\caption{Matrix elements of $\bdelta_k(L/a)$ with $c_{\rm sw}=0$ for the operator bases $\{ \cQ_2^-, \cQ_3^- \}$ (top) and $\{ \cQ_4^-, \cQ_5^- \}$ (bottom). Different colours distinguish the various choices of $\alpha$ and different symbols the various choices of $(s_1, s_2)$.}
\label{fig:cutoffcsw0m}
\end{figure}

In Tab.~\ref{tab:cutoff_all}  we gather numerical values for $\bdelta_k(L/a)$, defined in \req{eq:def-delta}.
We have calculated this quantity for a fermionic action with ($c_{\rm sw}=1$) and without ($c_{\rm sw}=0$) a Clover term.
These results are also
displayed in Figs.~\ref{fig:cutoffcsw1p},~\ref{fig:cutoffcsw1m},~\ref{fig:cutoffcsw0p}, and~\ref{fig:cutoffcsw0m} (the target scheme $\alpha=3/2$,
$(s_1,s_2)=(3,5)$ is plotted with a blue triangle). Notice that the element $(3,2)$ of $\bdelta_k$ is independent from $\alpha$ due to an accidental cancellation.
This is why all data-points in the corresponding figures are not in colour.
As expected, the Clover term has an important effect on the discretisation errors, which are significantly reduced when $c_{\rm sw}=1$.
The observed $\cO(a g_0^2)$ discretisation effects in Figs.~\ref{fig:cutoffcsw1p} and~\ref{fig:cutoffcsw1m} are only due to the unimproved operators, the action being tree-level improved.


\newpage

\newpage

\newpage

\begin{table}
\begin{scriptsize}
\begin{center}
\begin{tabular}{ccccc}
\toprule
$L/a$ & $[\delta_k]^{(+;c_{\rm sw}=0)}_{23}$ & $[\delta_k]^{(-;c_{\rm sw}=0)}_{23}$ & $[\delta_k]^{(+;c_{\rm sw}=1)}_{23}$ & $[\delta_k]^{(-;c_{\rm sw}=1)}_{23}$\\
\midrule
$4 $ & $\begin{pmatrix} -2.649644  & -1.146696 \\ -0.897521  & -0.562903 \end{pmatrix} $ & $\begin{pmatrix}-1.728316 & 0.769175\\ 1.068844 & -0.750880 \end{pmatrix} $  & $\begin{pmatrix} 0.187434  & 0.980468  \\  -1.397054  & -0.925080 \end{pmatrix} $ & $\begin{pmatrix} -0.239029 &  0.172755  \\ 1.224870 & -0.608259 \end{pmatrix} $\\
$6 $ & $\begin{pmatrix} -2.270147  & -1.115956\\  -0.633302  & -0.312431 \end{pmatrix} $ & $\begin{pmatrix}-1.712244 & 1.002450\\ 0.730167 & -0.401243 \end{pmatrix} $  & $\begin{pmatrix} 0.531248  & 0.846324  \\  -0.750504  & -0.584800 \end{pmatrix} $ & $\begin{pmatrix} 0.235172  & -0.183865  \\ 0.630602 & -0.384869 \end{pmatrix} $\\
$8 $ & $\begin{pmatrix}-1.869882  & -0.960863\\   -0.492422  & -0.204899 \end{pmatrix} $ & $\begin{pmatrix}-1.482077 & 0.931524\\ 0.556962 & -0.255909 \end{pmatrix} $  & $\begin{pmatrix} 0.453106  & 0.609325  \\  -0.463794  & -0.381264 \end{pmatrix} $ & $\begin{pmatrix} 0.255353  & -0.199539  \\ 0.382567 & -0.251823 \end{pmatrix} $\\
$10$ & $\begin{pmatrix}-1.562385 &  -0.819867\\   -0.404426  & -0.149062 \end{pmatrix} $ & $\begin{pmatrix}-1.271702 & 0.826113\\0.452405 & -0.182018 \end{pmatrix} $  & $\begin{pmatrix}  0.377981 &  0.462327 \\  -0.315184  & -0.267748 \end{pmatrix} $ & $\begin{pmatrix}  0.235681 &  -0.184042 \\ 0.258383 & -0.178388 \end{pmatrix} $\\
$12$ & $\begin{pmatrix}-1.333472 &  -0.707992\\   -0.343891  & -0.115961 \end{pmatrix} $  & $\begin{pmatrix}-1.105124 & 0.731935\\0.382266 & -0.138962 \end{pmatrix} $  & $\begin{pmatrix}  0.320224 &  0.367452 \\  -0.228612  & -0.199372 \end{pmatrix} $ & $\begin{pmatrix}  0.211516 &  -0.165182 \\ 0.187365 & -0.134313 \end{pmatrix} $\\
$14$ & $\begin{pmatrix}-1.159958 &  -0.620313\\   -0.299573  & -0.094449 \end{pmatrix} $  & $\begin{pmatrix}-0.974682 & 0.653858\\0.331822 & -0.111397 \end{pmatrix} $  & $\begin{pmatrix}  0.275829 &  0.302158 \\  -0.173803  & -0.155093 \end{pmatrix} $ & $\begin{pmatrix}  0.189124 &  -0.147796 \\ 0.142824 & -0.105725 \end{pmatrix} $\\
$16$ & $\begin{pmatrix}-1.024995 &  -0.550700\\   -0.265671  & -0.079507 \end{pmatrix} $  & $\begin{pmatrix}-0.871044 & 0.589677\\0.293718 & -0.092508 \end{pmatrix} $  & $\begin{pmatrix}  0.241149 &  0.254939 \\  -0.136903  & -0.124728 \end{pmatrix} $ & $\begin{pmatrix}  0.169756 &  -0.132789 \\ 0.112970 & -0.086034 \end{pmatrix} $\\
$18$ & $\begin{pmatrix}-0.917432 &  -0.494446\\   -0.238873  & -0.068599 \end{pmatrix} $  & $\begin{pmatrix}-0.787148 & 0.536519\\0.263869 & -0.078883 \end{pmatrix} $  & $\begin{pmatrix}  0.213561 &  0.219466 \\  -0.110860  & -0.102950 \end{pmatrix} $ & $\begin{pmatrix}  0.153325 &  -0.120063 \\ 0.091934 & -0.071829 \end{pmatrix} $\\
$20$ & $\begin{pmatrix}-0.829877 &  -0.448202\\   -0.217139  & -0.060320 \end{pmatrix} $  & $\begin{pmatrix}-0.718006 & 0.491978\\0.239821 & -0.068657 \end{pmatrix} $  & $\begin{pmatrix}  0.191226 &  0.191997 \\  -0.091780  & -0.086762 \end{pmatrix} $ & $\begin{pmatrix}  0.139416 &  -0.109285 \\ 0.076522 & -0.061199 \end{pmatrix} $\\
$22$ & $\begin{pmatrix}-0.757313 &  -0.409593\\   -0.199147  & -0.053840 \end{pmatrix} $  & $\begin{pmatrix}-0.660112 & 0.454212\\0.220008 & -0.060735 \end{pmatrix} $  & $\begin{pmatrix}  0.172850 &  0.170193 \\  -0.077372  & -0.074372 \end{pmatrix} $ & $\begin{pmatrix}  0.127590 &  -0.100114 \\ 0.064873 & -0.053007 \end{pmatrix} $\\
$24$ & $\begin{pmatrix}-0.696242 &  -0.376917\\   -0.183999  & -0.048639 \end{pmatrix} $  & $\begin{pmatrix}-0.610956 & 0.421828\\0.203386 & -0.054436 \end{pmatrix} $  & $\begin{pmatrix}  0.157511 &  0.152526 \\  -0.066218  & -0.064659 \end{pmatrix} $ & $\begin{pmatrix}  0.117464 &  -0.092254 \\ 0.055837 & -0.046540 \end{pmatrix} $ \\
\midrule
$L/a$ & $[\delta_k]^{(+;c_{\rm sw}=0)}_{45}$ & $[\delta_k]^{(-;c_{\rm sw}=0)}_{45}$ & $[\delta_k]^{(+;c_{\rm sw}=1)}_{45}$ & $[\delta_k]^{(-;c_{\rm sw}=1)}_{45}$\\
\toprule
$4 $ & $\begin{pmatrix} 0.304518  & 0.045248  \\ -2.225718 & -0.320010 \end{pmatrix}$ & $\begin{pmatrix} 0.023085 & 0.376880 \\ 3.211378 & 5.236303  \end{pmatrix}$ & $\begin{pmatrix} 0.393235    & 0.031881   \\ -1.865444  & -0.041070  \end{pmatrix}$   &  $\begin{pmatrix}   0.314036  & 0.632269  \\ 1.996419 & 3.051790 \end{pmatrix}$\\
$6 $ & $\begin{pmatrix} 0.360832  & 0.026906  \\ -0.907110 & -0.331808 \end{pmatrix}$ & $\begin{pmatrix} 0.083096 & 0.213740 \\ 2.639245 & 4.926161  \end{pmatrix}$ & $\begin{pmatrix} 0.086802    & 0.016542   \\ -1.156020  & 0.013281   \end{pmatrix}$   &  $\begin{pmatrix}   0.119120  & 0.373079  \\ 0.736291 & 0.802921 \end{pmatrix}$\\
$8 $ & $\begin{pmatrix} 0.340234  & 0.018988  \\ -0.466335 & -0.282722 \end{pmatrix}$ & $\begin{pmatrix} 0.097391 & 0.147874 \\ 2.149300 & 4.203452  \end{pmatrix}$ & $\begin{pmatrix} 0.022327    & 0.010258   \\ -0.776005  & 0.025993   \end{pmatrix}$   &  $\begin{pmatrix}   0.063469  & 0.240718  \\ 0.351217 & 0.234227 \end{pmatrix}$\\
$10$ & $\begin{pmatrix}  0.308242 &  0.014678 \\ -0.277130 & -0.237795 \end{pmatrix}$ & $\begin{pmatrix} 0.097575 & 0.113428 \\ 1.790539 & 3.583070  \end{pmatrix}$ & $\begin{pmatrix}  0.000572   & 0.007095  \\ -0.569481  & 0.030288    \end{pmatrix}$  &   $\begin{pmatrix}   0.039145 &  0.168897 \\ 0.180364 & 0.008998  \end{pmatrix}$\\
$12$ & $\begin{pmatrix}  0.278023 &  0.011975 \\ -0.180140 & -0.203058 \end{pmatrix}$ & $\begin{pmatrix} 0.093222 & 0.092219 \\ 1.527657 & 3.099067  \end{pmatrix}$ & $\begin{pmatrix}  -0.008433  & 0.005262 \\ -0.443270  & 0.030817     \end{pmatrix}$ &    $\begin{pmatrix}  0.026045 &  0.125824 \\ 0.092686 & -0.093092  \end{pmatrix}$\\
$14$ & $\begin{pmatrix}  0.251877 &  0.010121 \\ -0.124242 & -0.176481 \end{pmatrix}$ & $\begin{pmatrix} 0.087627 & 0.077803 \\ 1.329668 & 2.722092  \end{pmatrix}$ & $\begin{pmatrix}  -0.012424  & 0.004096 \\ -0.359273  & 0.029790     \end{pmatrix}$ &    $\begin{pmatrix}  0.018172 &  0.097923 \\ 0.043715 & -0.141307  \end{pmatrix}$\\
$16$ & $\begin{pmatrix}  0.229686 &  0.008772 \\ -0.089343 & -0.155782 \end{pmatrix}$ & $\begin{pmatrix} 0.081969 & 0.067351 \\ 1.176128 & 2.423499  \end{pmatrix}$ & $\begin{pmatrix}  -0.014147  & 0.003303 \\ -0.299932  & 0.028210     \end{pmatrix}$ &    $\begin{pmatrix}  0.013104 &  0.078774 \\ 0.014706 & -0.163650  \end{pmatrix}$\\
$18$ & $\begin{pmatrix}  0.210851 &  0.007745 \\ -0.066253 & -0.139304 \end{pmatrix}$ & $\begin{pmatrix} 0.076660 & 0.059420 \\ 1.053939 & 2.182368  \end{pmatrix}$ & $\begin{pmatrix}  -0.014760  & 0.002737 \\ -0.256116  & 0.026503     \end{pmatrix}$ &    $\begin{pmatrix}  0.009676 &  0.065031 \\ 0.003208 & -0.172802  \end{pmatrix}$\\
$20$ & $\begin{pmatrix}  0.194763 &  0.006937 \\ -0.050289 & -0.125917 \end{pmatrix}$ & $\begin{pmatrix} 0.071819 & 0.053193 \\ 0.954545 & 1.984093  \end{pmatrix}$ & $\begin{pmatrix}  -0.014807  & 0.002317 \\ -0.222634  & 0.024843     \end{pmatrix}$ &    $\begin{pmatrix}  0.007267 &  0.054810 \\ 0.014590 & -0.174959  \end{pmatrix}$\\
$22$ & $\begin{pmatrix}  0.180910 &  0.006285 \\ -0.038863 & -0.114844 \end{pmatrix}$ & $\begin{pmatrix} 0.067455 & 0.048172 \\ 0.872187 & 1.818430  \end{pmatrix}$ & $\begin{pmatrix}  -0.014560  & 0.001995 \\ -0.196335  & 0.023295     \end{pmatrix}$ &    $\begin{pmatrix}  0.005523 &  0.046986 \\ 0.021945 & -0.173333  \end{pmatrix}$\\
$24$ & $\begin{pmatrix}  0.168880 &  0.005748 \\ -0.030454 & -0.105542 \end{pmatrix}$ & $\begin{pmatrix} 0.063534 & 0.044036 \\ 0.802870 & 1.678076  \end{pmatrix}$ & $\begin{pmatrix}  -0.014162  & 0.001743 \\ -0.175208  & 0.021879     \end{pmatrix}$ &    $\begin{pmatrix}  0.004229 &  0.040850 \\ 0.026729 & -0.169666  \end{pmatrix}$\\
\bottomrule
\end{tabular}
\end{center}
\end{scriptsize}
\caption{Matrix elements of $\bdelta_k(L/a)$ for the SF scheme with $\alpha=3/2$ and $(s_1,s_2)=(3,5)$}
\label{tab:cutoff_all}
\end{table}